\documentclass[aps, pra, notitlepage, citeautoscript, twocoulumn,reprint, longbibliography,superscriptaddress]{revtex4-1}
\pdfoutput=1
\usepackage[noend]{algpseudocode}
\usepackage{algorithm}
\usepackage{natbib}
\usepackage[english]{babel}
\usepackage{letltxmacro}
\usepackage{latexsym}
\LetLtxMacro{\ORIGselectlanguage}{\selectlanguage}
\makeatletter
\DeclareRobustCommand{\selectlanguage}[1]{%
  \@ifundefined{alias@\string#1}
    {\ORIGselectlanguage{#1}}
    {\begingroup\edef\x{\endgroup
       \noexpand\ORIGselectlanguage{\@nameuse{alias@#1}}}\x}%
}
\newcommand{\definelanguagealias}[2]{%
  \@namedef{alias@#1}{#2}%
}
\makeatother
\definelanguagealias{en}{english}
\definelanguagealias{English}{english}
\usepackage{graphicx}
\usepackage{amsmath}
\usepackage{amsfonts}
\usepackage{amssymb}
\usepackage{bm}
\usepackage{color}
\usepackage[percent]{overpic}
\usepackage{soul} 
\usepackage{amssymb}
\usepackage{wasysym}
\usepackage{dsfont}
\usepackage{float}
\usepackage[mathscr]{euscript}
\usepackage{lipsum}
\usepackage{hyperref}
\hypersetup{
    pdfstartview={FitH},    
    colorlinks=true,       
    linkcolor=blue,          
    citecolor=blue,        
    filecolor=magenta,      
    urlcolor=blue           
}

\newcommand{\be}{\begin{equation}}
\newcommand{\ee}{\end{equation}}
\newcommand{\bea}{\begin{eqnarray}}
\newcommand{\eea}{\end{eqnarray}}

\usepackage{enumitem}

\renewcommand{\approx}{\simeq}

\renewcommand{\vec}[1]{\boldsymbol{#1}}

\newcommand{\equref}[1]{Eq.~(\ref{#1})}
\newcommand{\equsref}[2]{Eqs.~(\ref{#1}) and (\ref{#2})}
\newcommand{\secref}[1]{Sec.~\ref{#1}}
\newcommand{\figref}[1]{Fig.~\ref{#1}}
\newcommand{\refcite}[1]{Ref.~\onlinecite{#1}}

\newcommand{\tableref}[1]{Table~\ref{#1}}
\newcommand{\appref}[1]{Appendix~\ref{#1}}

\usepackage{graphicx}
\usepackage[colorinlistoftodos]{todonotes}
\usepackage{verbatim}
\usepackage[normalem]{ulem}
\usepackage{braket}

\newcommand{\sandwich}[3]{\mbox{$ \langle #1 | #2 | #3 \rangle $}}
\newcommand{\ev}[1]{\mbox{$ \langle #1 \rangle $}}
\newcommand{\abs}[1]{\lvert#1\rvert}

\newcommand{\commu}[2]{\left[#1, #2\right]}
\newcommand{\qtyset}[1]{\{#1\}}

\begin{document}

\title{Classifying topological neural network quantum states via diffusion maps}

\author{Yanting Teng}
\affiliation{Department of Physics, Harvard University, Cambridge MA 02138, USA}

\author{Subir Sachdev}
\affiliation{Department of Physics, Harvard University, Cambridge MA 02138, USA}

\author{Mathias S. Scheurer}
\affiliation{Institut f\"ur Theoretische Physik, Universit\"at Innsbruck, A-6020 Innsbruck, Austria}

\begin{abstract}
We discuss and demonstrate an unsupervised machine-learning procedure to detect topological order in quantum many-body systems. Using a restricted Boltzmann machine to define a variational ansatz for the low-energy spectrum, we sample wave functions with probability decaying exponentially with their variational energy; this defines our training dataset that we use as input to a diffusion map scheme. The diffusion map provides a low-dimensional embedding of the wave functions, revealing the presence or absence of superselection sectors and, thus, topological order. We show that for the diffusion map, the required similarity measure of quantum states can be defined in terms of the network parameters, allowing for an efficient evaluation within polynomial time. However, possible ``gauge redundancies’’ have to be carefully taken into account.
As an explicit example, we apply the method to the toric code.
\end{abstract}
\maketitle

\hypersetup{linkcolor=blue}

\section{Introduction}
In the last few years, machine learning (ML) techniques have been very actively studied as novel tools in many-body physics \cite{mehtaHighbiasLowvarianceIntroduction2019, carleoMachineLearningPhysical2019,2019PhT....72c..48D,RBMReview,CarrasquillaReview, carrasquillaHowUseNeural2021,dawidModernApplicationsMachine2022}. A variety of valuable applications of ML has been established, such as ML-based variational ans\"atze for many-body wave functions, application of ML to experimental data to extract information about the underlying physics, ML methods for more efficient Monte-Carlo sampling , and employment of ML to detect phase transitions, to name a few. Regarding the latter type of applications, a particular focus has recently been on topological phase transitions \cite{MelkoPT,PhysRevLett.120.066401,zhangQuantumLoopTopography2017,PhysRevB.97.045207,rodriguez-nievaIdentifyingTopologicalOrder2019,10.21468/SciPostPhys.11.2.043,TSENG2022105134,greplovaUnsupervisedIdentificationTopological2020,PhysRevB.96.245119,huangProvablyEfficientMachine2021,sadouneUnsupervisedInterpretableLearning2022,2020arXiv201200783C,sehayekPersistentHomologyMathbbZ2022,Kaeming_2021,Ho_2021,2022arXiv220914551L,PhysRevB.105.205139,PhysRevB.105.195115,PhysRevB.104.024506,PhysRevB.104.165108,2022arXiv220113306J,10.21468/SciPostPhys.11.3.073,2022arXiv220209281T,PhysRevB.104.235146}. This is motivated by the challenges associated with capturing topological phase transitions: by definition, topological features are related to the global connectivity of the dataset rather than local similarity of samples. Therefore, unless the dataset is sufficiently simple such that topologically connected pairs of samples also happen to be locally similar or features are used as input data that are closely related to the underlying topological invariant, the topological structure is hard to capture reliably with many standard ML techniques \cite{PhysRevB.97.045207,rodriguez-nievaIdentifyingTopologicalOrder2019}. 

In this regard, the ML approach proposed in \refcite{rodriguez-nievaIdentifyingTopologicalOrder2019}, which is based on diffusion maps (DM) \cite{coifmanGeometricDiffusionsTool2005,nadlerDiffusionMapsSpectral2005, nadlerDiffusionMapsSpectral2006, coifmanDiffusionMaps2006}, is a particularly promising route to learn topological phase transitions; it allows to embed high-dimensional data in a low-dimensional subspace such that pairs of samples that are smoothly connected in the dataset will be mapped close to each other, while disconnected pairs will be mapped to distant points. As such, the method captures the central notion of topology. In combination with the fact that it is unsupervised and thus does not require \textit{a priori} knowledge of the underlying topological invariants, it is ideally suited for the task of topological phase classification. As a result, there have been many recent efforts applying this approach to a variety of problems, such as different symmetry-protected, including non-Hermitian, topological systems \cite{scheurerUnsupervisedMachineLearning2020b,longUnsupervisedManifoldClustering2020,PhysRevLett.126.240402,2022npjQI...8..116Y,cheTopologicalQuantumPhase2020,scheurerUnsupervisedMachineLearning2020b,kuoUnsupervisedLearningSymmetry2021,longUnsupervisedManifoldClustering2020}, experimental data \cite{lustigIdentifyingTopologicalPhase2020,2022npjQI...8..116Y}, many-body localized states \cite{lidiakUnsupervisedMachineLearning2020}, and dynamics \cite{gyawaliRevealingMicrocanonicalPhase2022}; extensions based on combining DM with path finding \cite{scheurerUnsupervisedMachineLearning2020b} as well as with quantum computing schemes \cite{sornsaengQuantumDiffusionMap2021} for speed-up have also been studied. 

As alluded to above, another very actively pursued application of ML in physics are neural network quantum states: as proposed in \refcite{carleoSolvingQuantumManybody2017}, neural networks can be used to efficiently parameterize and, in many cases, optimize variational descriptions of wave functions of quantum many-body systems \cite{gaoEfficientRepresentationQuantum2017, carleoConstructingExactRepresentations2018, luEfficientRepresentationTopologically2019a, sharirNeuralTensorContractions2021,chenEquivalenceRestrictedBoltzmann2018,nomuraInvestigatingNetworkParameters2022,dengQuantumEntanglementNeural2017, jiaEntanglementAreaLaw2020, chengInformationPerspectiveProbabilistic2017,QSTNatPhys}. In particular, restricted Boltzmann machines (RBMs) \cite{RBMReview} represent a very popular neural-network structure in this context. For instance, the ground states of the toric code model \cite{kitaevFaulttolerantQuantumComputation2003} can be exactly expressed with a \textit{local} RBM ansatz \cite{dengMachineLearningTopological2017}, i.e., where only neighboring spins are connected to the same hidden neurons. When additional non-local extensions to the RBM ansatz of \refcite{dengMachineLearningTopological2017} are added, this has been shown to also provide a very accurate variational description of the toric code in the presence of a magnetic field \cite{valentiCorrelationEnhancedNeuralNetworks2021}.

\begin{figure*}[tb]
    \centering
    \includegraphics[width=0.9\linewidth]{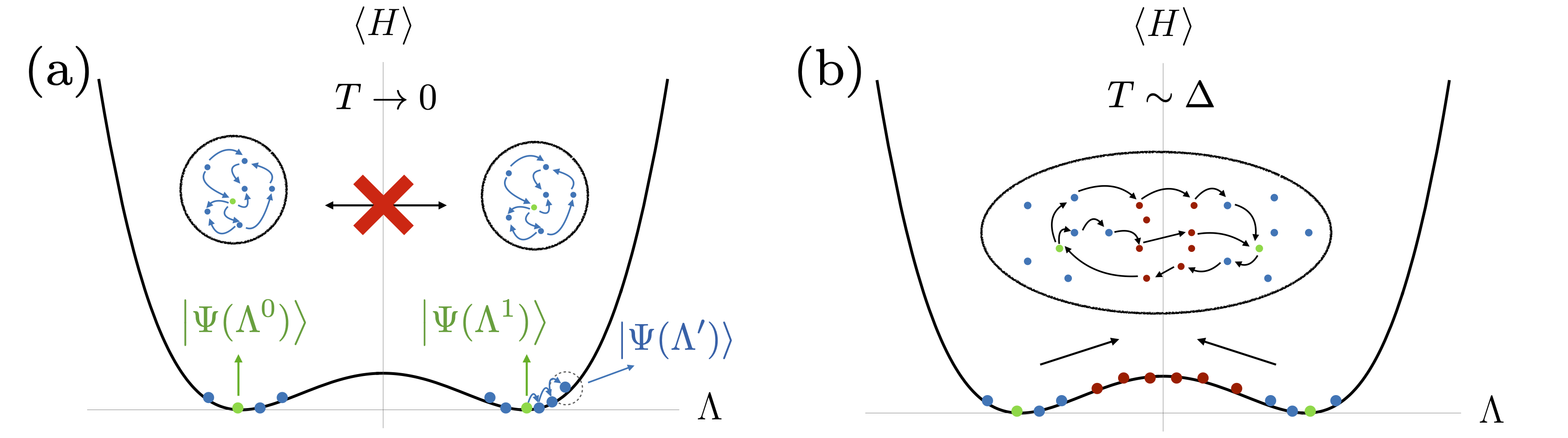}
    \caption{(a) An illustration of a ``low-energy'' ensemble. Two (or more) initial states, $\ket{\Psi(\Lambda^0)}$ and $\ket{\Psi(\Lambda^1)}$, from two distinct topological sectors are chosen as ``seeds'' (green dots). The dots denote the dataset (later fed into the DM), which are a set of quantum states labeled by network parameters $\Lambda$. This dataset is generated using the procedure outlined in \secref{DataGeneration} and Algorithm.~\ref{alg:data_generation}, where the next state $\Lambda^\prime$ (blue dots at each arrow) is proposed by a random local perturbation and accepted with probability based on the energy expectation $\ev{H}_{\Lambda^\prime}$. In the small-$T$ regime, the full dataset is not inter-connected by such local perturbations and cluster among each topological sectors (at left and right valley). (b) An illustration of a ``high-energy'' ensemble. The states are generated using the same algorithm as before, however with a large hyperparameter $T$ (compared to the energy gap $\Delta$). In this regime, the dataset include some of the low-energy states (blue dots), but also some high-energy states (red dots). Because the high-energy states are agnostic of the low-energy topological sectors, there exist paths (denoted by arrows among dots in the elliptical blob) such that the two initial seeds from distinct topological sectors effectively ``diffuse'' and form one connected cluster.}
    \label{fig:schematics}
\end{figure*}

In this work, we combine the DM approach of \refcite{rodriguez-nievaIdentifyingTopologicalOrder2019} with neural network quantum states with the goal of capturing topological order in an unsupervised way in interacting quantum many-body systems. We use a local network ansatz, with parameters $\Lambda$, as a variational description for the wave functions $\ket{\Psi(\Lambda)}$ of the low-energy subspace of a system with Hamiltonian $\hat{\mathcal{H}}$. While we also briefly mention other possible ways of generating ensembles of states, we primarily focus on an energetic principle: we sample wavefunctions such that the probability of $\ket{\Psi(\lambda)}$ is proportional to $\exp(- \braket{\hat{\mathcal{H}}}_\Lambda/T)$ where $\braket{\hat{\mathcal{H}}}_\Lambda=\bra{\Psi(\Lambda)}{\hat{\mathcal{H}}}\ket{\Psi(\Lambda)}$. As illustrated in \figref{fig:schematics}(a), the presence of superselection sectors in the low-energy spectrum of $\hat{\mathcal{H}}$ implies that the ensemble of states decays into disconnected subsets of states for sufficiently small $T$ (at least at fixed finite system size); these can be extracted, without need of prior labels, with dimensional reduction via DM (and subsequent $k$-means clustering), and thus allow to identify topological order. For sufficiently large $T$, more and more high-energy states are included and all sectors are connected, see \figref{fig:schematics}(b), as can also be readily revealed via DM-based embedding of the states.

Importantly, DM is a kernel technique in the sense that the input data $x_l$ (in our case the states $\ket{\Psi(\Lambda_l)}$) does not directly enter as a high-dimensional vector but only via a similarity measure $S(x_l,x_{l'})$, comparing how ``similar’’ two samples $l$ and $l’$ are. In the context of applying DM to the problem of topological classification, it defines what a smooth deformation (``homotopy'') of samples is. We discuss two possible such measures. The first one is just the quantum mechanical overlap, $S_{\rm q}(\Lambda_l, \Lambda_{l^\prime}) = \abs{\braket{\Psi(\Lambda_l)|\Psi(\Lambda_{l'})}}^2$, of the wave functions. Although conceptually straightforward, its evaluation is computationally costly on a classical computer as it requires importance sampling. The local nature of our network ansatz allows us to also construct an alternative similarity measure that is expressed as a simple function of the network parameters $\Lambda_l$ and $\Lambda_{l’}$ describing the two states to be compared. This can, however, lead to subtleties associated with the fact that two states with different $\Lambda$ can correspond to the same wave functions (modulo global phase). We discuss how these ``gauge redundancies’’ can be efficiently circumvented for generic states. 

We illustrate these aspects and explicitly demonstrate the success of this approach using the toric code \cite{kitaevFaulttolerantQuantumComputation2003}, a prototype model for topological order which has also been previously studied with other ML techniques with different focus \cite{dengMachineLearningTopological2017,greplovaUnsupervisedIdentificationTopological2020,PhysRevB.96.245119,huangProvablyEfficientMachine2021,sadouneUnsupervisedInterpretableLearning2022,valentiCorrelationEnhancedNeuralNetworks2021,2022arXiv220914302C}. We show that the DM algorithm learns the underlying loop operators wrapping around the torus without prior knowledge; at low $T$, this leads to four clusters corresponding to the four ground states. At larger $T$, these clusters start to merge, as expected. Interestingly, the DM still uncovers the underlying structure of the dataset related to the expectation value of the loop operators. Finally, we also show that applying a magnetic field leads to the disappearance of clusters in the DM, capturing the transition from topological order to the 
confined phase.

The remainder of the paper is organized as follows. In \secref{GeneralAlgorithm}, we describe our ML approach in general terms, including the local network quantum state description we use, the ensemble generation, a brief review of the DM scheme of \refcite{rodriguez-nievaIdentifyingTopologicalOrder2019}, and the similarity measure in terms of neural network parameters. Using the toric code model as an example, all of these general aspects are then discussed in detail and illustrated in \secref{sec:ToricCodeExample}. Finally, explicit numerical results can be found in \secref{NumericalResults} and a conclusion is provided in \secref{Conclusion}.

\section{General Algorithm}\label{GeneralAlgorithm}
\begin{figure*}[t!]
    \centering
    \includegraphics[width=0.7\linewidth]{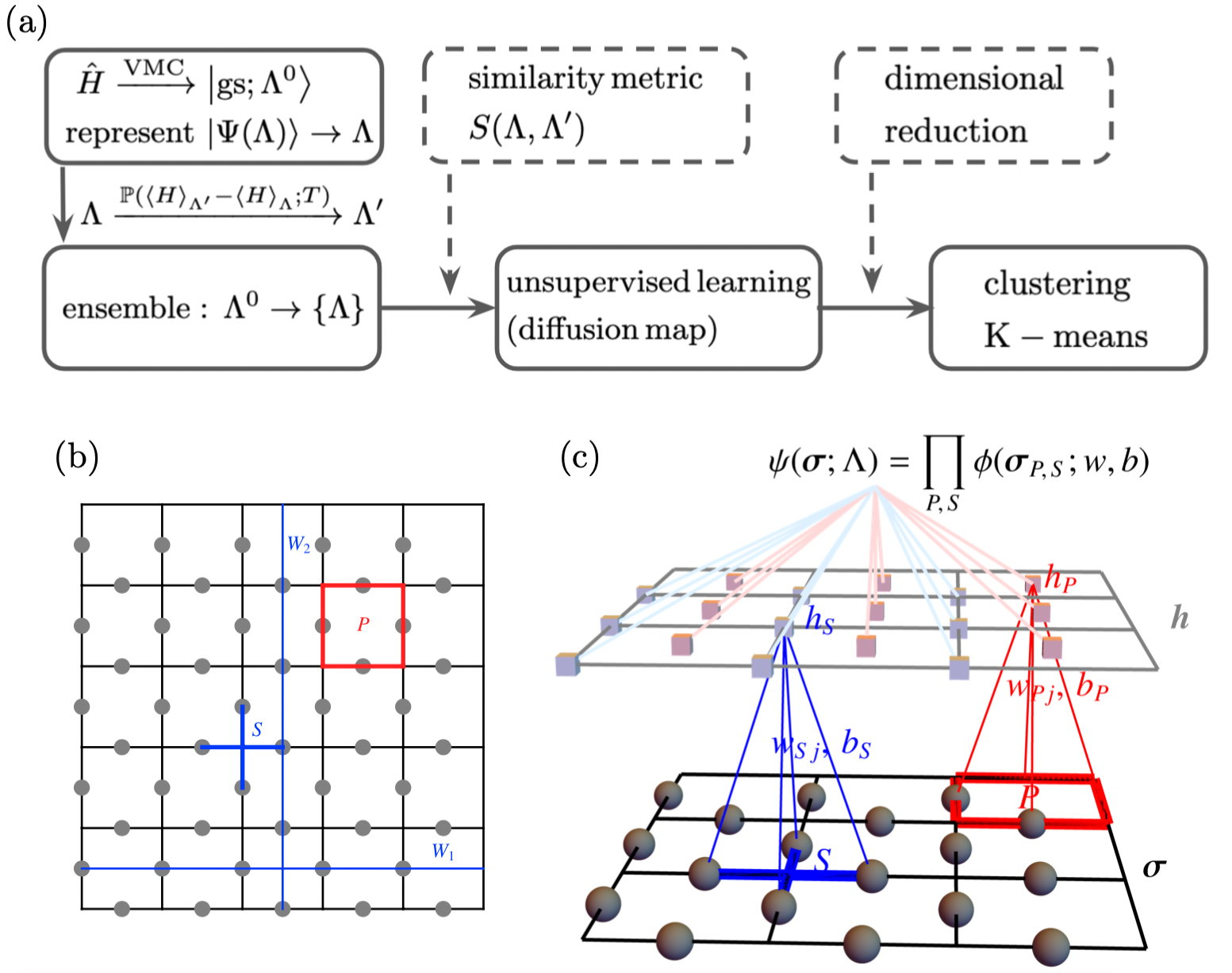}
    \caption{(a) Overview of the ML algorithm applied in this work: the ``seeds" $\{\Lambda^0\}$ are computed using variational Monte Carlo (see \appref{app:vmc}), a Markov-chain algorithm is used to generate the network parameter ensemble dataset (\secref{DataGeneration}), then a similarity metric is used for the definition of kernels in the DM method (\secref{sec:DM} and \secref{SimilarityMeasure}), and finally $k$-means is applied to the low-dimensional embedding in the subspace provided by the dominant DM eigenvector components.  (b) The square lattice geometry for the toric code model, where the qubits $\hat{s}_i$ are defined on the links of the lattice (grey dots). The Hamiltonian [given in \equref{eq:ham_tc}] is written in terms of the operators $\hat{\mathcal{P}}_P$ (supported by spins on plaquette $P$ denoted by the red square) and star $\hat{\mathcal{S}}_S$ (supported by spins on star $S$ denoted by the blue links). The two blue lines along $x(y)$ directions denote the Wilson loop operators $\hat{W}_{1, \bar{x}} (\hat{W}_{2,\bar{y}})$ along the straight paths $\bar{x}(\bar{y})$. (c) An illustration of the quasi-local ansatz in Eq.~\eqref{eq:rbm}. The ansatz is a product over local function $\phi$ of spins in plaquette (or star), which depends on parameters $\{w_{Xj}, b_X\}$ for $X=P(S)$ being plaquette (or star).}
    \label{fig:alg_ansatz}
\end{figure*}

Here, we first present and discuss our algorithm [see \figref{fig:alg_ansatz}(a)] in general terms before illustrating it using the toric code as an example in the subsequent sections. 
Consider a system of $N$ qubits or spins, with associated operators  $\{\hat{\boldsymbol{s}}\}=\{\hat{\boldsymbol{s}}_{i}, i=1, \cdots, N\}$, $\hat{\vec{s}}_i=(\hat{s}_i^x,\hat{s}_i^y,\hat{s}_i^z)$, and interactions governed by a local, gapped Hamiltonian $\hat{\mathcal{H}}=\mathcal{H}(\{\hat{\boldsymbol{s}}\})$. We represent the states $\ket{\Psi(\Lambda)}$ of this system using neural network quantum states~\cite{carleoSolvingQuantumManybody2017},
\begin{align}
    \ket{\Psi(\Lambda)} = \sum_{\boldsymbol{\sigma}} \psi(\boldsymbol{\sigma};\,\Lambda) \ket{\boldsymbol{\sigma}}, \label{WavefunctionAnsatz}
\end{align}
where $\boldsymbol{\sigma}$\,$=$\,$\{\sigma_1, \sigma_2, ..., \sigma_N \rvert \sigma_i= \pm 1\}$ enumerates configurations of the physical spin variables in a local computational basis (e.g.~$s^z$-basis) and $\Lambda$ is the set of parameters that the network $\psi$ depends on to output the wavefunction amplitude $\psi(\boldsymbol{\sigma};\,\Lambda)$\,$=$\,$\braket{\boldsymbol{\sigma}|\Psi(\Lambda)}$ for configuration $\ket{\boldsymbol{\sigma}}$. Because the physical Hilbert space scales exponentially with the system size, there is a trade-off between the expressivity versus efficiency when choosing a network architecture (or ansatz) $\psi$, so that the weights $\Lambda$ can approximate the state $\ket{\Psi(\Lambda)}$ to a reasonable degree and can at the same time be an efficient representation (with minimal number of parameters $\Lambda$ that scale as a polynomial in $N$). 
To reach the ground state or, more generally, the relevant low-energy sector of the Hamiltonian $\hat{\mathcal{H}}$ for the low-temperature physics, we minimize the energy in the variational subspace defined by \equref{WavefunctionAnsatz} using gradient descent with a learning rate $\lambda$,
\begin{align}
    \Lambda \rightarrow \Lambda - \lambda\, \partial_{\Lambda} \braket{\hat{\mathcal{H}}}_\Lambda, \quad \braket{\hat{\mathcal{H}}}_\Lambda=\bra{\Psi(\Lambda)}{\hat{\mathcal{H}}}\ket{\Psi(\Lambda)}. \label{GradientDescent}
\end{align}
Here, the quantum mechanical expectation value $\braket{\hat{\mathcal{H}}}_\Lambda$ is evaluated using importance sampling (see \appref{app:vmc}).

While there are exponentially many states in the Hilbert space, the low-energy sector of a local Hamiltonian is expected to occupy a small subspace where states obey area law entanglement~\cite{hastingsAreaLawOnedimensional2007, verstraeteCriticalityAreaLaw2006} whereas a typical state obeys volume law~\cite{wolfAreaLawsQuantum2008, eisertColloquiumAreaLaws2010}. Motivated by these considerations, we consider a class of networks that naturally describe quantum states that obey area-law entanglement. Pictorially, in such networks, the connections from the hidden neurons (representing the weights $\Lambda$) to the physical spins are \textit{quasi-local}~\cite{dengQuantumEntanglementNeural2017, chengInformationPerspectiveProbabilistic2017, chenEquivalenceRestrictedBoltzmann2018, jiaEntanglementAreaLaw2020}. In that case, it holds 
\begin{align}
     \psi(\boldsymbol{\sigma}, \Lambda) = \phi_1(\boldsymbol{\sigma}_1, \Lambda_1) \times \phi_2(\boldsymbol{\sigma}_2, \Lambda_2) \times \cdots, \label{LocalAnsatz}
\end{align}
where $\boldsymbol{\sigma}_{\j}$\,$=$\,$\{\sigma_k\}_{k\in \j}$ denote (overlapping) subsets of neighboring spins with $\cup_{{\j}} \boldsymbol{\sigma}_{\j} = \boldsymbol{\sigma}$ and $\Lambda_{\j}$ are the subsets of the network parameters (weights and biases) that are connected to the physical spins in $\j$.

\begin{algorithm}[H]
    \begin{algorithmic}
    \Procedure{}{$\{\Lambda\}_{n=1}^N$}
      \State init: optimized parameters $\Lambda$
      \For{$k$ independent times}:
            \For{$n$ sampling steps}:
              \State Propose new parameter $\Lambda_p = f(\Lambda_t)$
              \State Accept with probability determined by energy $\braket{\hat{\mathcal{H}}}_\Lambda$ and parameter $T$:
              \State $\Lambda_{t+1} = \mathbb{P}_{\rm accept}(\Lambda^\prime | \Lambda; T)$
            \EndFor
      \EndFor
      \State \textbf{return} the last $m$ states for each $k$: $\{\Lambda_i| i=n-m, ..., n\}_k$
    \EndProcedure
    \caption{Ensemble generation}
    \label{alg:data_generation}
    \end{algorithmic}
\end{algorithm}

\subsection{Dataset: network parameter ensembles}\label{DataGeneration}  
The dataset we use for unsupervised detection of topological order consists of an ensemble of wavefunctions $\{\ket{\Psi(\Lambda)}\}_{l}$, parameterized by the set of network parameters $\{\Lambda\}_{l}$. While, depending on the precise application, other choices are conceivable, we generate this ensemble such that the relative occurrence of a state $\ket{\Psi(\Lambda)}$ is given by $\rho_T(\Lambda)=\exp(- \braket{\hat{\mathcal{H}}}_\Lambda/T)/Z$, with appropriate normalization factor $Z$. As such, a small value of the ``temperature-like" hyperparameter $T$ corresponds to a ``low-energy'' ensemble while large $T$ parametrize ``high-energy'' ensembles.  

In practice, to generate this ensemble, we here first optimize the parameters $\Lambda$ via \equref{GradientDescent} to obtain wavefunctions with lowest energy expectation values. As \equref{WavefunctionAnsatz} does not contain all possible states, this will, in general, only yield approximations to the exact low-energy eigenstates of $\hat{\mathcal{H}}$. However, as long as it is able to capture all superselection sectors of the system as well as (a subset of) higher energy states connecting these sectors, \equref{WavefunctionAnsatz} will be sufficient for our purpose of detecting topological order or the absence thereof. We perform this optimization several times, $\Lambda \rightarrow \Lambda_l^0$, with different initial conditions, to obtain several ``seeds'', $\Lambda_l^0$; this is done to make sure we have a low-energy representative of all superselection sectors. Ideally the dataset is sampled directly from the the target probability distribution $\rho_T$, if for instance, one has access to an experimental system at finite temperature. Here, we adopt a Markov-chain-inspired procedure for generating the ensemble based on $\rho_T$ for each of these seeds. Specifically, starting from a state $\Lambda$, we propose updates on a randomly chosen local block of parameters connected to the spins at sites $\j$,
\begin{align}\label{eq:lambda_prime}
    \Lambda\,\rightarrow\,\Lambda^{\prime} =  \{\Lambda_{1}, \Lambda_{2}, \cdots, u(\Lambda_{\j}), \cdots, \Lambda_{N} \},
\end{align}
where the update $u$ only depends on $\Lambda_{\j}$. The proposed parameter $\Lambda^\prime$ given the current parameter $\Lambda$ is accepted with probability
\begin{align}\label{eq:p_accept_energy}
    \mathbb{P}_{\rm accept}(\Lambda^\prime | \Lambda; T)  = \min\Bigl(1,\, e^{ - \frac{\ev{\hat{\mathcal{H}}}_{\Lambda^\prime} - \ev{\hat{\mathcal{H}}}_{\Lambda}}{T}} \Bigr).
\end{align}
This means that if the proposed state $\Psi(\Lambda^\prime)$ has a lower energy expectation value than $\Psi(\Lambda)$, then the proposal will be accepted; otherwise, it will be accepted with a probability determined by the Boltzmann factor. The entire ensemble generation procedure is summarized in Algorithm~\ref{alg:data_generation}.

\subsection{Diffusion map}\label{sec:DM}
As proposed in \refcite{rodriguez-nievaIdentifyingTopologicalOrder2019}, DM is ideally suited as an unsupervised ML algorithm to identify the presence and number of superselection sectors in a collection of states, such as $\{\ket{\Psi(\Lambda)}\}_{l}$ defined above. To briefly review the key idea of the DM algorithm~\cite{coifmanGeometricDiffusionsTool2005,nadlerDiffusionMapsSpectral2005, nadlerDiffusionMapsSpectral2006, coifmanDiffusionMaps2006} and introduce notation, assume we are given a dataset $X = \{x_{l} |l = 1, 2, ..., M\}$, consisting of $M$ samples $x_{l}$. Below we will consider the cases $x_l = \Lambda_l$ and $x_l= \ket{\Psi(\Lambda_l)}$; in the first case, the samples are the network parameters parametrizing the wavefunction and, in the second, the samples are the wavefunctions themselves. 

To understand DM intuitively, let us define a diffusion process among states $x_{l} \in X$. The probability of state $x_{l}$ transitioning to $x_{l^\prime}$ is defined by the Markov transition matrix element $p_{l, l^{\prime}}$. To construct $p_{l, l^{\prime}}$, we introduce a symmetric and positive-definite kernel $k_{\epsilon}(x_l, x_{l'})$ between states $x_l$ and $x_{l^{\prime}}$. Then the transition probability matrix $p_{l, l^{\prime}}$ is defined as
\begin{align}\label{eq:transition_mat}
    & p_{l, l^{\prime}} = \frac{k_\epsilon(x_l, x_{l^{\prime}} )}{z_l}, \quad z_l = \sum_{l^{\prime}} k_\epsilon(x_l, x_{l^{\prime}}), 
\end{align}
where the factor $z_l$ ensures probability conservation, $\sum_{l^{\prime}}$\,$p_{l, l^{\prime}}$\,$=$\,$1$\,$\forall l$.
Then spectral analysis on the transition probability matrix leads to information on the \textit{global} connectivity of the dataset $X$, which, in our context of $X$ containing low-energy states, allows to identify superselection sectors and, thus, topological order \cite{rodriguez-nievaIdentifyingTopologicalOrder2019}. 
To quantify how strongly two samples $x_l$ and $x_{l'}$ are connected, one introduces the $2t$-step diffusion distance~\cite{coifmanGeometricDiffusionsTool2005,nadlerDiffusionMapsSpectral2005, nadlerDiffusionMapsSpectral2006, coifmanDiffusionMaps2006},
\begin{align}
    D_{2t}(l, l^{\prime}) = \sum_{l^{\prime\prime}} \frac{1}{z_{l^{\prime\prime}}}[(p^t)_{l^{}, l^{\prime\prime}}-(p^t)_{l^{\prime}, l^{\prime\prime}}]^2,
\end{align}
where $p^t$ denotes the $t$-th matrix power of the transition probability matrix $p$.
It was shown that $D_{2t}$ can be computed from the eigenvalues $\lambda_n$ and right eigenvectors $\psi_n$ of the transition matrix $p$: with $\sum_{l^{\prime}}$\,$p_{l, l^{\prime}}$\,$(\psi_n)_{l^{\prime}}$\,$=$\,$\lambda_n$\,$(\psi_n)_{l}$, and in descending ordering $\lambda_n > \lambda_{n+1}$, it follows
\begin{align}
    D_{2t}(l, l^{\prime}) = \sum_{n=1}^{M-1}\lambda_{n}^{2t}[(\psi_n)_{l} - (\psi_n)_{l^\prime}]^2
\end{align}
after straightforward algebra~\cite{coifmanDiffusionMaps2006}.
Geometrically, this means that the diffusion distance is represented as a Euclidean distance (weighted with $\lambda_n$) if we perform the non-linear coordinate transformation $x_l \rightarrow \{(\psi_n)_l,n=0,\dots M-1\}$. Furthermore, as the global connectivity is seen from the long-time limit, $t\rightarrow \infty$, of the diffusion distance, the largest eigenvalues are most important to describe the connectivity. To be more precise, let us choose a kernel $k_{\epsilon}$ of the form
\begin{align}\label{eq:k_mat}
    & k_{\epsilon}(x_l, x_{l^{\prime}}) = \exp\left(-\frac{1 -S(x_l, x_{l^\prime}) }{\epsilon}\right),
\end{align}
where $S$ is a \textit{local similarity measure} which obeys $S\in [0, 1]$, $S(x_l,x_{l'})=S(x_{l'},x_{l})$, and $S(x,x)=1$. Here ``local'' means that $S(x_l,x_{l'})=\sum_i \mathcal{S}_i(x_{l},x_{l'})$ where $\mathcal{S}_i(x_{l},x_{l'})$ only depend on the configuration of $x_l$ and $x_{l'}$ in the vicinity of site $i$. While we will discuss possible explicit forms of $S$ for our quantum mechanical $N$ spin/qubit system in \secref{SimilarityMeasure} below, a natural choice for a classical system of $N$ spins, $x_l=\{\vec{S}^l_i,(\vec{S}^l_i)^2=1, i=1,2,\dots ,N\}$, is $S_{\text{cl}} (x_{l},x_{l'}) = \sum_i \vec{S}^l_i\cdot \vec{S}^{l'}_i/N$. In \equref{eq:k_mat}, $\epsilon$ plays the role of a ``coarse graining" parameter that is necessary as we only deal with finite datasets $X$: for given $X$, we generically expect $k_{\epsilon}(x_l, x_{l^{\prime}}) = p_{l,l'} = \delta_{l,l'}$ as $\epsilon\rightarrow 0$, i.e., all samples are dissimilar if $\epsilon$ is sufficiently small and all eigenvalues $\lambda_n$ approach $1$. In turn, for $\epsilon \rightarrow \infty$ the coarse graining parameter is so large that all samples become connected, $k_{\epsilon}(x_l, x_{l^{\prime}})\rightarrow 1$; as $p_{l,l'}\rightarrow 1/M$, we will have $\lambda_{n>0} \rightarrow 0$, while the largest eigenvalue $\lambda_0$ is always $1$ (as a consequence of probability conservation). For values of $\epsilon$ in between these extreme limits, the DM spectrum contains information about $X$, including its topological structure: as shown in \refcite{rodriguez-nievaIdentifyingTopologicalOrder2019}, the presence of $k\in\mathbb{N}$ distinct topological equivalence classes in $X$ is manifested by a range of $\epsilon$ where $\lambda_1,\dots \lambda_{k-1}$ are all exponentially close (in $\epsilon$) to $1$, with a clear gap to $\lambda_{n\geq k}$. Furthermore, the different samples $l$ will cluster---with respect to the normal Euclidean measure, e.g., as can be captured with $k$-means---according to their topological equivalence class when plotted in the mapped $k-1$-dimensional space $\{(\psi_{1})_l,(\psi_{2})_l, \dots , (\psi_{k-1})_l\}$. In the following, we will use this procedure to identify the superselection sectors in the ensemble of wave functions defined in \secref{DataGeneration}. To this end, however, we first need to introduce a suitable similarity measure $S$, to be discussed next.

\subsection{Local similarity measure}
\label{SimilarityMeasure}
A natural generalization of the abovementioned classical similarity measure $S_{\text{cl}} = \sum_i \vec{S}^l_i\cdot \vec{S}^{l'}_i/N$, which can be thought of as the (Euclidean) inner product in the classical configuration space, is to take the inner product in the Hilbert space of the quantum system,
\begin{align} \label{eq:s_exact_overlap}
    S_{\rm q}(\Lambda_l, \Lambda_{l^\prime}) = \abs{\braket{\Psi(\Lambda_l)|\Psi(\Lambda_{l'})}}^2.
\end{align}
While this or other related fidelity measures for low-rank quantum states could be estimated efficiently with quantum simulation and computing setups~\cite{chenAlternativeFidelityMeasure2002, mendoncaAlternativeFidelityMeasure2008, puchalaBoundTraceDistance2009, miszczakSubSuperFidelity2008}, estimating $S_{\rm q}$ is generally a computationally expensive task on a classical computer, as it requires sampling over spin configurations for our variation procedure. 
To make the evaluation of the similarity measure more efficient, we here propose an alternative route that takes advantage of the fact that we use a local ansatz for $\psi(\vec{\sigma};\Lambda)$, see \equref{LocalAnsatz}. Our goal is to express the similarity measure directly as
\begin{align} \label{s_local}
    S_{\rm n}(\Lambda_l, \Lambda_{l^\prime}) = \frac{1}{N_{\j}}\sum_{\j} f((\Lambda_{l})_{\j}, (\Lambda_{l^\prime})_{\j}), 
\end{align}
where $f$ only compares a local block of parameters denoted by $\j$ and is a function that can be quickly evaluated, without having to sample spin configurations.
Furthermore, $S(x_l,x_{l'})=S(x_{l'},x_{l})$ can be ensured by choosing a function $f$ that is symmetric in its arguments and $S\in [0, 1]$ is also readily implemented by setting $N_{\j} = \sum_{\j}$ and appropriate rescaling of $f$ such that $f\in [0, 1]$. The most subtle condition is
\begin{equation}
    S_{\rm n}(\Lambda_l, \Lambda_{l^\prime}) = 1 \quad \Longleftrightarrow \quad \ket{\Psi(\Lambda_l)} \propto \ket{\Psi(\Lambda_l')},
\end{equation}
since, depending on the precise network architecture used for $\psi(\boldsymbol{\sigma};\,\Lambda)$, there are ``gauge transformations'' $g\in\mathcal{G}$ of the weights, $\Lambda_l \rightarrow g[\Lambda_l]$, with 
\begin{equation}
    \ket{\Psi(\Lambda_l)} = e^{i\vartheta_g}\ket{\Psi(g[\Lambda_l])} \label{GaugeTrafoOnWaveFunc}
\end{equation}
for some global phase $\vartheta_g$. We want to ensure that
\begin{equation}
     S_{\rm n}(\Lambda_{l}, \Lambda_{l^\prime}) = S_{\rm n}(\Lambda_{l}, g[\Lambda_{l^\prime}]) = S_{\rm n}(g[\Lambda_{l}], \Lambda_{l^\prime}) \label{GaugeConstraint}
\end{equation}
for all such gauge transformations $g\in\mathcal{G}$. A general way to guarantee \equref{GaugeConstraint} proceeds by replacing,
\begin{equation}
    S_{\rm n}(\Lambda_l, \Lambda_{l^\prime}) \quad \longrightarrow \quad \max_{g,g'\in\mathcal{G}} S_{\rm n}(g[\Lambda_{l}], g'[\Lambda_{l^\prime}]). \label{NaiveWayOfDeadlingWithGaugeTrafo} 
\end{equation}
However, in practice, it might not be required to iterate over all possible gauge transformations in $\mathcal{G}$ due to the locality of the similarity measure.
In the following, we will use the toric code and a specific RBM variational ansatz as an example to illustrate these gauge transformations and how an appropriate function $f$ in \equref{s_local} and gauge invariance (\ref{GaugeConstraint}) can be implemented efficiently.

Finally, note that, while we focus on applying DM in this work, a similarity measure in terms of neural network parameters can also be used for other kernel techniques such as kernel PCA. Depending on the structure of the underlying dataset, DM has clear advantage over kernel PCA: the former really captures the global connectivity of the dataset rather than the subspace with most variance that is extracted by the latter. This is why kernel PCA fails when identifying, e.g., winding numbers, in general datasets where DM still works well \cite{rodriguez-nievaIdentifyingTopologicalOrder2019}. Specifically for our case study of the toric code below, we find that kernel PCA can also identify topological sectors for small $T$ and without magnetic field, $h=0$, as a result of the simple data structure; however, only DM works well when $h$ is turned on, as we discuss below.

\begin{figure}[t!]
    \centering
    \includegraphics[width=\linewidth]{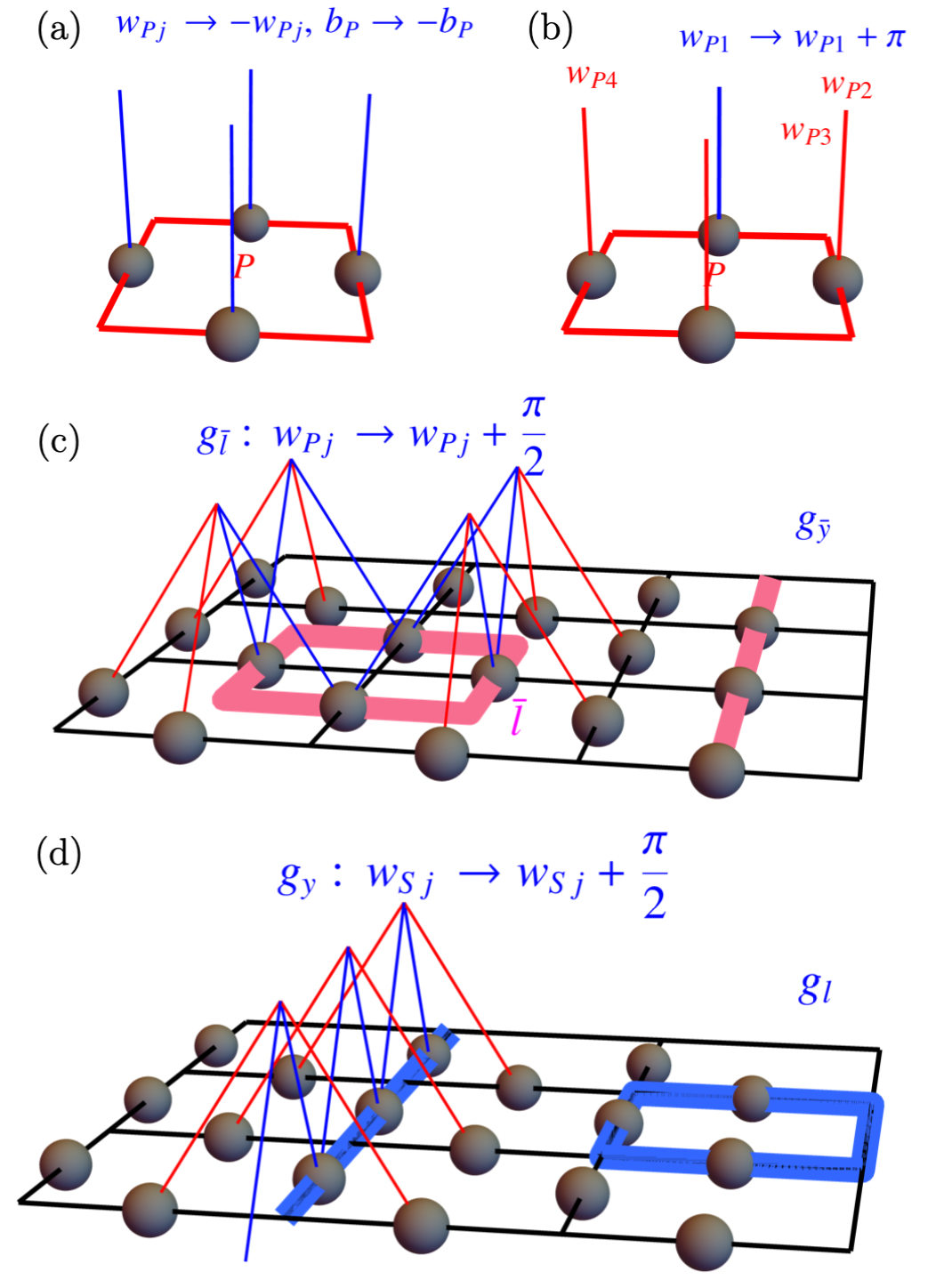}
    \caption{Gauge freedom of RBM ansatz in \equref{eq:rbm}. The following transformations only lead to a global phase: (a) Multiplying all the parameters of a plaquette (or star, not shown) by a minus sign, see \equref{gXminusGaugeTrafo};  (b) A $\pi$ shift of a single parameter, see \equsref{Xpib}{Xpij}; (c) A $\pi/2$ shift to the weights crossed by a string $\bar{l}$, defined by $g_{\bar{l}}$ in \equref{StringRedundancyP}. The straight pink line represents the transformation on a non-contractible loop denoted by $g_y$; (d) Same as (c) but for loops on the direct lattice and $g_l$ and $g_{\bar{y}}$, cf.~\equref{StringRedundancyS}.}
    \label{fig:gauge}
\end{figure}

\section{Example: toric code}\label{sec:ToricCodeExample}
Now we illustrate our DM-based ML algorithm using the toric code model \cite{kitaevFaulttolerantQuantumComputation2003}, defined on an $L_x \times L_y$ square lattice with spin-$1/2$ operators or qubits on every bond, see Fig.~\ref{fig:alg_ansatz}(b), leading to a total of $N=2L_xL_y$ spins; throughout this work, we will assume periodic boundary conditions. 
Referring to all four spins on the edges of an elementary square (vertex) of the lattice as plaquette $P$ (star $S$), the plaquette and star operators are defined as $\hat{\mathcal{P}}_P=\prod_{i \in P} \hat{s}^z_i$ and $\hat{\mathcal{S}}_S=\prod_{i \in S} \hat{s}^x_i$, respectively. The toric code Hamiltonian then reads as
\begin{equation}\label{eq:ham_tc}
    \hat{H}_{\rm tc} = - J_P \sum_P \hat{\mathcal{P}}_P - J_S \sum_S \hat{\mathcal{S}}_S,
\end{equation}
where the sums are over all plaquettes and stars of the lattice. All ``stabilizers'' $\hat{\mathcal{P}}_P$, $\hat{\mathcal{S}}_S$ commute among each other and with the Hamiltonian. Focusing on $J_P, J_S > 0$, the ground states are obtained as the eigenstates with eigenvalue $+1$ under all stabilizers. A counting argument, taking into account the constraint $\prod_S \hat{\mathcal{S}}_S$\,$=$\,$\prod_P \hat{\mathcal{P}}_P$\,$=$\,$\mathds{1}$, reveals that there are four, exactly degenerate ground states for periodic boundary conditions.

To describe the ground-states and low-energy subspace of the toric code model (\ref{eq:ham_tc}) variationally, we parameterize $\psi(\boldsymbol{\sigma};\,\Lambda)$ in \equref{WavefunctionAnsatz} using the ansatz 
\begin{align}\label{eq:rbm}
    \psi_{\rm{rbm}}(\boldsymbol{\sigma};\,\Lambda) =& \prod_P \cos( b_{P} + \sum_{ j \in P} w_{Pj} \sigma_j) \nonumber \\
     \times & \prod_S \cos(b_
    {S}+ \sum_{j \in S} w_{Sj} \sigma_j), 
\end{align}
proposed in \refcite{dengMachineLearningTopological2017}, where every plaquette $P$ (star $S$) is associated with a ``bias'' $b_P$ ($b_S$) and four weights $w_{P,j}$ ($w_{S,j}$), all of which are chosen to be real here, i.e., $\Lambda=\{b_P,b_S,w_{P,j},w_{S,j}\}$. This ansatz can be thought of as an RBM~\cite{carleoSolvingQuantumManybody2017} (see Appendix \ref{app:ansatz}), as illustrated in Fig.~\ref{fig:alg_ansatz}(c), with the same geometric properties as the underlying toric code model. It is clear that \equref{eq:rbm} defines a quasi-local ansatz as it is of the form of \equref{LocalAnsatz}, with $\j$ enumerating all plaquettes and stars (and thus $N_{\j}=2N$).
For this specific ansatz, the gauge transformations $g\in\mathcal{G}$, as introduced in \secref{SimilarityMeasure} above, are generated by the following set of operations on the parameters $b_P$, $b_S$, $w_{P,j}$, and $w_{S,j}$:
\begin{subequations}\label{gTrafos}
\begin{enumerate}[leftmargin=1.2\parindent]
\item For $X$ being any plaquette or star, multiplying all biases and weights of that plaquette or star by $-1$ [see \figref{fig:gauge}(a)],
\begin{equation}
    g_{X,-}: \, b_X \rightarrow - b_X, \,\,  w_{Xj}\rightarrow - w_{Xj}, \label{gXminusGaugeTrafo}
\end{equation}
leaves the wave function invariant [$\vartheta_g=0$ in \equref{GaugeTrafoOnWaveFunc}].
\item Adding $\pi$ to either the bias or any of the weights associated with the plaquette or star $X$ [see \figref{fig:gauge}(b)],
\begin{align}
    g_{X,\pi,b}: \, &b_X \rightarrow b_X + \pi, \label{Xpib}  \\
    g_{X,\pi,j}: \, &w_{Xj} \rightarrow w_{Xj} + \pi, \quad j \in X, \label{Xpij}
\end{align}
leads to an overall minus sign [$\vartheta_g=\pi$ in \equref{GaugeTrafoOnWaveFunc}].
\item For any closed loop $\ell$ (or $\bar{\ell}$) on the direct (or dual lattice), adding $\frac{\pi}{2}$ to all weights of the stars (plaquettes) that are connected to the spins crossed by the string [see \figref{fig:gauge}(c-d)],
\begin{align}
    g_{\ell}: \, w_{Sj} &\rightarrow  w_{Sj} + \frac{\pi}{2}, \quad Sj \in \ell, \label{StringRedundancyS} \\
    g_{\bar{\ell}}: \, w_{Pj} &\rightarrow  w_{Pj} + \frac{\pi}{2}, \quad Pj \in \bar{\ell}, \label{StringRedundancyP}
\end{align}
leads to $\vartheta_g=0$ or $\pi$ in \equref{GaugeTrafoOnWaveFunc} depending on the length of the string. Note that any loop configuration $\mathcal{L}$, which can contain an arbitrary number of loops, can be generated by the set $\{g_{S},g_{P},g_{x,y},g_{\bar{x},\bar{y}}\}$, where $g_{S}$ ($g_{P}$) creates an elementary loop on the dual (direct) lattice encircling the star $S$ (plaquette $P$), see \figref{fig:gauge}(c,d), and $g_{x,y}$ ($g_{\bar{x},\bar{y}}$) creates a non-contractible loop on the direct (dual) lattice along the $x,y$ direction. Since the length of any contractible loop is even, $\vartheta_g=0$ for any string transformations generated by $g_S$ and $g_P$. Meanwhile, on an odd lattice, the gauge transformations $g_{x, y}(g_{\bar{x},\bar{y}})$ involve an odd number of sites and thus lead to $\vartheta_g=\pi$.
\end{enumerate}
\label{GaugeTrafos}\end{subequations}
A highly inefficient way of dealing with this gauge redundancy would be to use a choice of $S_n$ in \equref{s_local} which is not invariant under any of the transformations in \equref{GaugeTrafos}; this would, for instance, be the case by just taking the Euclidean distance of the weights,
\begin{align*}
    &S_{\rm eu}(\Lambda_l,\Lambda_{l'}) \propto || \Lambda_l - \Lambda_{l'} ||^2 \\
    &\qquad = \sum_X \Bigl[(b^l_X-b^{l'}_X)^2 + \sum_{j\in X} (w^l_{Xj}-w^{l'}_{Xj})^2  \Bigr],
\end{align*}
where the sum over $X$ involves all plaquettes and stars.
Naively going through all possible gauge transformations to find the maximum in \equref{NaiveWayOfDeadlingWithGaugeTrafo} would in principle rectify the lack of gauge invariance. However, since the number of gauge transformations scales exponentially with system size $N$ (holds for each of the three classes, 1.-3., of transformations defined above), such an approach would become very expensive for large $N$. Luckily, locality of the ansatz and of the similarity measure allows us to construct similarity measures that can be evaluated much faster: as an example, consider 
\begin{align}\begin{split}
    &S_{n}(\Lambda_l, \Lambda_{l^\prime}) =  \frac{1}{2} +
    \frac{1}{10 N} \sum_X \max_{\tau_X = \pm}\Bigl[ \\ &\quad  \sum_{j \in X}  \cos 2(\tau_X w^l_{Xj} - w^{l'}_{Xj}) + \cos 2(\tau_X b^l_{X} - b^{l'}_{X}) \Bigr],\label{Srbm}
\end{split}\end{align}
which clearly obeys $S_{n}(\Lambda_l, \Lambda_{l^\prime})=S_{n}(\Lambda_{l'}, \Lambda_{l})$, $S_{n}(\Lambda_l, \Lambda_{l^\prime}) \in [0,1]$, and locality [it is of the form of \equref{s_local} with $\j$ enumerating all $X$].
Concerning gauge invariance, first note that the choice of $\cos(\cdot)$ immediately leads to invariance under \equref{gXminusGaugeTrafo}. Second, for each $X$ we only have to maximize over two values ($\tau_X$) to enforce invariance under \equsref{Xpib}{Xpij}, i.e., the maximization only doubles the computational cost. 

The ``string'' redundancy, see \equsref{StringRedundancyS}{StringRedundancyP}, however, is not yet taken into account in \equref{Srbm}. It can be formally taken care of  by maximizing over all possible loop configurations, denoted by $\mathcal{L}$,
\begin{align}\begin{split}
    &S_{\rm str}(\Lambda_l, \Lambda_{l^\prime}) =  \frac{1}{2} +
    \frac{1}{10 N} \max_{\mathcal{L}}\Bigl\{\sum_X \max_{\tau_X = \pm}\Bigl[ \\ &\quad  \sum_{j \in X}  \mu^{\mathcal{L}}_{Xj} \cos 2(\tau_X w^l_{Xj} - w^{l'}_{Xj}) + \cos 2(\tau_X b^l_{X} - b^{l'}_{X}) \Bigr] \Bigr\},\label{SrbmString}
\end{split}\end{align}
where $\mu^{\mathcal{L}}_{Xj}$\,$=$\,$-1$ if $Xj$ lives on a loop contained in ${\mathcal{L}}$ and $\mu_{Xj}^{\mathcal{L}}$\,$=$\,$1$ otherwise. While there is an exponential number of such strings, \refcite{rodriguez-nievaIdentifyingTopologicalOrder2019} has proposed an algorithm to efficiently find an approximate maximum value. In our case, this algorithm amounts to randomly choosing a plaquette $P$ or a star $S$ or a direction $d=x,y$ and then applying $g_{S}$ or $g_{P}$ or $g_{d=x,y}$ to $\Lambda_l$ in \equref{Srbm}. If this does not decrease the similarity, keep that transformation; if it decreases the similarity, discard the gauge transformation. Repeat this procedure $N_g$ times. In \refcite{rodriguez-nievaIdentifyingTopologicalOrder2019}, $N_g$ between $10^3$ and $10^4$ was found to be enough for a large system consisting $18 \times 18$ square-lattice sites (total of $N=2\times 18^2$ qubits). On top of this, $g_{S}$ and $g_{P}$ are local and, hence, the evaluation of the change of the similarity with the gauge transformation only requires $\mathcal{O}(N^0)$ amount of work.

In the numerical simulations below, using \equref{Srbm} without sampling over loop configurations $\mathcal{L}$ turned out to be sufficient. The reason is that, for our Markov-chain-inspired sampling procedure of $\Lambda_l$ (see \appref{app:ensemble}), updates that correspond to these loop transformations happen very infrequently. Furthermore, even if a few pairs of samples are incorrectly classified as distinct due to the string redundancy, the DM will still correctly capture the global connectivity and, hence, absence or presence of topological sectors.

\begin{figure}[th!]
    \centering
    \includegraphics[width=\linewidth]{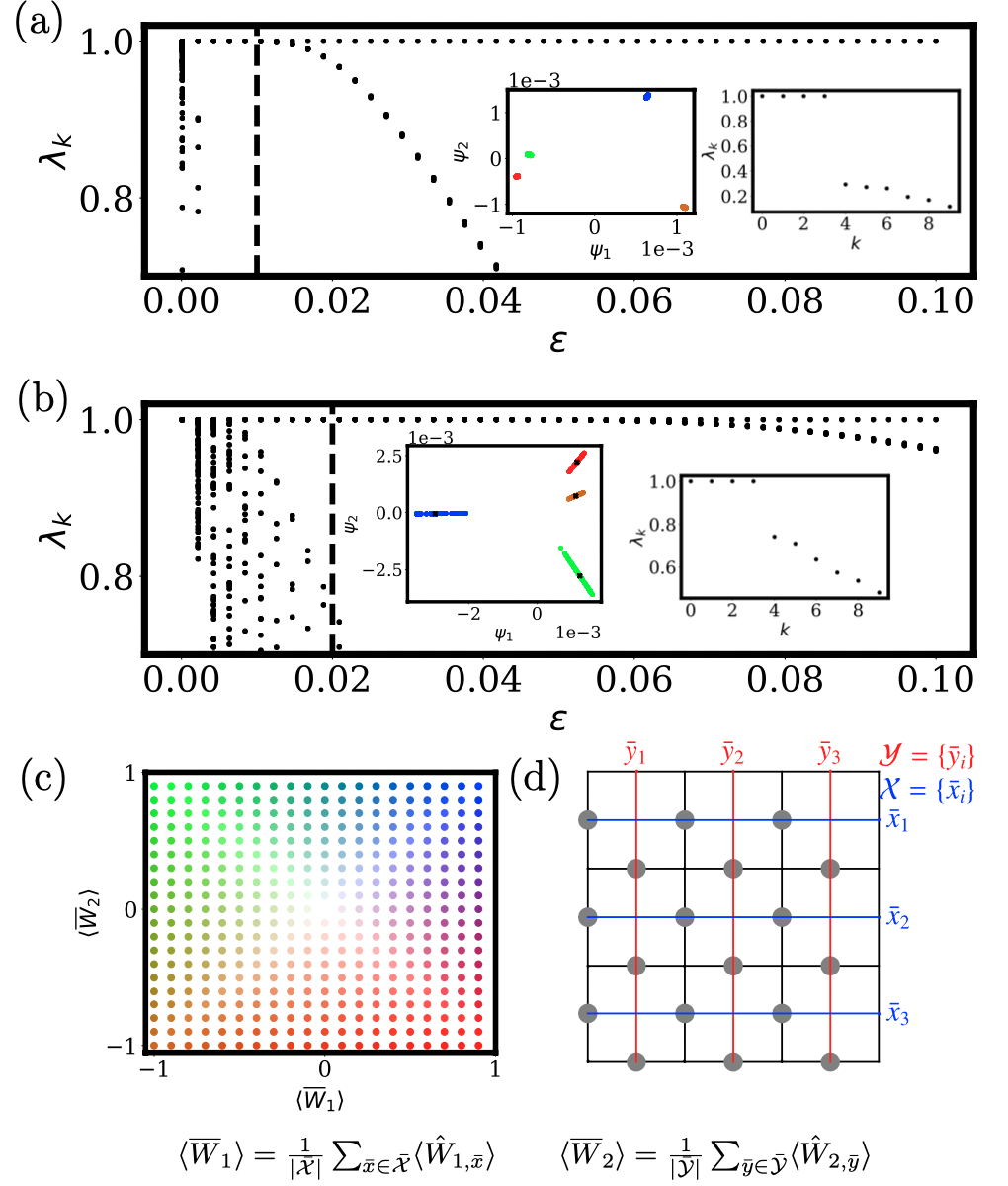}
    \caption{(a) DM spectrum for topological phase at $h=0$ and $T=0.1$ using the neutral network similarity measure in  \equref{Srbm}. Inset left: associated leading DM components; color represents the loop observable expectations values defined in (c-d). Inset right: DM spectrum in descending order at $\epsilon=0.01$ indicated by the dashed line. (b) Same as (a), but using exact overlaps $S_{\text{q}}$ in Eq.~\eqref{eq:s_exact_overlap} as metric. (c) Color map for the non-local loop values $\ev{\overline{W}_1}, \ev{\overline{W}_2}$ in the left insets of (a) and (b). (d) Different straight Wilson loops $\hat{W}_{1, \bar{x}_i}$ ($\hat{W}_{2, \bar{y}_i}$) along $x$ ($y$) direction, denoted by blue (red) lines. The loop values in the color map in (c) are spatial averages over all straight-loop expectation values (as in the equations for $\ev{\overline{W}_1}, \ev{\overline{W}_2}$).
    }
    \label{fig:topological}
\end{figure}

\section{Numerical results}\label{NumericalResults}
We next demonstrate explicitly how the general procedure outlined above can be used to probe and analyze topological order in the toric code. We start from the pure toric code Hamiltonian defined in \equref{eq:ham_tc} using the variational RBM ansatz in \equref{eq:rbm}. An ensemble of network parameters is generated by applying the procedure of \secref{DataGeneration} (see also Algorithm \ref{alg:data_generation}) for a system size of $N$\,$=$\,$18$ spins; the hyperparameters for ensemble generation and more details including the form of $u$ in \equref{eq:lambda_prime} are given in \appref{app:ensemble}. From now on, we measure all energies in units of $J_P$ and set $J_S=J_P=1$.

Let us first focus on the low-energy ensemble and choose $T=0.1$ in \equref{eq:p_accept_energy}. For the simple similarity measure in \equref{Srbm}, that can be exactly evaluated at a time linear in system size $N$, we find the DM spectrum shown in \figref{fig:topological}(a) as a function of $\epsilon$ in \equref{eq:k_mat}. We observe the hallmark feature of four superselection sectors \cite{rodriguez-nievaIdentifyingTopologicalOrder2019}: there is a finite range of $\epsilon$ where there are four eigenvalues exponentially close to $1$. The association of samples (in our case states) and these four sectors is thus expected to be visible in a scatter plot of a projected subspace spanned by the first three non-trivial eigenvectors $\psi_{1,2,3}$~\cite{rodriguez-nievaIdentifyingTopologicalOrder2019}; note the zeroth eigenvector $(\psi_{0})_l=C$ is always constant with eigenvalue $\lambda=1$ from probability conservation. In fact, we can see these clusters already in the first two components, see left inset in \figref{fig:topological}(a). Then a standard $k$-means algorithm is applied onto this projected subspace to identify the cluster number for each data point. To verify that the ML algorithm has correctly clustered the states according to the four physical sectors, we compute the expectation value for each state of the string operators,
\begin{align}
    \hat{W}_{1, \bar{x}} = \prod_{i \in \bar{x}} \hat{s}^x_i, \quad \hat{W}_{2, \bar{y}} = \prod_{i \in \bar{y}} \hat{s}^x_i, \label{WilsonLoopOperators}
\end{align}
where $\bar{x}(\bar{y})$ are loops defined on the dual lattice winding along the $x(y)$ direction, shown as blue lines in Fig.~\ref{fig:alg_ansatz}(b). We quantify the association of a state to physical sectors by the average of a set of straight loops $\mathcal{X}(\mathcal{Y})$ winding around the $x(y)$ direction, shown as blue (red) lines in Fig.~\ref{fig:topological}(d). Indicating this averaged expectation value $\ev{\overline{W}_1}, \ev{\overline{W}_2}$ in the inset of \figref{fig:topological}(a) using the color code defined in \figref{fig:topological}(c), we indeed see that the clustering is done correctly. 

To demonstrate that this is not a special feature of the similarity measure in \equref{Srbm}, we have done the same analysis, with result shown in \figref{fig:topological}(b), using the full quantum mechanical overlap measure in \equref{eq:s_exact_overlap}. Quantitative details change but, as expected, four superselection sectors are clearly identified and the clustering is done correctly. We reiterate that the evaluation of the neural-network similarity measure in \equref{Srbm} [exact evaluation $\mathcal{O}(N)$] is much fast than that in \equref{eq:s_exact_overlap} [exact evaluation $\mathcal{O}(2^N)$, but we can compute it approximately with importance sampling] on a classical computer. Note, however, that once $S_n$ is computed for all samples, the actual DM-based clustering takes the same amount of computational time for both approaches. Consequently, suppose there is a quantum simulator that can efficiently measure the quantum overlap in \equref{eq:s_exact_overlap} or any other viable similarity measure for that matter, then we can equivalently use the ``measured" similarity for an efficient clustering of the superselection sectors via the DM scheme.
\begin{figure}[t!]
    \centering
    \includegraphics[width=\linewidth]{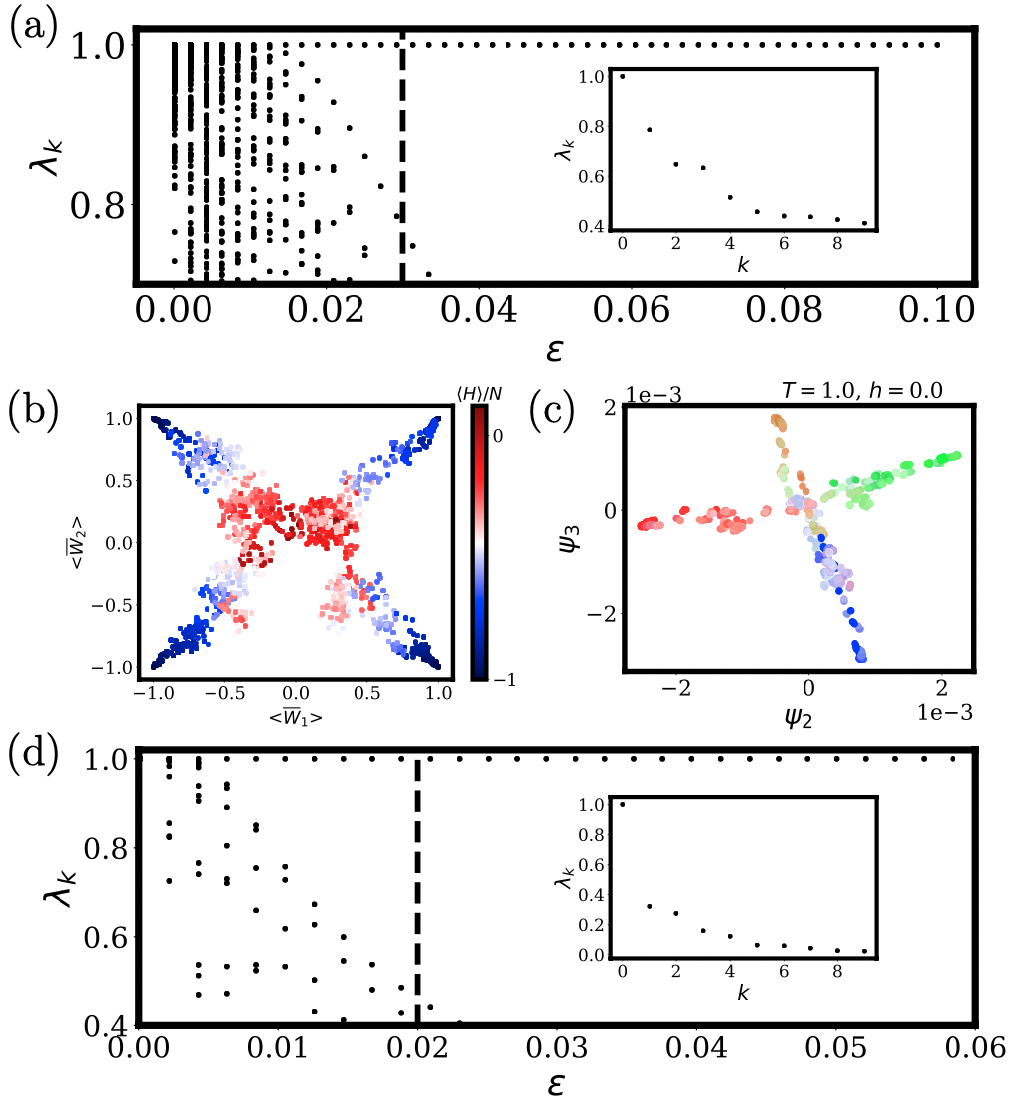}
    \caption{
    (a) DM spectrum for the high-energy ensemble at $h$\,$=$\,$0$ and $T$\,$=$\,$1$. The inset is the spectrum at $\epsilon=0.03$ indicated by the dashed line in the main panel; (b) Spatially averaged straight Wilson loops $\ev{\overline{W}_{1( 2)}}$ [see \figref{fig:topological}(c-d)] along two directions for the states in (a), where the color encodes energy density $\ev{H}/N$; (c) Leading DM components where the color of the dots encodes $\ev{\overline{W}_{1( 2)}}$ using the color map in Fig.~\ref{fig:topological}(d); (d) DM spectrum for the trivial phase at $h$\,$=$\,$1.0$ and $T$\,$=$\,$0.1$ using the quantum metric $S_{\text{q}}$. }
    \label{fig:diffused_states}
\end{figure}
As a next step, we demonstrate that the superselection sectors are eventually connected if we take into account states with sufficiently high energy. To this end, we repeat the same analysis but for an ensemble with $T=1$. As can be seen in the resulting DM spectrum in \figref{fig:diffused_states}(a), there is no value of $\epsilon$ where more than one eigenvalue is (exponentially) close to $1$ and separated from the rest of the spectrum by a clear gap. Here we used again the simplified measure in \equref{Srbm}, but have checked nothing changes qualitatively when using the overlap measure. To verify that this is the correct answer for the given dataset, we again computed the expectation value of the loop operators in \equref{WilsonLoopOperators} for each state in the ensemble. This is shown in \figref{fig:diffused_states}(b), where we also use color to indicate the energy expectation value for each state. We can clearly see the four low-energy (blue) sectors (with $|W_{1,2}|\approx 1$) are connected via high-energy (red) states (with $|W_{1,2}|\ll 1$). This agrees with the DM result that all states are connected within the ensemble (topological order is lost). We can nonetheless investigate the clustering in the leading three non-trivial DM components $\psi_{1,2,3}$. Focusing on a 2D projection in \figref{fig:diffused_states}(c) for simplicity of the presentation, we can see that the DM reveals very interesting structure in the data: the four lobes roughly correspond to the four colors blue, red, orange, and green associated with the four superselection sectors and the states closer to $|W_{1,2}|= 1$ (darker color) appear closer to the tips. Finally, note that the colors are arranged such that the red and green [orange and blue] lobes are on opposite ends, as expected since they correspond to $(W_1,W_2) \approx (1,-1)$ and $(-1,1)$ [$(-1,-1)$ and $(1,1)$].

\begin{figure}[t!]
    \centering
    \includegraphics[width=\linewidth]{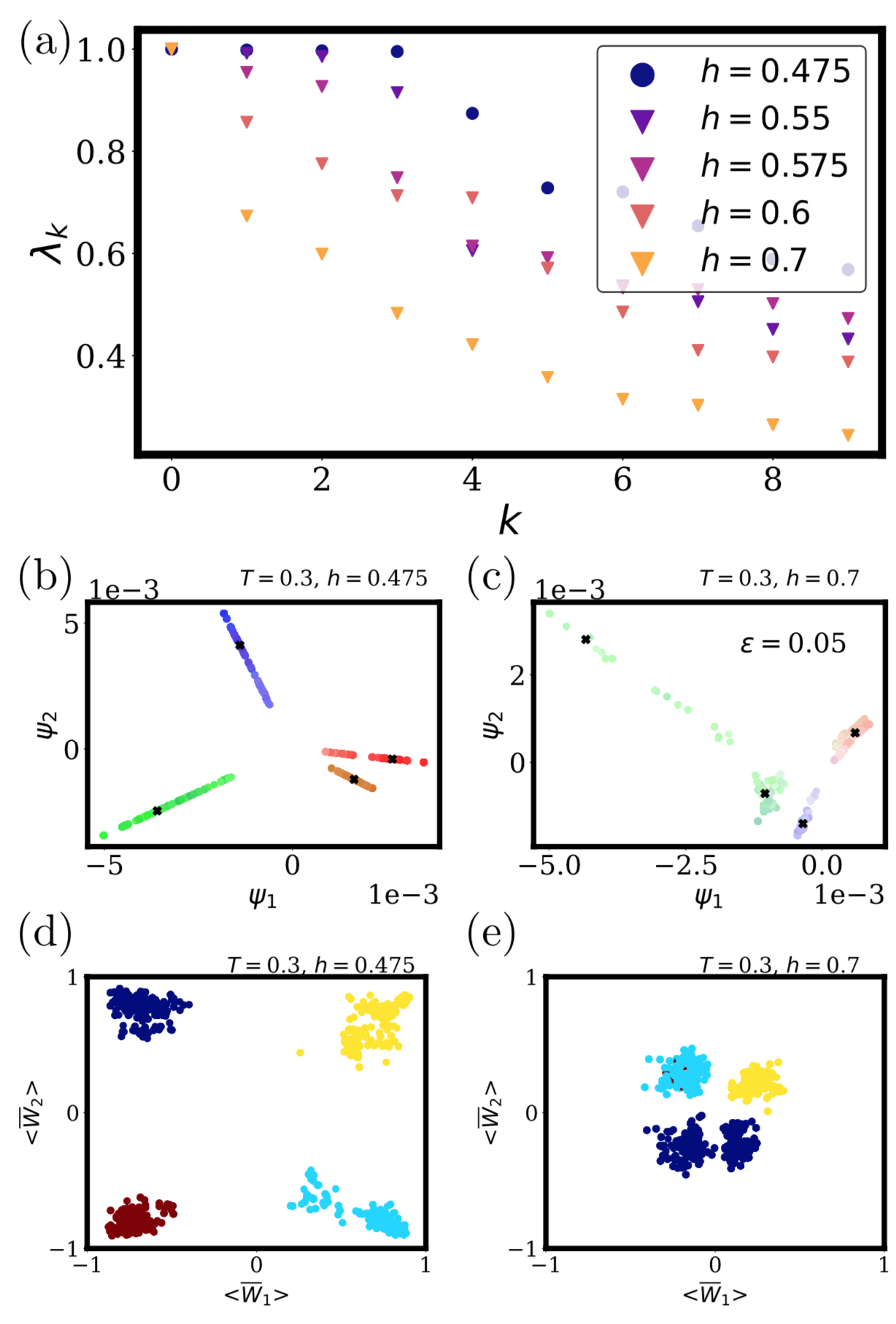}
    \caption{DM spectra for low-energy ensembles with $T$\,$=$\,$0.3$ at finite field $h$. (a) First 10 eigenvalues for various field values $h$\,$=$\,$0.475, 0.55, 0.575, 0.6, 0.7$ at $\epsilon$\,$=$\,$0.05$. The dot marker ($h$\,$=$\,$0.475$) shows that the eigenvalue spectra have four-fold degeneracy, indicating signature for topological order. In comparison, for spectra marked by the the triangular markers ($h\geq 0.55$), such degeneracy is absent. A transition field value $h_t\approx0.55$ is identified by observing that a gap opens in the degenerate eigenvalue spectra. This is consistent with what we have observed in the fidelity using the same dataset [see \appref{app:vmc_fidelity}]. (b) Projected eigenvectors onto the first two components for $h$\,$=$\,$0.475$. The color encodes $\ev{\overline{W}_{1( 2)}}$ with the color scheme of \figref{fig:topological}(c). The black cross marks the $k$-means centers. (c) Same as (b) for $h=0.7$. (d) Expectation for averaged straight Wilson loops $\ev{\overline{W}_{1( 2)}}$ along two directions for the states in (b). The color encodes the clustering results from $k$-means in the projected subspace of the eigenvectors shown in (b).  (e) Same as (d) for ensemble shown in (c).}
    \label{fig:field}
\end{figure}
Another route to destroying topological order proceeds via application of a magnetic field. To study this, we extend the toric code Hamiltonian according to
\begin{equation}\label{eq:htc_field}
    \hat{H}'_{\rm tc} = \hat{H}_{\rm tc} - h\sum_i  \hat{s}^z_i.
\end{equation}
Clearly, in the limit of $h\rightarrow \infty$, the ground state is just a state where all spins are polarized along $\hat{s}^z$ and topological order is lost. Starting from the pure toric model ($h=0$) and turning on $h$ reduces the gap of the ``charge excitations'' defined by flipping $\hat{\mathcal{S}}_S$ from $+1$ in the toric code groundstate to $-1$. Their condensation leads to a second-order quantum phase transition \cite{tupitsynTopologicalMulticriticalPoint2010, trebstBreakdownTopologicalPhase2007, wuPhaseDiagramToric2012, Lauchlitoriccode}.

Before addressing the transition, let us study the large-$h$ limit. We first note that our ansatz in \equref{eq:rbm} does not need to be changed as it can capture the polarized phase as well. For instance, denoting the ``northmost'' (and ``southmost'') spin of the plaquette $P$ (and star $S$) by $j_0(P)$ (and $j_0(S)$), respectively, the spin polarized state is realized for [see also \figref{fig:polarized}(a) in the Appendix]
\begin{equation}\label{eq:polarizedWeights}
    b_P=b_S=-\frac{\pi}{4}, \,\,\,\, w_{Xj} = \begin{cases} \frac{\pi}{4},\,\,\,\, j=j_0(X), \\
    0, \,\,\,\, \text{otherwise}.
    \end{cases}
\end{equation}
In fact, the spin polarized state has many representations within our RBM ansatz in \equref{eq:rbm}, including representations that are not just related by the gauge transformations in \equref{GaugeTrafos}. For instance, the association $j\rightarrow j_0(X)$ of a spin to a plaquette and star can be changed, e.g., by using the ``easternmost'' spin. As discussed in more detail in \appref{app:ansatz_redundancy}, this redundancy is a consequence of the product from of $\psi_{\text{rbm}}(\vec{\sigma})$ in \equref{eq:rbm} and the fact that $\psi_{\text{rbm}}(\vec{\sigma})$ is \textit{exactly} zero if there is a single $j$ with $\sigma_{j} = -1$; consequently, it is a special feature of the simple product nature of the spin-polarized ground state. While in general there can still be additional redundancies besides the aforementioned gauge transformations, we do not expect such a structured set of redundancy to hold for generic states. There are various ways of resolving this issue. The most straightforward one is to replace the simple overlap measure $S_{\text{n}}$ in \equref{s_local} by the direct overlap $S_{\text{q}}$ in \equref{eq:s_exact_overlap} for a certain fraction of pairs of samples $l$ and $l'$. If this fraction is large enough, the DM algorithm will be able recognize that clusters of network parameters that might be distinct according to $S_{\text{n}}$ actually correspond to identical wave functions. We refer to \appref{ResolveRedundancies} where this is explicitly demonstrated. 
We note, however, that kernel PCA will not work anymore in this case; it will incorrectly classify connected samples as distinct as it's based on the variance of the data rather than connectivity. For simplicity of the presentation, we use $S_{\text{q}}$ for all states in the main text and focus on DM.  

The DM spectrum for large magnetic field, $h=1$, and low temperatures, $T=0.1$, is shown in \figref{fig:diffused_states}(d). Clearly, there is no value of $\epsilon$ for which there is more than one eigenvalue close to $1$ while exhibiting a gap to the rest of the spectrum. This shows that, as expected, the magnetic field $h$ has lead to the loss of topological order. 

To study with our DM algorithm the associated phase transition induced by $h$, we repeat the same procedure for various different values of $h$. The resulting spectra for selected $h$ are shown in \figref{fig:field}(a). We see that there are still four sectors for $h=0.55$ in the data that are absent for $h=0.575$ and larger values. While the associated critical value of $h$ is larger than expected \cite{tupitsynTopologicalMulticriticalPoint2010, trebstBreakdownTopologicalPhase2007, wuPhaseDiagramToric2012}, this is not a shortcoming of the DM algorithm but rather a consequence of our simple local variational ansatz in \equref{eq:rbm}. By computing the fidelity as well as loop-operator expectation values, we can see that a critical value around $h=0.55$ is the expected answer for our dataset (see \appref{app:vmc_fidelity}). More sophisticated ans\"atze for the wavefunction are expected to yield better values, but this is not the main focus of this work.
More importantly, we see in \figref{fig:field}(b) that the DM clustering of the states correctly reproduces the clustering according to the averaged loop operator expectation values $\ev{\overline{W}_j}$ (again indicated with color). Alternatively, this can be seen in \figref{fig:field}(d) where $\ev{\overline{W}_j}$ is indicated for the individual samples. Using four different colors for the four different clusters identified by the DM, we see that all states are clustered correctly. As expected based on the eigenvalues, there are no clear clusters anymore for larger $h$, \figref{fig:field}(c); nonetheless, naively applying $k$-means clustering in $\psi_{1,2,3}$ manages to discover some residual structure of the wavefunctions related to $\ev{\overline{W}_j}$ as demonstrated in \figref{fig:field}(e).   

\section{Summary and discussion}\label{Conclusion}
In this work, we have described an unsupervised ML algorithm for quantum phases with topological order. We use neural network parameters to efficiently represent an ensemble of quantum states, which are sampled according to their energy expectation values. To uncover the structure of the superselection sectors in the quantum states, we used the dimensional reduction technique of diffusion map and provided a kernel defined in terms of network parameters. As opposed to a kernel based on the overlap of wavefunctions (or other quantum mechanical similarity measures of states for that matter), this metric can be evaluated efficiently (within polynomial time) on a classical computer.

We illustrated our general algorithm using a quasi-local restricted Boltzmann machine (RBM) and the toric code model in an external field; the choice of network ansatz was inspired by previous works \cite{dengMachineLearningTopological2017,valentiCorrelationEnhancedNeuralNetworks2021} showing the existence of efficient representations of the low-energy spectrum in terms of RBMs. Allowing for spatially inhomogeneous RBM networks, we identified the ``gauge symmetries’’ of the ansatz, i.e., the set of changes in the network parameters that do not change the wavefunction, apart from trivial global phase factors. We carefully designed a similarity measure that is gauge invariant---a key property as, otherwise, identical wavefunctions represented in different gauges would be falsely identified as being distinct.
We showed that the resultant unsupervised diffusion-map-based embedding of the wavefunctions is consistent with the expectation values of loop operators; it correctly captures the presence of superselection sectors and topological order at low energies and fields, as well as the lack thereof when higher-energy states are involved and/or the magnetic field is increased. We also verified our results using the full quantum mechanical overlap of wavefunctions as similarity measure. 

On a more general level, our analysis highlights the importance of the following two key properties of diffusion maps: first, in the presence of different topological sectors, the leading eigenvectors of diffusion maps capture the connectivity rather than, e.g., the variance as is the case for PCA. For this reason, the clustering is still done correctly even if a fraction of pairs of wavefunctions are incorrectly classified as being distinct due to the usage of an approximate similarity measure. This is why complementing the neural-network similarity measure, which has additional, state-specific redundancies in the large-field limit, by direct quantum mechanical overlaps for a certain fraction of pairs of states is sufficient to yield the correct classification. The second key property is that diffusion map is a kernel technique. This means that the actual machine learning procedure does not require the full wavefunctions as input; instead, only (some measure of) the kernel of all pairs of wavefunctions in the dataset is required. We have used this to effectively remove the gauge redundancy in the RBM parametrization of the states by proper definition of the network similarity measure in \equref{SrbmString}. Since the evaluation of full quantum mechanical similarity measures, like the wavefunction overlap, are very expensive on classical computers, an interesting future direction would be to use the emerging quantum-computing resources to evaluate a similarity measure quantum mechanically. This could then be used as input for a diffusion-map-based clustering.

We finally point out that the ensemble of states we used in this work, which was based on sampling states according to their energy with respect to a Hamiltonian, is only one of many possibilities. The proposed technique of applying diffusion map clustering using a gauge-invariant kernel in terms of network parameters of a variational description of quantum many-body wavefunctions can be applied more generally, in principle, to any ensemble of interest. For instance, to consider arbitrary local perturbations, one could generate an ensemble using finite depth local unitary circuits. Alternatively, one could generate an ensemble based on (Lindbladian) time-evolution to probe the stability of topological order against time-dependent perturbations or the coupling to a bath. We leave the investigation of such possibilities for future works.

\section{Code and data availability}\label{sec:code}
The Monte Carlo simulations in this work were implemented in JAX~\cite{jax2018github}. Python code and data will be available at \href{https://github.com/teng10/ml_toric_code/}{https://github.com/teng10/ml\_toric\_code/}.
\section*{acknowledgements}
Y.T. acknowledges useful discussions with Dmitrii Kochkov, Juan Carrasquilla, Khadijeh Sona Najafi, Maine Christos and Rhine Samajdar. Y.T. and S.S.~acknowledge funding by the U.S. Department of Energy under Grant DE-SC0019030. M.S.S. thanks Joaquin F. Rodriguez-Nieva for a previous collaboration on DM \cite{rodriguez-nievaIdentifyingTopologicalOrder2019}. The computations in this paper were run on the FASRC Cannon cluster supported by the FAS Division of Science Research Computing Group at Harvard University.

\appendix
\section{Variational Ansatz: Restricted Boltzmann Machine}\label{app:ansatz}
The variational ansatz in Eq.~\eqref{eq:rbm} is a \textit{further-restricted} restricted Boltzmann machine (RBM), first introduced by~\refcite{dengMachineLearningTopological2017}. RBM is a restricted class of Boltzmann machine with an ``energy" function $E_{\rm RBM}(\boldsymbol{\sigma}, \boldsymbol{h};\Lambda)$ dependent on the network parameters $\Lambda$, where $\boldsymbol{\sigma}$ are physical spins and $\boldsymbol{h}=\{h_1, h_2, \cdots, h_N \mid h_i=\pm 1\}$ are hidden spins (or hidden neurons) that are Ising variables. The parameters $\Lambda$ define the coupling strength among the physical and hidden spins. The restriction in RBM is that the couplings are only between the physical spin $\sigma_i$ and hidden spin $h_j$ with strength $-w_{ij}$, so that the ``energy" function takes the form $E_{\rm RBM}(\boldsymbol{\sigma}, \boldsymbol{h};\Lambda)$\,$=$\,$-\sum_i a_i \sigma_i$\,$-\sum_i b_i h_i$\,$-\sum_{ij} w_{ij} \sigma_i h_j$. It is a generative neural network that aims to model a probability distribution $\mathbb{P}$ based on the Boltzmann factor,
\begin{subequations}
\begin{align}\label{eq:}
    \mathbb{P}(\boldsymbol{\sigma}; \Lambda) &= \frac{1}{Z} \sum_{\boldsymbol{h}} e^{-E_{\rm RBM}(\boldsymbol{\sigma}, \boldsymbol{h};\Lambda)}, \\
    \text{normalization}\quad Z &= \sum_{\boldsymbol{\sigma}, \boldsymbol{h}} e^{-E_{\rm RBM}(\boldsymbol{\sigma}, \boldsymbol{h};\Lambda)}.
\end{align}
\end{subequations}
For the task of modeling a quantum wavefunction amplitude $\psi(\boldsymbol{\sigma}; \Lambda)$, RBMs can be used as a variational ansatz by extending the parameters $\Lambda$ to complex numbers. 

Further restricting parameters to the interlayer connections to the plaquette and star geometry in the toric code model [cf.~\figref{fig:alg_ansatz}(c)] and taking all parameters $\Lambda$ to be purely imaginary, we recover the ansatz in \equref{eq:rbm} (up to normalization factor $\widetilde{Z}$),
\begin{align}\label{eq:rbm_psi}
    \psi(\boldsymbol{\sigma}; \Lambda) &=\frac{1}{\widetilde{Z}}\sum_{X=P, S}\sum_{h_X=\pm 1}  e^{-i \sum_X (w_{Xj}\sigma_j + b_X) h_X}, \nonumber \\
    &= \frac{1}{\widetilde{Z}}\prod_{X=P, S} \cos(\sum_{j \in X} w_{Xj}\sigma_j + b_X).
\end{align}
 The $\cos(\cdot)$ factors come from summing over the hidden neurons and the ansatz factorizes into the product of individual plaquette (star) terms because of the restricted connections. The estimation of physical observables of a wave function based on the RBM ansatz requires Monte Carlo sampling procedure which we discuss in \appref{app:vmc}. 

\begin{figure}[t]
    \centering
    \includegraphics[width=\linewidth]{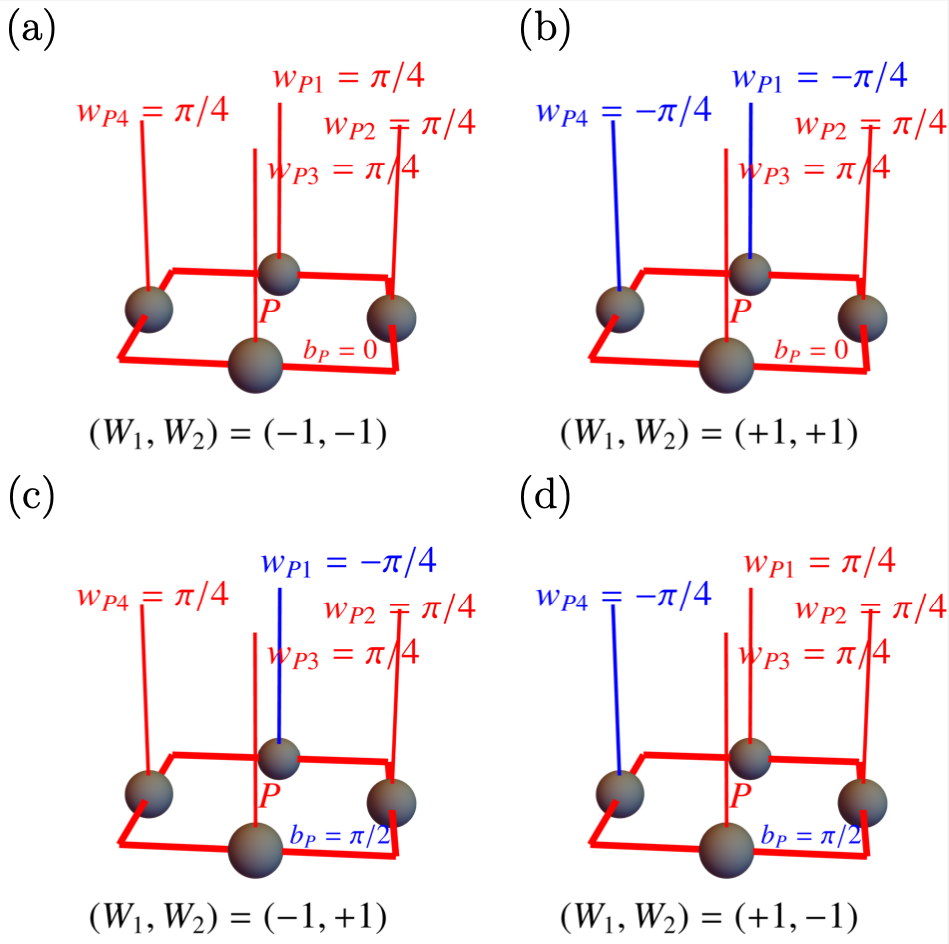}
    \caption{RBM representations of the four toric code ground states in the eigenbasis [\equref{eq:top_states_rep}] of loop operators $\hat{W}_{1}, \hat{W}_{2}$ in \equref{WOPeratorsDefinition}.}
    \label{fig:top_states_weights}
\end{figure}

\begin{figure*}[t!]
    \centering
    \includegraphics[width=\linewidth]{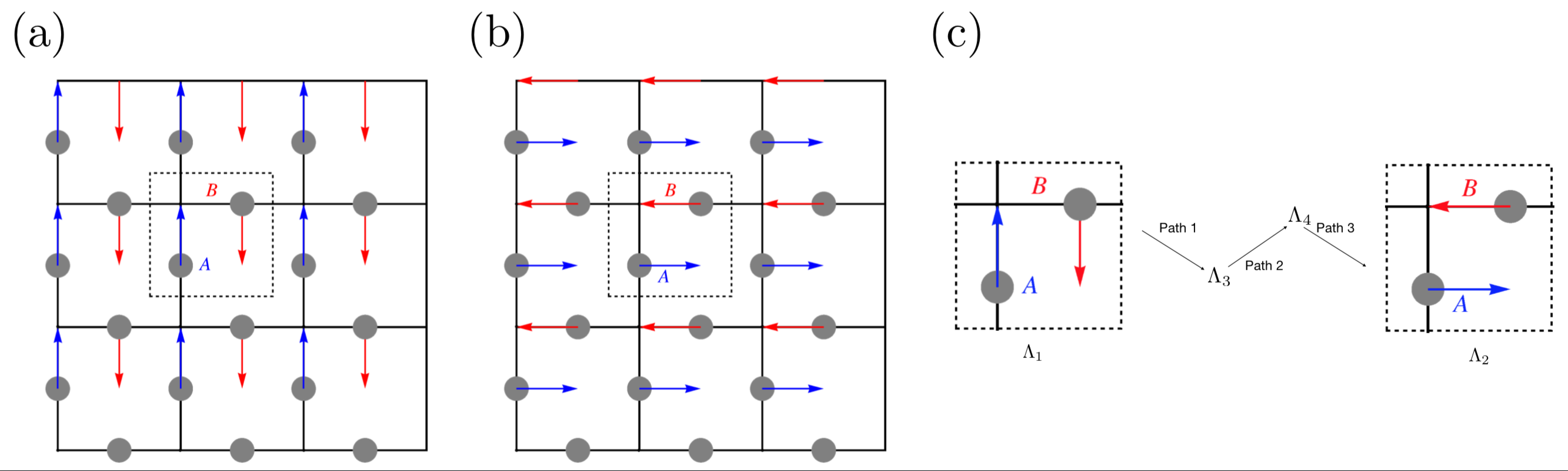}
    \caption{(a-b) Two RBM representations \equref{eq:rbm_polarized} of the polarized state. (c) A path that connects the presentation for two spins in (a-b), which is explicitly shown in Table.~\ref{tb:path}. }
    \label{fig:polarized}
\end{figure*}

\subsection{Ground states representation in different topological sectors}\label{app:ansatz_top}
Placing the toric code model in \equref{eq:ham_tc} on the torus geometry, it is useful to define the loop operators,
\begin{subequations}\begin{align}
\hat{W}_1 &= \prod_{i \in \bar{l}_x} \hat{s}^x_i, \quad \hat{W}_2 = \prod_{i \in \bar{l}_y} \hat{s}^x_i, \label{WOPeratorsDefinition}\\
\hat{V}_1 &= \prod_{i \in l_x} \hat{s}^z_i, \quad \hat{V}_2 = \prod_{i \in l_y} \hat{s}^z_i,
\end{align}\label{LoopOperators}\end{subequations}
where $l_{x, y}$ is a non-contractible loop along $x$, $y$ direction, and $\bar{l}_{x, y}$ is similar on the dual lattice. Note the loop operators along two directions do not commute with each other as $\commu{\hat{W}_{1}}{\hat{V}_2}\neq0$ and $\commu{\hat{W}_{2}}{\hat{V}_1}\neq0$. However, since the hamiltonian commute with these loop operators $\commu{\hat{W}_{1,2}}{\hat{H}_{\rm tc}}$\,$=$\,$\commu{\hat{V}_{1,2}}{\hat{H}_{\rm tc}}$\,$=$\,$0$, it follows that the ground state subspace is four-fold degenerate and spanned by the eigenvectors of the loop operators. 

Suppose we work in the eigenbasis of $\hat{W}_{1,2}$; we define the four orthogonal ground states $\ket{\psi_i} (i=0,1,2, 3)$ that span $\mathcal{L}$ as,
\begin{subequations}\label{eq:top_states_rep}
\begin{align}
    \hat{W}_1 \ket{\psi_0} &=  \ket{\psi_0}, \quad \hat{W}_2 \ket{\psi_0} =  \ket{\psi_0}, \\ 
    \hat{W}_1 \ket{\psi_1} &=  \ket{\psi_1}, \quad \hat{W}_2 \ket{\psi_1} =  -\ket{\psi_1}, \\ 
    \hat{W}_1 \ket{\psi_2} &=  \ket{\psi_2}, \quad \hat{W}_2 \ket{\psi_2} =  -\ket{\psi_2}, \\ 
    \hat{W}_1 \ket{\psi_3} &=  -\ket{\psi_3}, \quad \hat{W}_2 \ket{\psi_3} =  -\ket{\psi_3}.
\end{align}
\end{subequations}
The RBM ansatz in \equref{eq:rbm_psi} can represent eigenstates of $\hat{W}_{1,2}$ with eigenvalues $(W_1, W_2) = (\pm 1, \pm 1)$. Ref.~\cite{dengMachineLearningTopological2017} gave an representation of $\ket{\psi_3}$ with parameters,
\begin{subequations}\label{eq:toricCode_weights}
\begin{align}
    w_{Pj} &= \frac{\pi}{4}, \quad b_{P} = 0, \quad w_{Sj} = \frac{\pi}{2}, \quad b_{S} = 0.
\end{align}
\end{subequations}
On a system with odd number of sites along $x$ and $y$ direction, the other three degenerate states can be realized analogously by fixing the weights associated to stars to be $w_{Sj}$\,$=$\,$0$,\,$ b_{S}$\,$=$\,$0$.
Then the four states can be chosen by changing the $w_{Pj}$ and $b_P$ as shown in Fig.~\ref{fig:top_states_weights}.

\begin{widetext}
\subsection{Network parameter redundancies in polarized phase}\label{app:ansatz_redundancy}
In \secref{sec:ToricCodeExample}, we identified a set of gauge transformations \equref{gTrafos} that leave a generic wavefunction parameterized by the RBM ansatz in \equref{eq:rbm} invariant up to a global phase [\equref{GaugeTrafoOnWaveFunc}]. Such gauge transformations should be taken into consideration when evaluating the similarity measure $S_n$. Moreover, we have numerically verified that for states generated close to the exact toric code wave functions, $S_n$ is a good proxy for the quantum measure $S_{\text{q}}$ after explicit removals of such redundancies via $S_{n}$ in \equref{Srbm}. However, as alluded to in the discussions of the large-$h$ limit, there are state-specific redundancies that are generally not related by the gauge transformations in \equref{gTrafos}. 

Let us illustrate such redundancies here for the polarized state $\ket{\Psi} =  \ket{1, \cdots, 1}_z$ which has all spin pointing up in the $z$-basis. Notice that there is the same number of $\cos(\cdot)$ factors in the wavefunction ansatz as the number of spins. As a result, we can define a ``\textit{covering}'' by assigning each individual spin to a single factor, and choosing the weights to ensure all spins are pointing up. Any such ``covering" is a valid representation of the polarized state. For example, one representation is given by,
\begin{equation}\label{eq:polarizedWeights}
    b_P=b_S=-\frac{\pi}{4}, \,\,\,\, w_{Sj} = \begin{cases} \frac{\pi}{4},\,\,\,\, j=j_s(S),  \\
    0, \,\,\,\, \text{otherwise},
    \end{cases} \rm{and} \,\,\,\, w_{Pj} = \begin{cases} \frac{\pi}{4},\,\,\,\, j=j_n(P),  \\
    0, \,\,\,\, \text{otherwise}.
    \end{cases}
\end{equation}
where $j_s(S)$ denotes the ``southmost'' spin in the star $S$ and $j_n(P)$ denotes the ``northmost'' spin in the plaquette $P$ [see \figref{fig:polarized}(a)]. Any such coverings of the spins will correspond to a polarized state. For example, performing a ``rotation'' leads to a different covering in \figref{fig:polarized}(b). Actually, because most amplitudes in local-$z$ basis are $0$ so there are so few constraints in the wave function amplitudes, a continuous set of weights exist to represent the polarized state, so there are an infinite amount of redundancies for completely polarized state.

To illustrate this, let us consider the simplest example of just two spins [the boxed region in \figref{fig:polarized}(c)] with the same RBM ansatz, which can be easily generalized to more spins.
 For two spins, such ansatz is given by,
\begin{align}
    \psi_{\Lambda}(\sigma_A, \sigma_B) = \cos(b_{S} + w_{SA} \sigma_A + w_{SB} \sigma_B) \cos(b_{P} + w_{PA}  \sigma_A + w_{PB} \sigma_B),
\end{align}
where the weights $\Lambda = \{\Lambda_S = \{b_{S}, w_{SA}, w_{SB}\},\, \Lambda_P=\{b_{P}, w_{PA} , w_{PB}\}\}$ with $\Lambda_{Xj} \in [0, \pi)$ for $X= S$ or $P$ fully determine the two-qubits physical state. For example, the following two choices of weights [$\Lambda_1$ and $\Lambda_2$ pictorially in \figref{fig:polarized}(c)]  both parametrize the polarized state:
\begin{subequations}\label{eq:rbm_polarized}
\begin{align}
     \Lambda_1 = \{b_{S} &=-\frac{\pi}{4}, w_{SA}=0, w_{SB}=\frac{\pi}{4}, b_{P}=-\frac{\pi}{4}, w_{PA} =\frac{\pi}{4}, w_{PB}=0\}, \\
    \Lambda_2 = \{b_{S} &=-\frac{\pi}{4}, w_{SA}=\frac{\pi}{4}, w_{SB}=0, b_{P}=-\frac{\pi}{4}, w_{PA} =0, w_{PB}=\frac{\pi}{4}\}, \\
     \psi_{\Lambda_{1, 2}} &=\begin{cases}
			1, & \sigma_A=\sigma_B=1,\\
            0, & \text{otherwise}.
		 \end{cases}
\end{align}
\end{subequations}

Now to illustrate the continuous redundancies, we construct a path in the parameter space to go from $\Lambda_{1}$ to $\Lambda_{2}$. The path is composed of three steps [\figref{fig:polarized}(c)],
\begin{align}
    \Lambda_1 \xrightarrow{\text{path}\, 1} \Lambda_3 \xrightarrow{\text{path}\, 2} \Lambda_4 \xrightarrow{\text{path}\, 3} \Lambda_2, 
\end{align}
where the intermediate parameters are given by, 
\begin{align}
    \Lambda_3 = \{b_{S} &=0, w_{SA}=\frac{\pi}{4}, w_{SB}=-\frac{\pi}{4}, b_{P}=-\frac{\pi}{4}, w_{PA} =\frac{\pi}{4}, w_{PB}=0\}, \\
    \Lambda_4 = \{b_{S} &=0, w_{SA}=\frac{\pi}{4}, w_{SB}=-\frac{\pi}{4}, b_{P}=-\frac{\pi}{4}, w_{PA} =0, w_{PB}=\frac{\pi}{4}\}.
\end{align}
Along each path component, referred to as path $1$ through $3$ in \tableref{tb:path}, the parameters of $S$ (or $P$) are varied and the other held fixed, while remaining in the exactly polarized state. The path is continuous except at a singular point on path $1$ where the wave function vanishes at $\Lambda_{\rm singular}=\{b_{S}= 0, w_{SA}= \frac{\pi}{4}, w_{SB}=-\frac{\pi}{4}, b_{P}=-\frac{\pi}{4}, w_{PA} =\frac{\pi}{4}, w_{PB} = 0\}$. 

\begin{table*}[t!]
\begin{tabular}{ |c||c|c|c|c|c||  }
 \hline
 Path 1& $w_{SB} = b_{S} + w_{SA} - \frac{\pi}{2}$ & $\Lambda_P$ fixed & product $\psi = \psi_S \times \psi_P$\\
 $\Lambda_1 \rightarrow \Lambda_3$ & $w_{SA}: [0,  \frac{\pi}{4}), w_{SB}: [\frac{\pi}{4}, -\frac{\pi}{4}), b_{S}:[- \frac{\pi}{4}, 0)$ & $w_{PA} =\frac{\pi}{4}, w_{PB} = 0, b_{P}=-\frac{\pi}{4}$ &\\
 \hline\hline
 $\cos(b_X + w_{XA} + w_{XB})$ & $\neq 0$ if $b_{S} + w_{SA} \neq \frac{n}{2}\pi, n\in \mathbb{Z} \color{red} \rightarrow 0 \rightarrow 1$    & $1$ &   $\color{red} \rightarrow 0 \rightarrow 1$\\
 \hline
 $\cos(b_X + w_{XA} - w_{XB})$ &   0  &    &0\, \checkmark\\
 \hline
 $\cos(b_X - w_{XA} + w_{XB})$ &$\cos(2 b_{S} - \frac{\pi}{2}) \rightarrow 0 $ & 0&  0\, \checkmark\\
 \hline
 $\cos(b_X - w_{XA} - w_{XB})$ & & 0&  0\, \checkmark\\

 \hline\hline
 Path 2& $\Lambda_S$ fixed & $w_{PB}=b_{P} - w_{PA}  + \frac{\pi}{2}$ &\\
  $\Lambda_3 \rightarrow \Lambda_4$& $w_{SA}= \frac{\pi}{4}, w_{SB}=-\frac{\pi}{4}, b_{S}= 0$ & $w_{PA} : [\frac{\pi}{4},  0], w_{PB}: [0, \frac{\pi}{4}], b_{P}=- \frac{\pi}{4}$ &\\
 \hline
  $\cos(b_X + w_{XA} + w_{XB})$ & 1 & 1 & 1\\
  \hline
  $\cos(b_X + w_{XA} - w_{XB})$ &  0   &   $\cos(2 w_{PA}  - \frac{\pi}{2}) \rightarrow 0$ &0\, \checkmark\\
 \hline
 $\cos(b_X - w_{XA} + w_{XB})$ & 0 & &  0\, \checkmark\\
 \hline
 $\cos(b_X - w_{XA} - w_{XB})$ & & 0&  0\, \checkmark\\
 \hline\hline
  Path 3 & $w_{SB} = -b_{S} + w_{SA} + \frac{\pi}{2}$ & $\Lambda_P$ fixed &\\
  $\Lambda_4 \rightarrow \Lambda_2$ & $w_{SA}=\frac{\pi}{4}, w_{SB}: (-\frac{\pi}{4}, 0], b_{S}:(0, -\frac{\pi}{4}]$&  $w_{PA} =0, w_{PB}=\frac{\pi}{4}, b_{P}=-\frac{\pi}{4}$  &\\
 \hline
  $\cos(b_X + w_{XA} + w_{XB})$ & 1 & 1 &  $1$\\
  \hline
  $\cos(b_X + w_{XA} - w_{XB})$ &     &   0 &0\, \checkmark\\
 \hline
 $\cos(b_X - w_{XA} + w_{XB})$ & $0$ & &  0\, \checkmark\\
 \hline
 $\cos(b_X - w_{XA} - w_{XB})$ &  & 0& 0\, \checkmark\\
 \hline
\end{tabular}
\caption{A path going from $\Lambda_1$ to $\Lambda_2$ is composed of three steps. Path $1$ ($\Lambda_1 \rightarrow \Lambda_3$) is smooth except at the point $w_{SA}= \frac{\pi}{4}, w_{SB}=-\frac{\pi}{4}, b_{S}= 0$, where the wavefunction vanishes. This is further denoted by the red arrows first decreasing to $0$ before increasing to $1$ in the first row. Path $2$ and $3$ are both smooth. The last column illustrates that the wavefunction $\psi$ remains in the polarized state along the path.}
\label{tb:path}
\end{table*}

\end{widetext}

\clearpage
\subsection{Resolving the special redundancies}\label{ResolveRedundancies}
In \appref{app:ansatz_redundancy}, we explicitly showed that there can be a large set of redundancies given a polarized state. Hence, for simplicity in the main text, we have used the direct overlap $S_{\text{q}}$ in \equref{eq:s_exact_overlap} as the relevant measure at finite field values. As discussed in the main text, a straightforward way to alleviate the redundancies in the similarity measure $S_n$ in \equref{Srbm} of the network parameters is to complement it with the direct overlap. By using a combination of both measures, we are able to reduce the amount of computational cost of the direct overlap by a fraction as the similarity is easy to compute. More specifically, we define a mixed measure $S_m$ by replacing a random fraction (given by $f$) of the similarity measure pairs $\{l, l^\prime\}$ by a rescaled overlap measure $\widetilde{S}_q$ such that,
\begin{align}\label{eq:S_mixed}
    S_m(l, l^\prime) = \begin{cases}
    \widetilde{S}_q(l, l^\prime) \quad \text{with probability}\,\, f, \\
     S_n(l, l^\prime) \quad \text{with probability}\,\, 1-f.
    \end{cases}
\end{align}
The following rescaling of the overlap measure $S_q$ is necessary as we want to include the two measures on an equal-footing given by,
\begin{subequations}
\begin{align}
    \widetilde{S}_{q} &= \frac{S_q - n_q}{m_q- n_q} \cdot \left( m_n- n_n\right) + n_n,\\
    m_q & = \max(S_q), \quad n_q = \min(S_q),\\
    m_n & = \max(S_n), \quad n_n = \min(S_n).
\end{align}
\end{subequations}
For example, we see that the minimum of the rescaled overlap is the same as the minimum of the similarity $\min(\widetilde{S}_q) = \min(S_n)$.

\begin{figure}[h!]
    \centering
    \includegraphics[width=\linewidth]{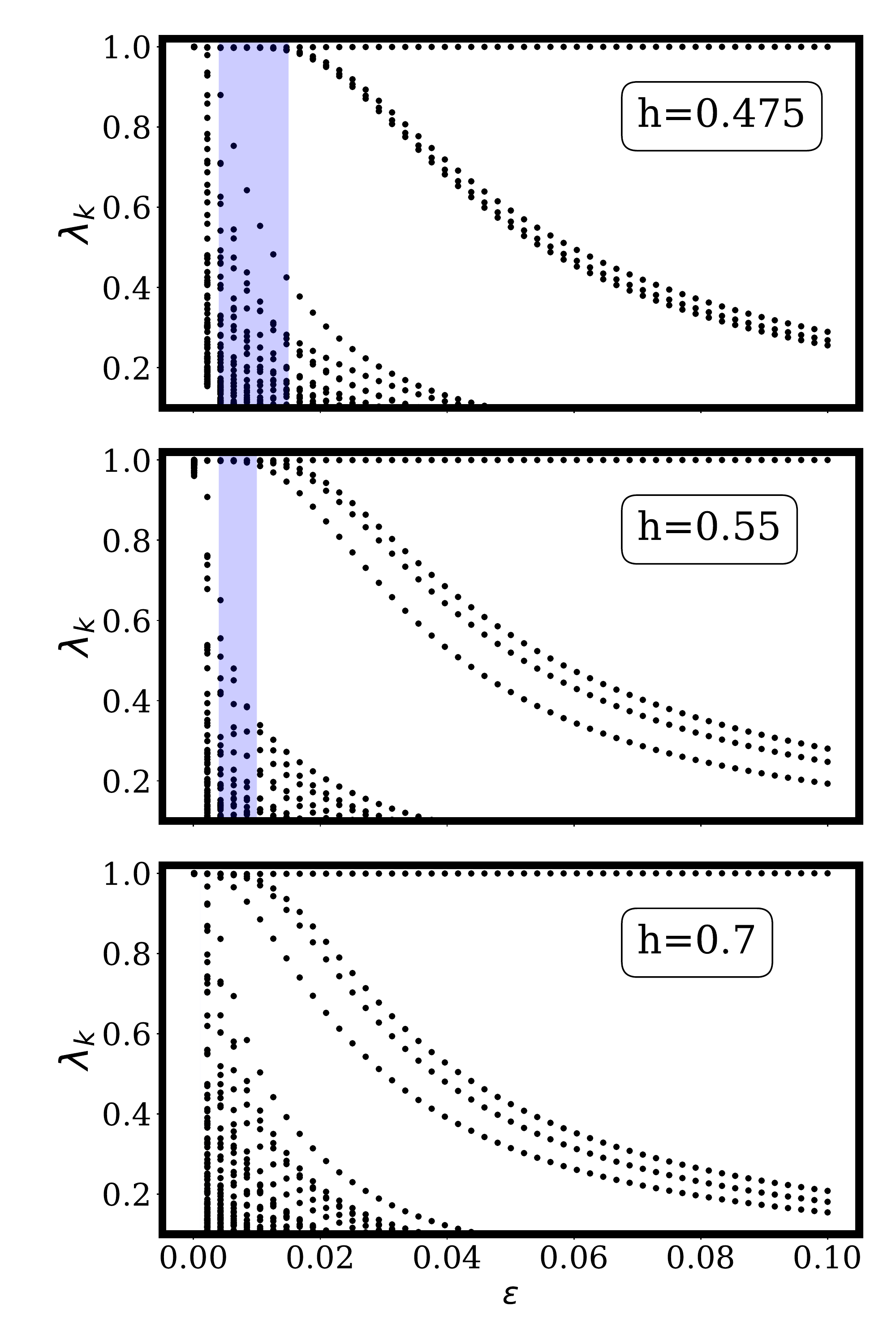}
    \caption{DM spectra for different field values $h=0.475, 0.55, 0.7$ at $T=0.3$ using a mixed similarity measure $S_m$ with a fraction $f=0.4$ in \equref{eq:S_mixed}. The blue shaded regions highlight the existence of a range of $\epsilon$ with spectral gap between the degenerate eigenvalues and the decaying eigenvalues, indicating underlying superselection sectors. As the field value approaches the transition field $h_c$, the range of such region shrinks and disappears at high field $h=0.7$, indicating the absence of sectors.  }
    \label{fig:mixed_measure}
\end{figure}

In \figref{fig:mixed_measure}, we demonstrate that by using a mixed measure with a fraction of $f=0.4$ replacement, our algorithm with DM is able to identify the presence (indicated by the shaded blue region for smaller field values $h=0.475$ and $h=0.55$) and absence ($h=0.7$) of superselection sectors across various field values, consistent with the predictions of the algorithm using direct overlap (shown in \figref{fig:field}). We note that in the case with a mixed measure, DM is a natural technique as the algorithm looks for connectivity; whereas kernel PCA would fail to identify such transition (since a fraction of pairs of wave functions are incorrectly considered to be dissimilar by $S_n$, the leading kernel PCA components still show four separated clusters up to the largest magnetic field, $h=1$).

\section{Optimization with Variational Monte Carlo}\label{app:vmc}
To find the ground state $\ket{\Psi(\Lambda^0)} \propto \sum_{\boldsymbol{\sigma}}\psi(\boldsymbol{\sigma};\Lambda^0) \ket{\boldsymbol{\sigma}}$, we wish to minimize the energy expectation $\ev{E} = \bra{\Psi} \hat{H} \ket{\Psi} /\braket{\Psi|\Psi}$ (omitting the variational parameters $\Lambda^0$ in this section), which is bounded by the ground state energy by the variational principle. An exact computation $\ev{E}_{\rm exact}$
is costly as the summation enumerates over exponentially many spin configurations $\boldsymbol{\sigma}$ as the system size increases. Here we use variational Monte Carlo (VMC) importance sampling algorithm to estimate such expectation values. The idea is to compute relative probability between different configurations and sample from the true wavefunction probability density $\abs{\psi(\boldsymbol{\sigma})}^2$, without having to compute $\abs{\psi(\boldsymbol{\sigma})}^2$ for all $\vec{\sigma}$. To perform this algorithm, we initialize $M$ random configurations $\{\boldsymbol{\sigma}_i\}_{i=1}^M$ and continue each with random walks based on previous configurations, hence forming $M$ Markov chains. 

In particular, the Metropolis–Rosenbluth algorithm~\cite{metropolisEquationStateCalculations1953} is used to propose the next configuration $\boldsymbol{\sigma}_i^\prime$ that is locally connected to $c_i$ according to function $g(\boldsymbol{\sigma}^\prime | \boldsymbol{\sigma})$. For the toric code model, we use two types of proposals: spin flips and vertex flips. Here, we will assume a probability of $p$ for proposing spin flips and analogously $1-p$ for vertex flips that are equally likely at all sites: 
\begin{align}
    g(\boldsymbol{\sigma}^\prime | \boldsymbol{\sigma}) = 
    \begin{cases}
    \frac{p}{n_s}, & \text{for }\text{spin flips} \\ 
    \frac{1-p}{n_v}, & \text{for }\text{vertex flips}
    \end{cases}
\end{align}
where $n_s$ and $n_v$ are the number of all possible spin and vertex flips. The acceptance of $\boldsymbol{\sigma}^\prime$ is determined by a probability, 
\begin{align}
    \mathbb{P}_{\rm accept}(\boldsymbol{\sigma} \rightarrow \boldsymbol{\sigma}^\prime) = \min\left(\abs{\frac{\psi(\boldsymbol{\sigma}^\prime)}{\psi(\boldsymbol{\sigma})}}^2,\,1\right).
\end{align}

The random walks will be repeated long enough so that the final configurations at the tail of the chains $\Sigma_{\rm MC} = \qtyset{\boldsymbol{\sigma}_f}_{i=b}^M$ approximate samples drawn from the probability distribution $\abs{\psi(\boldsymbol{\sigma})}^2$. A certain number $b$ of walkers in each chain are discarded to reduce the biases from initialization of the chains. Then the expectation of an observable $\hat{O}$ is given by,
\begin{subequations}
\begin{align}
    \ev{\hat{O}}_{\rm MC} & = \frac{\sum_{\boldsymbol{\sigma}} \psi(\boldsymbol{\sigma})^* \sandwich{\boldsymbol{\sigma}}{\hat{O}}{\Psi}}{\sum_{\boldsymbol{\sigma}} \abs{\psi(\boldsymbol{\sigma})}^2},\\
    &= \frac{\sum_{\boldsymbol{\sigma}} \abs{\psi(\boldsymbol{\sigma})}^2 \frac{\sandwich{\boldsymbol{\sigma}}{\hat{O}}{\Psi}}{\psi(\boldsymbol{\sigma})}}{\sum_{\boldsymbol{\sigma}} \abs{\psi(\boldsymbol{\sigma})}^2}, \\
    & = \frac{1}{M}\sum_{\boldsymbol{\sigma}\in\Sigma_{\rm MC}} \frac{\sandwich{\boldsymbol{\sigma}}{\hat{O}}{\Psi}}{\psi(\boldsymbol{\sigma})}.
\end{align}
\end{subequations}
Defining a local value of the operator $\hat{O}$ as,
\begin{align}
    O_{\rm loc} = \frac{\sandwich{\boldsymbol{\sigma}}{\hat{O}}{\Psi}}{\psi(\boldsymbol{\sigma})},
\end{align}
then the Monte Carlo estimation is the average of the local values in the Markov chain: $\ev{\hat{O}}_{\rm MC} = \frac{1}{M}\sum_{\boldsymbol{\sigma}\in\Sigma_{\rm MC}} O_{\rm loc}$.

Next, to minimize $\ev{E}$, we can compute its gradient with respect to the weights $\Lambda^0$ in terms of the local energy $E_{\rm loc}$ and wavefunction amplitude derivative $D_i$:
\begin{subequations}
\begin{align}
    \partial_{\Lambda_i} \ev{E} &= \ev{E_{\rm loc} D_i} - \ev{E_{\rm loc}} \ev{D_i} \\
    E_{\rm loc} &= \frac{\bra{\boldsymbol{\sigma}} H \ket{\Psi}}{\psi(\boldsymbol{\sigma})}, \quad D_i = \frac{\partial_{\Lambda_i} \psi(\boldsymbol{\sigma})}{\psi(\boldsymbol{\sigma})}
\end{align}
\end{subequations}
Finally, we use gradient descent with learning rate $\lambda$,
\begin{align}
    \Lambda_i \rightarrow \Lambda_i - \lambda \partial_{\Lambda_i} \ev{E},
\end{align}
to minimize the energy expectation value.
The gradient descent is performed by using an adaptive Adam optimizer~\cite{kingma2014adam}. We repeat this training step until empirical convergence. 

Note that the RBM ansatz can get stuck in local minima. To find the toric code ground state, we initialize the network parameters close to the analytic solutions in \equref{eq:toricCode_weights}.

\subsection{Fidelity}\label{app:vmc_fidelity}
\begin{figure}[t!]
    \centering
    \includegraphics[width=\linewidth]{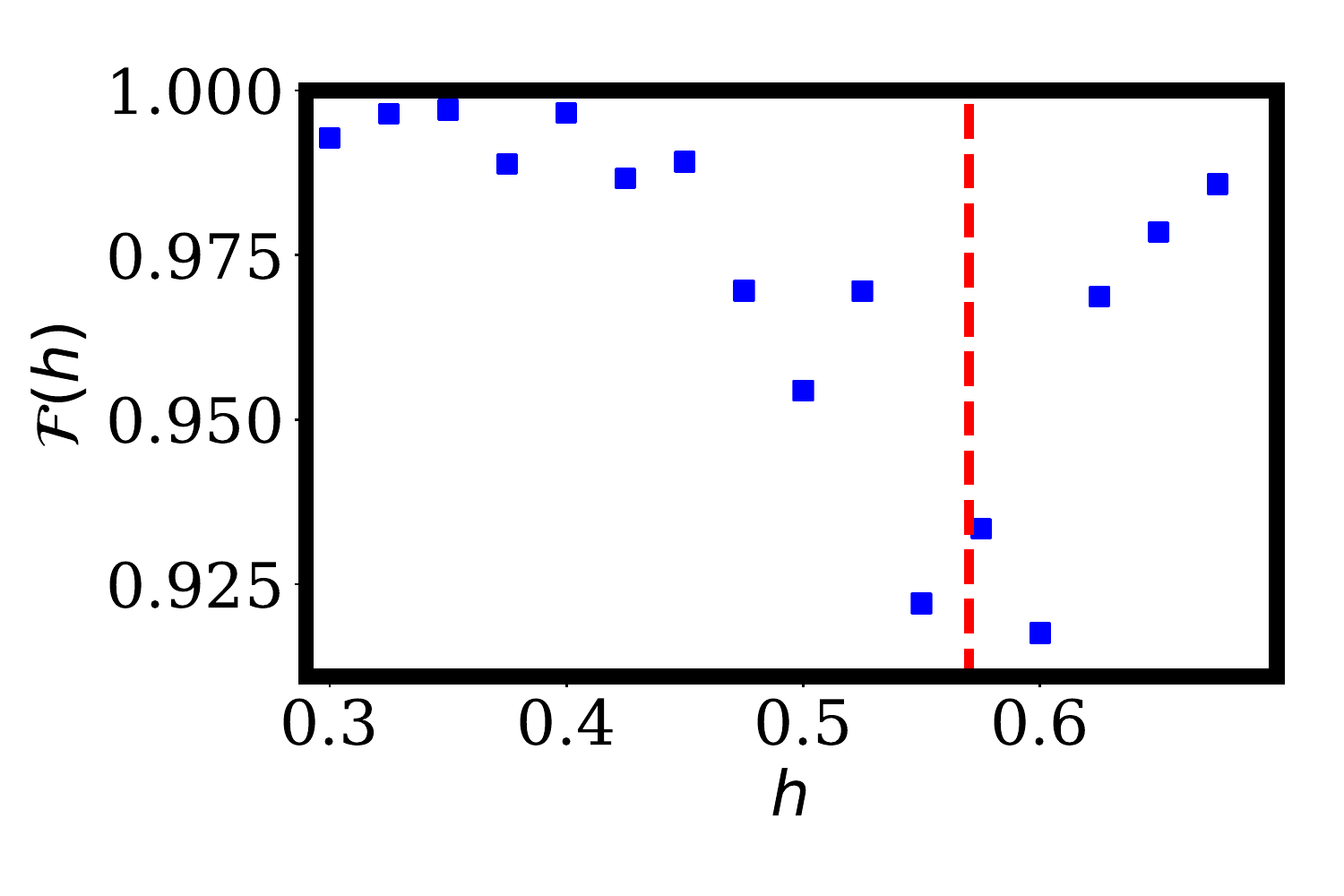}
    \caption{Fidelity $\mathcal{F}$ as a function of field $h$. The red dashed line is drawn to guide the eye, where the dip in fidelity indicates the critical field value $h_c\approx 0.57$.}
    \label{fig:fidelity}
\end{figure}

To find the approximate ground states at finite field values $h$ with step size $\Delta h$, we initialize the weights to be those from the previous field value $h- \Delta h$, and then use the current optimized weights as the initialization for the next step $h + \Delta h$. A good indication of a quantum phase transition is by inspecting the fidelity $\mathcal{F}(h)$ defined as,  
\begin{align}\label{eq:fidelity}
    \mathcal{F}(h) = \abs{\braket{\psi(h) | \psi(h + \Delta h)}}^2.
\end{align}
The critical field $h_c$ is identified as a dip in the fidelity, indicating an abrupt change in the ground state wavefunction. A field value of $h_c \approx 0.57$ (at dashed line in \figref{fig:fidelity}) is found for the RBM ansatz. Note that one can get more accurate field value by including loop expectations in the ansatz as done in \refcite{valentiCorrelationEnhancedNeuralNetworks2021}.

\begin{figure*}[t!]
    \centering
    \includegraphics[width=1.1\linewidth]{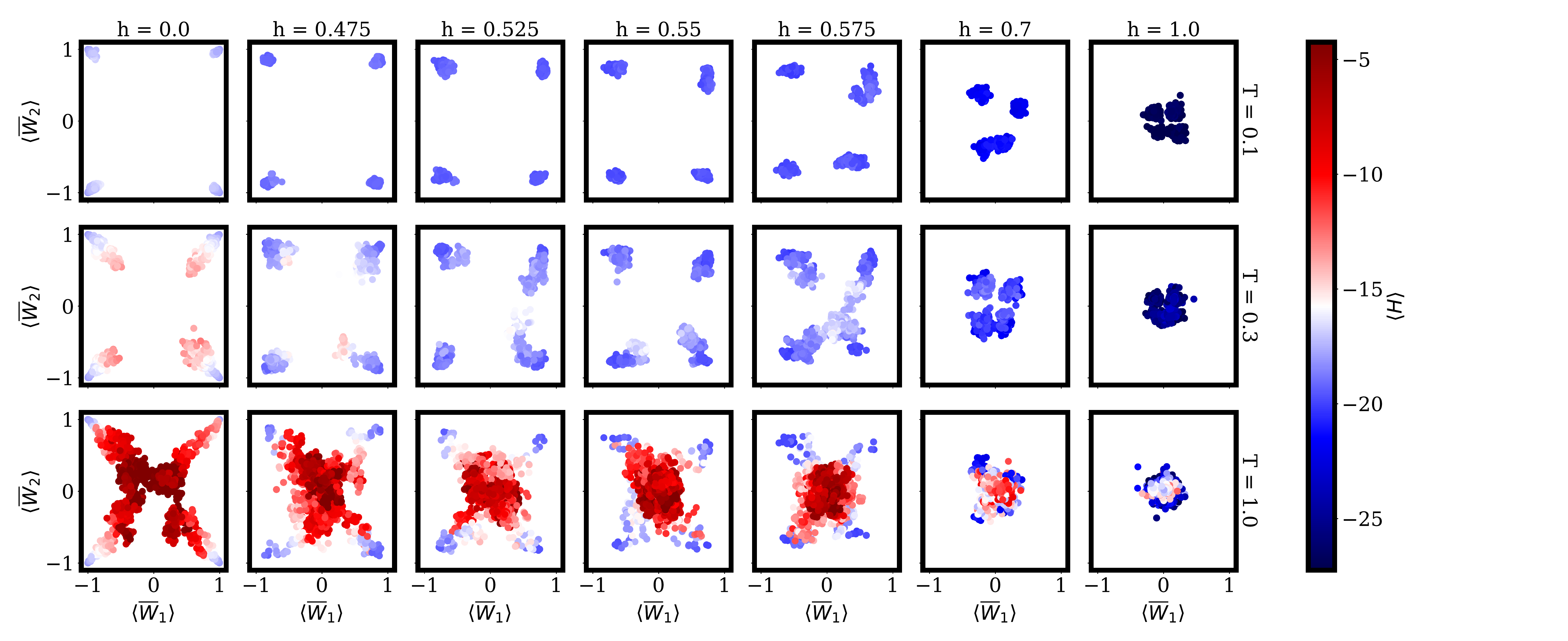}
    \caption{Illustration of the diffusion processes for different parameter $T$ and field $h$ at $N=18$ spins. The loop expectation values $\ev{\overline{W}_{1,2}}$ form four distinct clusters in the two-dimensional plane for small $T$ and $h$. For large $T=1.$ at all fields and intermediate $T=0.3$ at higher fields $h>0.57$, the clusters ``diffuse" and topological order is lost. Such ``diffusion" process can be visualized by color coding the energy expectation $\ev{H}$.}
    \label{fig:ensmble_field}
\end{figure*}
\section{Ensemble generation}\label{app:ensemble}
Using the algorithm outlined in Sec.~\ref{alg:data_generation}, we can generate ensembles that deviate from the initial optimized parameters by setting hyper-parameter $T=0.1, 0.3, 1$. The other choices of hyper-parameters for the ensembles are number of independent chains $k=2$, length of each chain $n=250$, and number of samples kept $m=n$. The parameter proposal function we use consists of with probability $p_m$ randomly apply minus sign or randomly adding local noise at a single spin site $\jmath$.  More precisely, 
\begin{subequations}\label{eq:proposal_param_fn}
\begin{align}
    f(\Lambda, \xi) &=
    \begin{cases}
    f_{-, \jmath}, \quad \text{with probability}: p_m, \\
    f_{\rm local, \jmath}, \quad \text{with probability}: 1- p_m, 
    \end{cases}\\
    f_{-, \jmath} &= \begin{cases}
    - (\Lambda )_i, \quad i \in \jmath \\
    (\Lambda )_i, \quad i \not\in \jmath
    \end{cases}\\
    f_{\rm local, \jmath} &= \begin{cases}
    \rm{uniform}(0, \xi) + (\Lambda )_i, \quad i \in \jmath \\
    (\Lambda )_i, \quad i \not\in \jmath
    \end{cases}
\end{align}
\end{subequations}

In the exact toric code state, $f_{-, \jmath}$ corresponds to act $\sigma_x$ operator at site $\jmath$ to create a pair of m-particles. In the trivial phase, depending on the parametrization of the state, $f_{-, \jmath}$ could correspond to a single spin flip at site $\jmath$. The hyperparameters are chosen to be $p_m=0.3$ and $\xi=0.2$. In \figref{fig:ensmble_field}, we visualize the ensembles by computing their loop expectations $\ev{\overline{W}_j}$ at different field values.

\bibliography{Refs.bib}

\begin{thebibliography}{75}%
\makeatletter
\providecommand \@ifxundefined [1]{%
 \@ifx{#1\undefined}
}%
\providecommand \@ifnum [1]{%
 \ifnum #1\expandafter \@firstoftwo
 \else \expandafter \@secondoftwo
 \fi
}%
\providecommand \@ifx [1]{%
 \ifx #1\expandafter \@firstoftwo
 \else \expandafter \@secondoftwo
 \fi
}%
\providecommand \natexlab [1]{#1}%
\providecommand \enquote  [1]{``#1''}%
\providecommand \bibnamefont  [1]{#1}%
\providecommand \bibfnamefont [1]{#1}%
\providecommand \citenamefont [1]{#1}%
\providecommand \href@noop [0]{\@secondoftwo}%
\providecommand \href [0]{\begingroup \@sanitize@url \@href}%
\providecommand \@href[1]{\@@startlink{#1}\@@href}%
\providecommand \@@href[1]{\endgroup#1\@@endlink}%
\providecommand \@sanitize@url [0]{\catcode `\\12\catcode `\$12\catcode
  `\&12\catcode `\#12\catcode `\^12\catcode `\_12\catcode `\%12\relax}%
\providecommand \@@startlink[1]{}%
\providecommand \@@endlink[0]{}%
\providecommand \url  [0]{\begingroup\@sanitize@url \@url }%
\providecommand \@url [1]{\endgroup\@href {#1}{\urlprefix }}%
\providecommand \urlprefix  [0]{URL }%
\providecommand \Eprint [0]{\href }%
\providecommand \doibase [0]{http://dx.doi.org/}%
\providecommand \selectlanguage [0]{\@gobble}%
\providecommand \bibinfo  [0]{\@secondoftwo}%
\providecommand \bibfield  [0]{\@secondoftwo}%
\providecommand \translation [1]{[#1]}%
\providecommand \BibitemOpen [0]{}%
\providecommand \bibitemStop [0]{}%
\providecommand \bibitemNoStop [0]{.\EOS\space}%
\providecommand \EOS [0]{\spacefactor3000\relax}%
\providecommand \BibitemShut  [1]{\csname bibitem#1\endcsname}%
\let\auto@bib@innerbib\@empty
\bibitem [{\citenamefont {Mehta}\ \emph {et~al.}(2019)\citenamefont {Mehta},
  \citenamefont {Bukov}, \citenamefont {Wang}, \citenamefont {Day},
  \citenamefont {Richardson}, \citenamefont {Fisher},\ and\ \citenamefont
  {Schwab}}]{mehtaHighbiasLowvarianceIntroduction2019}%
  \BibitemOpen
  \bibfield  {author} {\bibinfo {author} {\bibfnamefont {Pankaj}\ \bibnamefont
  {Mehta}}, \bibinfo {author} {\bibfnamefont {Marin}\ \bibnamefont {Bukov}},
  \bibinfo {author} {\bibfnamefont {Ching-Hao}\ \bibnamefont {Wang}}, \bibinfo
  {author} {\bibfnamefont {Alexandre G.~R.}\ \bibnamefont {Day}}, \bibinfo
  {author} {\bibfnamefont {Clint}\ \bibnamefont {Richardson}}, \bibinfo
  {author} {\bibfnamefont {Charles~K.}\ \bibnamefont {Fisher}}, \ and\ \bibinfo
  {author} {\bibfnamefont {David~J.}\ \bibnamefont {Schwab}},\ }\bibfield
  {title} {\enquote {\bibinfo {title} {A high-bias, low-variance introduction
  to {{Machine Learning}} for physicists},}\ }\href {\doibase
  10.1016/j.physrep.2019.03.001} {\bibfield  {journal} {\bibinfo  {journal}
  {Physics Reports}\ }\bibinfo {series} {A High-Bias, Low-Variance Introduction
  to {{Machine Learning}} for Physicists},\ \textbf {\bibinfo {volume} {810}},\
  \bibinfo {pages} {1--124} (\bibinfo {year} {2019})}\BibitemShut {NoStop}%
\bibitem [{\citenamefont {Carleo}\ \emph {et~al.}(2019)\citenamefont {Carleo},
  \citenamefont {Cirac}, \citenamefont {Cranmer}, \citenamefont {Daudet},
  \citenamefont {Schuld}, \citenamefont {Tishby}, \citenamefont
  {{Vogt-Maranto}},\ and\ \citenamefont
  {Zdeborov{\'a}}}]{carleoMachineLearningPhysical2019}%
  \BibitemOpen
  \bibfield  {author} {\bibinfo {author} {\bibfnamefont {Giuseppe}\
  \bibnamefont {Carleo}}, \bibinfo {author} {\bibfnamefont {Ignacio}\
  \bibnamefont {Cirac}}, \bibinfo {author} {\bibfnamefont {Kyle}\ \bibnamefont
  {Cranmer}}, \bibinfo {author} {\bibfnamefont {Laurent}\ \bibnamefont
  {Daudet}}, \bibinfo {author} {\bibfnamefont {Maria}\ \bibnamefont {Schuld}},
  \bibinfo {author} {\bibfnamefont {Naftali}\ \bibnamefont {Tishby}}, \bibinfo
  {author} {\bibfnamefont {Leslie}\ \bibnamefont {{Vogt-Maranto}}}, \ and\
  \bibinfo {author} {\bibfnamefont {Lenka}\ \bibnamefont {Zdeborov{\'a}}},\
  }\bibfield  {title} {\enquote {\bibinfo {title} {Machine learning and the
  physical sciences},}\ }\href {\doibase 10.1103/RevModPhys.91.045002}
  {\bibfield  {journal} {\bibinfo  {journal} {Rev. Mod. Phys.}\ }\textbf
  {\bibinfo {volume} {91}},\ \bibinfo {pages} {045002} (\bibinfo {year}
  {2019})}\BibitemShut {NoStop}%
\bibitem [{\citenamefont {{Das Sarma}}\ \emph {et~al.}(2019)\citenamefont {{Das
  Sarma}}, \citenamefont {{Deng}},\ and\ \citenamefont
  {{Duan}}}]{2019PhT....72c..48D}%
  \BibitemOpen
  \bibfield  {author} {\bibinfo {author} {\bibfnamefont {Sankar}\ \bibnamefont
  {{Das Sarma}}}, \bibinfo {author} {\bibfnamefont {Dong-Ling}\ \bibnamefont
  {{Deng}}}, \ and\ \bibinfo {author} {\bibfnamefont {Lu-Ming}\ \bibnamefont
  {{Duan}}},\ }\bibfield  {title} {\enquote {\bibinfo {title} {{Machine
  learning meets quantum physics}},}\ }\href {\doibase 10.1063/PT.3.4164}
  {\bibfield  {journal} {\bibinfo  {journal} {Physics Today}\ }\textbf
  {\bibinfo {volume} {72}},\ \bibinfo {pages} {48--54} (\bibinfo {year}
  {2019})},\ \Eprint {http://arxiv.org/abs/1903.03516} {arXiv:1903.03516
  [physics.pop-ph]} \BibitemShut {NoStop}%
\bibitem [{\citenamefont {Melko}\ \emph {et~al.}(2019)\citenamefont {Melko},
  \citenamefont {Carleo}, \citenamefont {Carrasquilla},\ and\ \citenamefont
  {Cirac}}]{RBMReview}%
  \BibitemOpen
  \bibfield  {author} {\bibinfo {author} {\bibfnamefont {Roger~G.}\
  \bibnamefont {Melko}}, \bibinfo {author} {\bibfnamefont {Giuseppe}\
  \bibnamefont {Carleo}}, \bibinfo {author} {\bibfnamefont {Juan}\ \bibnamefont
  {Carrasquilla}}, \ and\ \bibinfo {author} {\bibfnamefont {J.~Ignacio}\
  \bibnamefont {Cirac}},\ }\bibfield  {title} {\enquote {\bibinfo {title}
  {Restricted boltzmann machines in quantum physics},}\ }\href {\doibase
  10.1038/s41567-019-0545-1} {\bibfield  {journal} {\bibinfo  {journal} {Nature
  Physics}\ }\textbf {\bibinfo {volume} {15}},\ \bibinfo {pages} {887--892}
  (\bibinfo {year} {2019})}\BibitemShut {NoStop}%
\bibitem [{\citenamefont {Carrasquilla}(2020)}]{CarrasquillaReview}%
  \BibitemOpen
  \bibfield  {author} {\bibinfo {author} {\bibfnamefont {Juan}\ \bibnamefont
  {Carrasquilla}},\ }\bibfield  {title} {\enquote {\bibinfo {title} {Machine
  learning for quantum matter},}\ }\href {\doibase
  10.1080/23746149.2020.1797528} {\bibfield  {journal} {\bibinfo  {journal}
  {Advances in Physics: X}\ }\textbf {\bibinfo {volume} {5}},\ \bibinfo {pages}
  {1797528} (\bibinfo {year} {2020})}\BibitemShut {NoStop}%
\bibitem [{\citenamefont {Carrasquilla}\ and\ \citenamefont
  {Torlai}(2021)}]{carrasquillaHowUseNeural2021}%
  \BibitemOpen
  \bibfield  {author} {\bibinfo {author} {\bibfnamefont {Juan}\ \bibnamefont
  {Carrasquilla}}\ and\ \bibinfo {author} {\bibfnamefont {Giacomo}\
  \bibnamefont {Torlai}},\ }\bibfield  {title} {\enquote {\bibinfo {title} {How
  {{To Use Neural Networks To Investigate Quantum Many-Body Physics}}},}\
  }\href {\doibase 10.1103/PRXQuantum.2.040201} {\bibfield  {journal} {\bibinfo
   {journal} {PRX Quantum}\ }\textbf {\bibinfo {volume} {2}},\ \bibinfo {pages}
  {040201} (\bibinfo {year} {2021})}\BibitemShut {NoStop}%
\bibitem [{\citenamefont {Dawid}\ \emph {et~al.}(2022)\citenamefont {Dawid},
  \citenamefont {Arnold}, \citenamefont {Requena}, \citenamefont {Gresch},
  \citenamefont {P{\l}odzie{\'n}}, \citenamefont {Donatella}, \citenamefont
  {Nicoli}, \citenamefont {Stornati}, \citenamefont {Koch}, \citenamefont
  {B{\"u}ttner}, \citenamefont {Oku{\l}a}, \citenamefont {{Mu{\~n}oz-Gil}},
  \citenamefont {{Vargas-Hern{\'a}ndez}}, \citenamefont {{Cervera-Lierta}},
  \citenamefont {Carrasquilla}, \citenamefont {Dunjko}, \citenamefont
  {Gabri{\'e}}, \citenamefont {Huembeli}, \citenamefont {{van Nieuwenburg}},
  \citenamefont {Vicentini}, \citenamefont {Wang}, \citenamefont {Wetzel},
  \citenamefont {Carleo}, \citenamefont {Greplov{\'a}}, \citenamefont {Krems},
  \citenamefont {Marquardt}, \citenamefont {Tomza}, \citenamefont
  {Lewenstein},\ and\ \citenamefont
  {Dauphin}}]{dawidModernApplicationsMachine2022}%
  \BibitemOpen
  \bibfield  {author} {\bibinfo {author} {\bibfnamefont {Anna}\ \bibnamefont
  {Dawid}}, \bibinfo {author} {\bibfnamefont {Julian}\ \bibnamefont {Arnold}},
  \bibinfo {author} {\bibfnamefont {Borja}\ \bibnamefont {Requena}}, \bibinfo
  {author} {\bibfnamefont {Alexander}\ \bibnamefont {Gresch}}, \bibinfo
  {author} {\bibfnamefont {Marcin}\ \bibnamefont {P{\l}odzie{\'n}}}, \bibinfo
  {author} {\bibfnamefont {Kaelan}\ \bibnamefont {Donatella}}, \bibinfo
  {author} {\bibfnamefont {Kim~A.}\ \bibnamefont {Nicoli}}, \bibinfo {author}
  {\bibfnamefont {Paolo}\ \bibnamefont {Stornati}}, \bibinfo {author}
  {\bibfnamefont {Rouven}\ \bibnamefont {Koch}}, \bibinfo {author}
  {\bibfnamefont {Miriam}\ \bibnamefont {B{\"u}ttner}}, \bibinfo {author}
  {\bibfnamefont {Robert}\ \bibnamefont {Oku{\l}a}}, \bibinfo {author}
  {\bibfnamefont {Gorka}\ \bibnamefont {{Mu{\~n}oz-Gil}}}, \bibinfo {author}
  {\bibfnamefont {Rodrigo~A.}\ \bibnamefont {{Vargas-Hern{\'a}ndez}}}, \bibinfo
  {author} {\bibfnamefont {Alba}\ \bibnamefont {{Cervera-Lierta}}}, \bibinfo
  {author} {\bibfnamefont {Juan}\ \bibnamefont {Carrasquilla}}, \bibinfo
  {author} {\bibfnamefont {Vedran}\ \bibnamefont {Dunjko}}, \bibinfo {author}
  {\bibfnamefont {Marylou}\ \bibnamefont {Gabri{\'e}}}, \bibinfo {author}
  {\bibfnamefont {Patrick}\ \bibnamefont {Huembeli}}, \bibinfo {author}
  {\bibfnamefont {Evert}\ \bibnamefont {{van Nieuwenburg}}}, \bibinfo {author}
  {\bibfnamefont {Filippo}\ \bibnamefont {Vicentini}}, \bibinfo {author}
  {\bibfnamefont {Lei}\ \bibnamefont {Wang}}, \bibinfo {author} {\bibfnamefont
  {Sebastian~J.}\ \bibnamefont {Wetzel}}, \bibinfo {author} {\bibfnamefont
  {Giuseppe}\ \bibnamefont {Carleo}}, \bibinfo {author} {\bibfnamefont {Eli{\v
  s}ka}\ \bibnamefont {Greplov{\'a}}}, \bibinfo {author} {\bibfnamefont
  {Roman}\ \bibnamefont {Krems}}, \bibinfo {author} {\bibfnamefont {Florian}\
  \bibnamefont {Marquardt}}, \bibinfo {author} {\bibfnamefont {Micha{\l}}\
  \bibnamefont {Tomza}}, \bibinfo {author} {\bibfnamefont {Maciej}\
  \bibnamefont {Lewenstein}}, \ and\ \bibinfo {author} {\bibfnamefont
  {Alexandre}\ \bibnamefont {Dauphin}},\ }\href
  {http://arxiv.org/abs/2204.04198} {\enquote {\bibinfo {title} {Modern
  applications of machine learning in quantum sciences},}\ } (\bibinfo {year}
  {2022}),\ \Eprint {http://arxiv.org/abs/2204.04198} {arXiv:2204.04198
  [cond-mat, physics:quant-ph]} \BibitemShut {NoStop}%
\bibitem [{\citenamefont {Carrasquilla}\ and\ \citenamefont
  {Melko}(2017)}]{MelkoPT}%
  \BibitemOpen
  \bibfield  {author} {\bibinfo {author} {\bibfnamefont {Juan}\ \bibnamefont
  {Carrasquilla}}\ and\ \bibinfo {author} {\bibfnamefont {Roger~G.}\
  \bibnamefont {Melko}},\ }\bibfield  {title} {\enquote {\bibinfo {title}
  {Machine learning phases of matter},}\ }\href {\doibase 10.1038/nphys4035}
  {\bibfield  {journal} {\bibinfo  {journal} {Nature Physics}\ }\textbf
  {\bibinfo {volume} {13}},\ \bibinfo {pages} {431--434} (\bibinfo {year}
  {2017})}\BibitemShut {NoStop}%
\bibitem [{\citenamefont {Zhang}\ \emph {et~al.}(2018)\citenamefont {Zhang},
  \citenamefont {Shen},\ and\ \citenamefont {Zhai}}]{PhysRevLett.120.066401}%
  \BibitemOpen
  \bibfield  {author} {\bibinfo {author} {\bibfnamefont {Pengfei}\ \bibnamefont
  {Zhang}}, \bibinfo {author} {\bibfnamefont {Huitao}\ \bibnamefont {Shen}}, \
  and\ \bibinfo {author} {\bibfnamefont {Hui}\ \bibnamefont {Zhai}},\
  }\bibfield  {title} {\enquote {\bibinfo {title} {Machine learning topological
  invariants with neural networks},}\ }\href {\doibase
  10.1103/PhysRevLett.120.066401} {\bibfield  {journal} {\bibinfo  {journal}
  {Phys. Rev. Lett.}\ }\textbf {\bibinfo {volume} {120}},\ \bibinfo {pages}
  {066401} (\bibinfo {year} {2018})}\BibitemShut {NoStop}%
\bibitem [{\citenamefont {Zhang}\ and\ \citenamefont
  {Kim}(2017)}]{zhangQuantumLoopTopography2017}%
  \BibitemOpen
  \bibfield  {author} {\bibinfo {author} {\bibfnamefont {Yi}~\bibnamefont
  {Zhang}}\ and\ \bibinfo {author} {\bibfnamefont {Eun-Ah}\ \bibnamefont
  {Kim}},\ }\bibfield  {title} {\enquote {\bibinfo {title} {Quantum {{Loop
  Topography}} for {{Machine Learning}}},}\ }\href {\doibase
  10.1103/PhysRevLett.118.216401} {\bibfield  {journal} {\bibinfo  {journal}
  {Phys. Rev. Lett.}\ }\textbf {\bibinfo {volume} {118}},\ \bibinfo {pages}
  {216401} (\bibinfo {year} {2017})}\BibitemShut {NoStop}%
\bibitem [{\citenamefont {Beach}\ \emph {et~al.}(2018)\citenamefont {Beach},
  \citenamefont {Golubeva},\ and\ \citenamefont {Melko}}]{PhysRevB.97.045207}%
  \BibitemOpen
  \bibfield  {author} {\bibinfo {author} {\bibfnamefont {Matthew J.~S.}\
  \bibnamefont {Beach}}, \bibinfo {author} {\bibfnamefont {Anna}\ \bibnamefont
  {Golubeva}}, \ and\ \bibinfo {author} {\bibfnamefont {Roger~G.}\ \bibnamefont
  {Melko}},\ }\bibfield  {title} {\enquote {\bibinfo {title} {Machine learning
  vortices at the kosterlitz-thouless transition},}\ }\href {\doibase
  10.1103/PhysRevB.97.045207} {\bibfield  {journal} {\bibinfo  {journal} {Phys.
  Rev. B}\ }\textbf {\bibinfo {volume} {97}},\ \bibinfo {pages} {045207}
  (\bibinfo {year} {2018})}\BibitemShut {NoStop}%
\bibitem [{\citenamefont {{Rodriguez-Nieva}}\ and\ \citenamefont
  {Scheurer}(2019)}]{rodriguez-nievaIdentifyingTopologicalOrder2019}%
  \BibitemOpen
  \bibfield  {author} {\bibinfo {author} {\bibfnamefont {Joaquin~F.}\
  \bibnamefont {{Rodriguez-Nieva}}}\ and\ \bibinfo {author} {\bibfnamefont
  {Mathias~S.}\ \bibnamefont {Scheurer}},\ }\bibfield  {title} {\enquote
  {\bibinfo {title} {Identifying topological order through unsupervised machine
  learning},}\ }\href {\doibase 10.1038/s41567-019-0512-x} {\bibfield
  {journal} {\bibinfo  {journal} {Nature Physics}\ }\textbf {\bibinfo {volume}
  {15}},\ \bibinfo {pages} {790--795} (\bibinfo {year} {2019})}\BibitemShut
  {NoStop}%
\bibitem [{\citenamefont {Singh}\ \emph {et~al.}(2021)\citenamefont {Singh},
  \citenamefont {Scheurer},\ and\ \citenamefont
  {Arora}}]{10.21468/SciPostPhys.11.2.043}%
  \BibitemOpen
  \bibfield  {author} {\bibinfo {author} {\bibfnamefont {Japneet}\ \bibnamefont
  {Singh}}, \bibinfo {author} {\bibfnamefont {Mathias~S.}\ \bibnamefont
  {Scheurer}}, \ and\ \bibinfo {author} {\bibfnamefont {Vipul}\ \bibnamefont
  {Arora}},\ }\bibfield  {title} {\enquote {\bibinfo {title} {{Conditional
  generative models for sampling and phase transition indication in spin
  systems}},}\ }\href {\doibase 10.21468/SciPostPhys.11.2.043} {\bibfield
  {journal} {\bibinfo  {journal} {SciPost Phys.}\ }\textbf {\bibinfo {volume}
  {11}},\ \bibinfo {pages} {043} (\bibinfo {year} {2021})}\BibitemShut
  {NoStop}%
\bibitem [{\citenamefont {Tseng}\ and\ \citenamefont
  {Jiang}(2022)}]{TSENG2022105134}%
  \BibitemOpen
  \bibfield  {author} {\bibinfo {author} {\bibfnamefont {Y.-H.}\ \bibnamefont
  {Tseng}}\ and\ \bibinfo {author} {\bibfnamefont {F.-J.}\ \bibnamefont
  {Jiang}},\ }\bibfield  {title} {\enquote {\bibinfo {title}
  {Berezinskii–kosterlitz–thouless transition – a universal neural
  network study with benchmarks},}\ }\href {\doibase
  https://doi.org/10.1016/j.rinp.2021.105134} {\bibfield  {journal} {\bibinfo
  {journal} {Results in Physics}\ }\textbf {\bibinfo {volume} {33}},\ \bibinfo
  {pages} {105134} (\bibinfo {year} {2022})}\BibitemShut {NoStop}%
\bibitem [{\citenamefont {Greplova}\ \emph {et~al.}(2020)\citenamefont
  {Greplova}, \citenamefont {Valenti}, \citenamefont {Boschung}, \citenamefont
  {Sch{\"a}fer}, \citenamefont {L{\"o}rch},\ and\ \citenamefont
  {Huber}}]{greplovaUnsupervisedIdentificationTopological2020}%
  \BibitemOpen
  \bibfield  {author} {\bibinfo {author} {\bibfnamefont {Eliska}\ \bibnamefont
  {Greplova}}, \bibinfo {author} {\bibfnamefont {Agnes}\ \bibnamefont
  {Valenti}}, \bibinfo {author} {\bibfnamefont {Gregor}\ \bibnamefont
  {Boschung}}, \bibinfo {author} {\bibfnamefont {Frank}\ \bibnamefont
  {Sch{\"a}fer}}, \bibinfo {author} {\bibfnamefont {Niels}\ \bibnamefont
  {L{\"o}rch}}, \ and\ \bibinfo {author} {\bibfnamefont {Sebastian~D}\
  \bibnamefont {Huber}},\ }\bibfield  {title} {\enquote {\bibinfo {title}
  {Unsupervised identification of topological phase transitions using
  predictive models},}\ }\href {\doibase 10.1088/1367-2630/ab7771} {\bibfield
  {journal} {\bibinfo  {journal} {New J. Phys.}\ }\textbf {\bibinfo {volume}
  {22}},\ \bibinfo {pages} {045003} (\bibinfo {year} {2020})}\BibitemShut
  {NoStop}%
\bibitem [{\citenamefont {Zhang}\ \emph {et~al.}(2017)\citenamefont {Zhang},
  \citenamefont {Melko},\ and\ \citenamefont {Kim}}]{PhysRevB.96.245119}%
  \BibitemOpen
  \bibfield  {author} {\bibinfo {author} {\bibfnamefont {Yi}~\bibnamefont
  {Zhang}}, \bibinfo {author} {\bibfnamefont {Roger~G.}\ \bibnamefont {Melko}},
  \ and\ \bibinfo {author} {\bibfnamefont {Eun-Ah}\ \bibnamefont {Kim}},\
  }\bibfield  {title} {\enquote {\bibinfo {title} {Machine learning
  ${\mathbb{z}}_{2}$ quantum spin liquids with quasiparticle statistics},}\
  }\href {\doibase 10.1103/PhysRevB.96.245119} {\bibfield  {journal} {\bibinfo
  {journal} {Phys. Rev. B}\ }\textbf {\bibinfo {volume} {96}},\ \bibinfo
  {pages} {245119} (\bibinfo {year} {2017})}\BibitemShut {NoStop}%
\bibitem [{\citenamefont {Huang}\ \emph {et~al.}(2022)\citenamefont {Huang},
  \citenamefont {Kueng}, \citenamefont {Torlai}, \citenamefont {Albert},\ and\
  \citenamefont {Preskill}}]{huangProvablyEfficientMachine2021}%
  \BibitemOpen
  \bibfield  {author} {\bibinfo {author} {\bibfnamefont {Hsin-Yuan}\
  \bibnamefont {Huang}}, \bibinfo {author} {\bibfnamefont {Richard}\
  \bibnamefont {Kueng}}, \bibinfo {author} {\bibfnamefont {Giacomo}\
  \bibnamefont {Torlai}}, \bibinfo {author} {\bibfnamefont {Victor~V.}\
  \bibnamefont {Albert}}, \ and\ \bibinfo {author} {\bibfnamefont {John}\
  \bibnamefont {Preskill}},\ }\bibfield  {title} {\enquote {\bibinfo {title}
  {Provably efficient machine learning for quantum many-body problems},}\
  }\href {\doibase 10.1126/science.abk3333} {\bibfield  {journal} {\bibinfo
  {journal} {Science}\ }\textbf {\bibinfo {volume} {377}},\ \bibinfo {pages}
  {eabk3333} (\bibinfo {year} {2022})}\BibitemShut {NoStop}%
\bibitem [{\citenamefont {Sadoune}\ \emph {et~al.}(2022)\citenamefont
  {Sadoune}, \citenamefont {Giudici}, \citenamefont {Liu},\ and\ \citenamefont
  {Pollet}}]{sadouneUnsupervisedInterpretableLearning2022}%
  \BibitemOpen
  \bibfield  {author} {\bibinfo {author} {\bibfnamefont {Nicolas}\ \bibnamefont
  {Sadoune}}, \bibinfo {author} {\bibfnamefont {Giuliano}\ \bibnamefont
  {Giudici}}, \bibinfo {author} {\bibfnamefont {Ke}~\bibnamefont {Liu}}, \ and\
  \bibinfo {author} {\bibfnamefont {Lode}\ \bibnamefont {Pollet}},\ }\href
  {http://arxiv.org/abs/2208.08850} {\enquote {\bibinfo {title} {Unsupervised
  {{Interpretable Learning}} of {{Phases From Many-Qubit Systems}}},}\ }
  (\bibinfo {year} {2022}),\ \Eprint {http://arxiv.org/abs/2208.08850}
  {arXiv:2208.08850 [cond-mat, physics:quant-ph]} \BibitemShut {NoStop}%
\bibitem [{\citenamefont {{Cole}}\ \emph {et~al.}(2020)\citenamefont {{Cole}},
  \citenamefont {{Loges}},\ and\ \citenamefont {{Shiu}}}]{2020arXiv201200783C}%
  \BibitemOpen
  \bibfield  {author} {\bibinfo {author} {\bibfnamefont {Alex}\ \bibnamefont
  {{Cole}}}, \bibinfo {author} {\bibfnamefont {Gregory~J.}\ \bibnamefont
  {{Loges}}}, \ and\ \bibinfo {author} {\bibfnamefont {Gary}\ \bibnamefont
  {{Shiu}}},\ }\bibfield  {title} {\enquote {\bibinfo {title} {{Interpretable
  Phase Detection and Classification with Persistent Homology}},}\ }\href@noop
  {} {\bibfield  {journal} {\bibinfo  {journal} {arXiv e-prints}\ ,\ \bibinfo
  {eid} {arXiv:2012.00783}} (\bibinfo {year} {2020})},\ \Eprint
  {http://arxiv.org/abs/2012.00783} {arXiv:2012.00783 [cond-mat.stat-mech]}
  \BibitemShut {NoStop}%
\bibitem [{\citenamefont {Sehayek}\ and\ \citenamefont
  {Melko}(2022)}]{sehayekPersistentHomologyMathbbZ2022}%
  \BibitemOpen
  \bibfield  {author} {\bibinfo {author} {\bibfnamefont {Dan}\ \bibnamefont
  {Sehayek}}\ and\ \bibinfo {author} {\bibfnamefont {Roger~G.}\ \bibnamefont
  {Melko}},\ }\bibfield  {title} {\enquote {\bibinfo {title} {Persistent
  {{Homology}} of $\mathbb{Z}_2$ {{Gauge Theories}}},}\ }\href {\doibase
  10.1103/PhysRevB.106.085111} {\bibfield  {journal} {\bibinfo  {journal}
  {Phys. Rev. B}\ }\textbf {\bibinfo {volume} {106}},\ \bibinfo {pages}
  {085111} (\bibinfo {year} {2022})},\ \Eprint
  {http://arxiv.org/abs/2201.09856} {arXiv:2201.09856 [cond-mat,
  physics:hep-th]} \BibitemShut {NoStop}%
\bibitem [{\citenamefont {Käming}\ \emph {et~al.}(2021)\citenamefont
  {Käming}, \citenamefont {Dawid}, \citenamefont {Kottmann}, \citenamefont
  {Lewenstein}, \citenamefont {Sengstock}, \citenamefont {Dauphin},\ and\
  \citenamefont {Weitenberg}}]{Kaeming_2021}%
  \BibitemOpen
  \bibfield  {author} {\bibinfo {author} {\bibfnamefont {Niklas}\ \bibnamefont
  {Käming}}, \bibinfo {author} {\bibfnamefont {Anna}\ \bibnamefont {Dawid}},
  \bibinfo {author} {\bibfnamefont {Korbinian}\ \bibnamefont {Kottmann}},
  \bibinfo {author} {\bibfnamefont {Maciej}\ \bibnamefont {Lewenstein}},
  \bibinfo {author} {\bibfnamefont {Klaus}\ \bibnamefont {Sengstock}}, \bibinfo
  {author} {\bibfnamefont {Alexandre}\ \bibnamefont {Dauphin}}, \ and\ \bibinfo
  {author} {\bibfnamefont {Christof}\ \bibnamefont {Weitenberg}},\ }\bibfield
  {title} {\enquote {\bibinfo {title} {Unsupervised machine learning of
  topological phase transitions from experimental data},}\ }\href {\doibase
  10.1088/2632-2153/abffe7} {\bibfield  {journal} {\bibinfo  {journal} {Machine
  Learning: Science and Technology}\ }\textbf {\bibinfo {volume} {2}},\
  \bibinfo {pages} {035037} (\bibinfo {year} {2021})}\BibitemShut {NoStop}%
\bibitem [{\citenamefont {Ho}\ and\ \citenamefont {Wang}(2021)}]{Ho_2021}%
  \BibitemOpen
  \bibfield  {author} {\bibinfo {author} {\bibfnamefont {Chi-Ting}\
  \bibnamefont {Ho}}\ and\ \bibinfo {author} {\bibfnamefont {Daw-Wei}\
  \bibnamefont {Wang}},\ }\bibfield  {title} {\enquote {\bibinfo {title}
  {Robust identification of topological phase transition by self-supervised
  machine learning approach},}\ }\href {\doibase 10.1088/1367-2630/ac1709}
  {\bibfield  {journal} {\bibinfo  {journal} {New Journal of Physics}\ }\textbf
  {\bibinfo {volume} {23}},\ \bibinfo {pages} {083021} (\bibinfo {year}
  {2021})}\BibitemShut {NoStop}%
\bibitem [{\citenamefont {{Lin}}\ \emph {et~al.}(2022)\citenamefont {{Lin}},
  \citenamefont {{Li}},\ and\ \citenamefont {{Huang}}}]{2022arXiv220914551L}%
  \BibitemOpen
  \bibfield  {author} {\bibinfo {author} {\bibfnamefont {Min-Ruei}\
  \bibnamefont {{Lin}}}, \bibinfo {author} {\bibfnamefont {Wan-Ju}\
  \bibnamefont {{Li}}}, \ and\ \bibinfo {author} {\bibfnamefont {Shin-Ming}\
  \bibnamefont {{Huang}}},\ }\bibfield  {title} {\enquote {\bibinfo {title}
  {{Quaternion-based machine learning on topological quantum systems}},}\
  }\href@noop {} {\bibfield  {journal} {\bibinfo  {journal} {arXiv e-prints}\ }
  (\bibinfo {year} {2022})},\ \Eprint {http://arxiv.org/abs/2209.14551}
  {arXiv:2209.14551 [quant-ph]} \BibitemShut {NoStop}%
\bibitem [{\citenamefont {Margalit}\ \emph {et~al.}(2022)\citenamefont
  {Margalit}, \citenamefont {Lesser}, \citenamefont {Pereg-Barnea},\ and\
  \citenamefont {Oreg}}]{PhysRevB.105.205139}%
  \BibitemOpen
  \bibfield  {author} {\bibinfo {author} {\bibfnamefont {Gilad}\ \bibnamefont
  {Margalit}}, \bibinfo {author} {\bibfnamefont {Omri}\ \bibnamefont {Lesser}},
  \bibinfo {author} {\bibfnamefont {T.}~\bibnamefont {Pereg-Barnea}}, \ and\
  \bibinfo {author} {\bibfnamefont {Yuval}\ \bibnamefont {Oreg}},\ }\bibfield
  {title} {\enquote {\bibinfo {title} {Renormalization-group-inspired neural
  networks for computing topological invariants},}\ }\href {\doibase
  10.1103/PhysRevB.105.205139} {\bibfield  {journal} {\bibinfo  {journal}
  {Phys. Rev. B}\ }\textbf {\bibinfo {volume} {105}},\ \bibinfo {pages}
  {205139} (\bibinfo {year} {2022})}\BibitemShut {NoStop}%
\bibitem [{\citenamefont {Park}\ \emph {et~al.}(2022)\citenamefont {Park},
  \citenamefont {Hwang},\ and\ \citenamefont {Yang}}]{PhysRevB.105.195115}%
  \BibitemOpen
  \bibfield  {author} {\bibinfo {author} {\bibfnamefont {Sungjoon}\
  \bibnamefont {Park}}, \bibinfo {author} {\bibfnamefont {Yoonseok}\
  \bibnamefont {Hwang}}, \ and\ \bibinfo {author} {\bibfnamefont {Bohm-Jung}\
  \bibnamefont {Yang}},\ }\bibfield  {title} {\enquote {\bibinfo {title}
  {Unsupervised learning of topological phase diagram using topological data
  analysis},}\ }\href {\doibase 10.1103/PhysRevB.105.195115} {\bibfield
  {journal} {\bibinfo  {journal} {Phys. Rev. B}\ }\textbf {\bibinfo {volume}
  {105}},\ \bibinfo {pages} {195115} (\bibinfo {year} {2022})}\BibitemShut
  {NoStop}%
\bibitem [{\citenamefont {Chung}\ \emph {et~al.}(2021)\citenamefont {Chung},
  \citenamefont {Cheng}, \citenamefont {Huang},\ and\ \citenamefont
  {Tsai}}]{PhysRevB.104.024506}%
  \BibitemOpen
  \bibfield  {author} {\bibinfo {author} {\bibfnamefont {Ming-Chiang}\
  \bibnamefont {Chung}}, \bibinfo {author} {\bibfnamefont {Tsung-Pao}\
  \bibnamefont {Cheng}}, \bibinfo {author} {\bibfnamefont {Guang-Yu}\
  \bibnamefont {Huang}}, \ and\ \bibinfo {author} {\bibfnamefont {Yuan-Hong}\
  \bibnamefont {Tsai}},\ }\bibfield  {title} {\enquote {\bibinfo {title} {Deep
  learning of topological phase transitions from the point of view of
  entanglement for two-dimensional chiral $p$-wave superconductors},}\ }\href
  {\doibase 10.1103/PhysRevB.104.024506} {\bibfield  {journal} {\bibinfo
  {journal} {Phys. Rev. B}\ }\textbf {\bibinfo {volume} {104}},\ \bibinfo
  {pages} {024506} (\bibinfo {year} {2021})}\BibitemShut {NoStop}%
\bibitem [{\citenamefont {Tsai}\ \emph {et~al.}(2021)\citenamefont {Tsai},
  \citenamefont {Chiu}, \citenamefont {Lai}, \citenamefont {Su}, \citenamefont
  {Yang}, \citenamefont {Cheng}, \citenamefont {Huang},\ and\ \citenamefont
  {Chung}}]{PhysRevB.104.165108}%
  \BibitemOpen
  \bibfield  {author} {\bibinfo {author} {\bibfnamefont {Yuan-Hong}\
  \bibnamefont {Tsai}}, \bibinfo {author} {\bibfnamefont {Kuo-Feng}\
  \bibnamefont {Chiu}}, \bibinfo {author} {\bibfnamefont {Yong-Cheng}\
  \bibnamefont {Lai}}, \bibinfo {author} {\bibfnamefont {Kuan-Jung}\
  \bibnamefont {Su}}, \bibinfo {author} {\bibfnamefont {Tzu-Pei}\ \bibnamefont
  {Yang}}, \bibinfo {author} {\bibfnamefont {Tsung-Pao}\ \bibnamefont {Cheng}},
  \bibinfo {author} {\bibfnamefont {Guang-Yu}\ \bibnamefont {Huang}}, \ and\
  \bibinfo {author} {\bibfnamefont {Ming-Chiang}\ \bibnamefont {Chung}},\
  }\bibfield  {title} {\enquote {\bibinfo {title} {Deep learning of topological
  phase transitions from entanglement aspects: An unsupervised way},}\ }\href
  {\doibase 10.1103/PhysRevB.104.165108} {\bibfield  {journal} {\bibinfo
  {journal} {Phys. Rev. B}\ }\textbf {\bibinfo {volume} {104}},\ \bibinfo
  {pages} {165108} (\bibinfo {year} {2021})}\BibitemShut {NoStop}%
\bibitem [{\citenamefont {{Jos{\'e} Ur{\'\i}a-{\'A}lvarez}}\ \emph
  {et~al.}(2022)\citenamefont {{Jos{\'e} Ur{\'\i}a-{\'A}lvarez}}, \citenamefont
  {{Molpeceres-Mingo}},\ and\ \citenamefont {{Jos{\'e}
  Palacios}}}]{2022arXiv220113306J}%
  \BibitemOpen
  \bibfield  {author} {\bibinfo {author} {\bibfnamefont {Alejandro}\
  \bibnamefont {{Jos{\'e} Ur{\'\i}a-{\'A}lvarez}}}, \bibinfo {author}
  {\bibfnamefont {Daniel}\ \bibnamefont {{Molpeceres-Mingo}}}, \ and\ \bibinfo
  {author} {\bibfnamefont {Juan}\ \bibnamefont {{Jos{\'e} Palacios}}},\
  }\bibfield  {title} {\enquote {\bibinfo {title} {{Deep learning for
  disordered topological insulators through entanglement spectrum}},}\
  }\href@noop {} {\bibfield  {journal} {\bibinfo  {journal} {arXiv e-prints}\
  ,\ \bibinfo {eid} {arXiv:2201.13306}} (\bibinfo {year} {2022})},\ \Eprint
  {http://arxiv.org/abs/2201.13306} {arXiv:2201.13306 [cond-mat.dis-nn]}
  \BibitemShut {NoStop}%
\bibitem [{\citenamefont {Molignini}\ \emph {et~al.}(2021)\citenamefont
  {Molignini}, \citenamefont {Zegarra}, \citenamefont {van Nieuwenburg},
  \citenamefont {Chitra},\ and\ \citenamefont
  {Chen}}]{10.21468/SciPostPhys.11.3.073}%
  \BibitemOpen
  \bibfield  {author} {\bibinfo {author} {\bibfnamefont {Paolo}\ \bibnamefont
  {Molignini}}, \bibinfo {author} {\bibfnamefont {Antonio}\ \bibnamefont
  {Zegarra}}, \bibinfo {author} {\bibfnamefont {Evert}\ \bibnamefont {van
  Nieuwenburg}}, \bibinfo {author} {\bibfnamefont {R.}~\bibnamefont {Chitra}},
  \ and\ \bibinfo {author} {\bibfnamefont {Wei}\ \bibnamefont {Chen}},\
  }\bibfield  {title} {\enquote {\bibinfo {title} {{A supervised learning
  algorithm for interacting topological insulators based on local
  curvature}},}\ }\href {\doibase 10.21468/SciPostPhys.11.3.073} {\bibfield
  {journal} {\bibinfo  {journal} {SciPost Phys.}\ }\textbf {\bibinfo {volume}
  {11}},\ \bibinfo {pages} {073} (\bibinfo {year} {2021})}\BibitemShut
  {NoStop}%
\bibitem [{\citenamefont {{Tibaldi}}\ \emph {et~al.}(2022)\citenamefont
  {{Tibaldi}}, \citenamefont {{Magnifico}}, \citenamefont {{Vodola}},\ and\
  \citenamefont {{Ercolessi}}}]{2022arXiv220209281T}%
  \BibitemOpen
  \bibfield  {author} {\bibinfo {author} {\bibfnamefont {Simone}\ \bibnamefont
  {{Tibaldi}}}, \bibinfo {author} {\bibfnamefont {Giuseppe}\ \bibnamefont
  {{Magnifico}}}, \bibinfo {author} {\bibfnamefont {Davide}\ \bibnamefont
  {{Vodola}}}, \ and\ \bibinfo {author} {\bibfnamefont {Elisa}\ \bibnamefont
  {{Ercolessi}}},\ }\bibfield  {title} {\enquote {\bibinfo {title}
  {{Unsupervised and supervised learning of interacting topological phases from
  single-particle correlation functions}},}\ }\href@noop {} {\bibfield
  {journal} {\bibinfo  {journal} {arXiv e-prints}\ } (\bibinfo {year}
  {2022})},\ \Eprint {http://arxiv.org/abs/2202.09281} {arXiv:2202.09281
  [cond-mat.supr-con]} \BibitemShut {NoStop}%
\bibitem [{\citenamefont {Tirelli}\ and\ \citenamefont
  {Costa}(2021)}]{PhysRevB.104.235146}%
  \BibitemOpen
  \bibfield  {author} {\bibinfo {author} {\bibfnamefont {Andrea}\ \bibnamefont
  {Tirelli}}\ and\ \bibinfo {author} {\bibfnamefont {Natanael~C.}\ \bibnamefont
  {Costa}},\ }\bibfield  {title} {\enquote {\bibinfo {title} {Learning quantum
  phase transitions through topological data analysis},}\ }\href {\doibase
  10.1103/PhysRevB.104.235146} {\bibfield  {journal} {\bibinfo  {journal}
  {Phys. Rev. B}\ }\textbf {\bibinfo {volume} {104}},\ \bibinfo {pages}
  {235146} (\bibinfo {year} {2021})}\BibitemShut {NoStop}%
\bibitem [{\citenamefont {Coifman}\ \emph {et~al.}(2005)\citenamefont
  {Coifman}, \citenamefont {Lafon}, \citenamefont {Lee}, \citenamefont
  {Maggioni}, \citenamefont {Nadler}, \citenamefont {Warner},\ and\
  \citenamefont {Zucker}}]{coifmanGeometricDiffusionsTool2005}%
  \BibitemOpen
  \bibfield  {author} {\bibinfo {author} {\bibfnamefont {R.~R.}\ \bibnamefont
  {Coifman}}, \bibinfo {author} {\bibfnamefont {S.}~\bibnamefont {Lafon}},
  \bibinfo {author} {\bibfnamefont {A.~B.}\ \bibnamefont {Lee}}, \bibinfo
  {author} {\bibfnamefont {M.}~\bibnamefont {Maggioni}}, \bibinfo {author}
  {\bibfnamefont {B.}~\bibnamefont {Nadler}}, \bibinfo {author} {\bibfnamefont
  {F.}~\bibnamefont {Warner}}, \ and\ \bibinfo {author} {\bibfnamefont {S.~W.}\
  \bibnamefont {Zucker}},\ }\bibfield  {title} {\enquote {\bibinfo {title}
  {Geometric diffusions as a tool for harmonic analysis and structure
  definition of data: {{Diffusion}} maps},}\ }\href {\doibase
  10.1073/pnas.0500334102} {\bibfield  {journal} {\bibinfo  {journal}
  {Proceedings of the National Academy of Sciences}\ }\textbf {\bibinfo
  {volume} {102}},\ \bibinfo {pages} {7426--7431} (\bibinfo {year}
  {2005})}\BibitemShut {NoStop}%
\bibitem [{\citenamefont {Nadler}\ \emph {et~al.}(2005)\citenamefont {Nadler},
  \citenamefont {Lafon}, \citenamefont {Kevrekidis},\ and\ \citenamefont
  {Coifman}}]{nadlerDiffusionMapsSpectral2005}%
  \BibitemOpen
  \bibfield  {author} {\bibinfo {author} {\bibfnamefont {Boaz}\ \bibnamefont
  {Nadler}}, \bibinfo {author} {\bibfnamefont {Stephane}\ \bibnamefont
  {Lafon}}, \bibinfo {author} {\bibfnamefont {Ioannis}\ \bibnamefont
  {Kevrekidis}}, \ and\ \bibinfo {author} {\bibfnamefont {Ronald}\ \bibnamefont
  {Coifman}},\ }\bibfield  {title} {\enquote {\bibinfo {title} {Diffusion
  {{Maps}}, {{Spectral Clustering}} and {{Eigenfunctions}} of {{Fokker-Planck
  Operators}}},}\ }in\ \href
  {https://proceedings.neurips.cc/paper/2005/hash/2a0f97f81755e2878b264adf39cba68e-Abstract.html}
  {\emph {\bibinfo {booktitle} {Advances in {{Neural Information Processing
  Systems}}}}},\ Vol.~\bibinfo {volume} {18}\ (\bibinfo  {publisher} {{MIT
  Press}},\ \bibinfo {year} {2005})\BibitemShut {NoStop}%
\bibitem [{\citenamefont {Nadler}\ \emph {et~al.}(2006)\citenamefont {Nadler},
  \citenamefont {Lafon}, \citenamefont {Coifman},\ and\ \citenamefont
  {Kevrekidis}}]{nadlerDiffusionMapsSpectral2006}%
  \BibitemOpen
  \bibfield  {author} {\bibinfo {author} {\bibfnamefont {Boaz}\ \bibnamefont
  {Nadler}}, \bibinfo {author} {\bibfnamefont {St{\'e}phane}\ \bibnamefont
  {Lafon}}, \bibinfo {author} {\bibfnamefont {Ronald~R.}\ \bibnamefont
  {Coifman}}, \ and\ \bibinfo {author} {\bibfnamefont {Ioannis~G.}\
  \bibnamefont {Kevrekidis}},\ }\bibfield  {title} {\enquote {\bibinfo {title}
  {Diffusion maps, spectral clustering and reaction coordinates of dynamical
  systems},}\ }\href {\doibase 10.1016/j.acha.2005.07.004} {\bibfield
  {journal} {\bibinfo  {journal} {Applied and Computational Harmonic Analysis}\
  }\textbf {\bibinfo {volume} {21}},\ \bibinfo {pages} {113--127} (\bibinfo
  {year} {2006})}\BibitemShut {NoStop}%
\bibitem [{\citenamefont {Coifman}\ and\ \citenamefont
  {Lafon}(2006)}]{coifmanDiffusionMaps2006}%
  \BibitemOpen
  \bibfield  {author} {\bibinfo {author} {\bibfnamefont {Ronald~R.}\
  \bibnamefont {Coifman}}\ and\ \bibinfo {author} {\bibfnamefont
  {St{\'e}phane}\ \bibnamefont {Lafon}},\ }\bibfield  {title} {\enquote
  {\bibinfo {title} {Diffusion maps},}\ }\href {\doibase
  10.1016/j.acha.2006.04.006} {\bibfield  {journal} {\bibinfo  {journal}
  {Applied and Computational Harmonic Analysis}\ }\textbf {\bibinfo {volume}
  {21}},\ \bibinfo {pages} {5--30} (\bibinfo {year} {2006})}\BibitemShut
  {NoStop}%
\bibitem [{\citenamefont {Scheurer}\ and\ \citenamefont
  {Slager}(2020)}]{scheurerUnsupervisedMachineLearning2020b}%
  \BibitemOpen
  \bibfield  {author} {\bibinfo {author} {\bibfnamefont {Mathias~S.}\
  \bibnamefont {Scheurer}}\ and\ \bibinfo {author} {\bibfnamefont {Robert-Jan}\
  \bibnamefont {Slager}},\ }\bibfield  {title} {\enquote {\bibinfo {title}
  {Unsupervised {{Machine Learning}} and {{Band Topology}}},}\ }\href {\doibase
  10.1103/PhysRevLett.124.226401} {\bibfield  {journal} {\bibinfo  {journal}
  {Phys. Rev. Lett.}\ }\textbf {\bibinfo {volume} {124}},\ \bibinfo {pages}
  {226401} (\bibinfo {year} {2020})}\BibitemShut {NoStop}%
\bibitem [{\citenamefont {Long}\ \emph {et~al.}(2020)\citenamefont {Long},
  \citenamefont {Ren},\ and\ \citenamefont
  {Chen}}]{longUnsupervisedManifoldClustering2020}%
  \BibitemOpen
  \bibfield  {author} {\bibinfo {author} {\bibfnamefont {Yang}\ \bibnamefont
  {Long}}, \bibinfo {author} {\bibfnamefont {Jie}\ \bibnamefont {Ren}}, \ and\
  \bibinfo {author} {\bibfnamefont {Hong}\ \bibnamefont {Chen}},\ }\bibfield
  {title} {\enquote {\bibinfo {title} {Unsupervised {{Manifold Clustering}} of
  {{Topological Phononics}}},}\ }\href {\doibase
  10.1103/PhysRevLett.124.185501} {\bibfield  {journal} {\bibinfo  {journal}
  {Phys. Rev. Lett.}\ }\textbf {\bibinfo {volume} {124}},\ \bibinfo {pages}
  {185501} (\bibinfo {year} {2020})}\BibitemShut {NoStop}%
\bibitem [{\citenamefont {Yu}\ and\ \citenamefont
  {Deng}(2021)}]{PhysRevLett.126.240402}%
  \BibitemOpen
  \bibfield  {author} {\bibinfo {author} {\bibfnamefont {Li-Wei}\ \bibnamefont
  {Yu}}\ and\ \bibinfo {author} {\bibfnamefont {Dong-Ling}\ \bibnamefont
  {Deng}},\ }\bibfield  {title} {\enquote {\bibinfo {title} {Unsupervised
  learning of non-hermitian topological phases},}\ }\href {\doibase
  10.1103/PhysRevLett.126.240402} {\bibfield  {journal} {\bibinfo  {journal}
  {Phys. Rev. Lett.}\ }\textbf {\bibinfo {volume} {126}},\ \bibinfo {pages}
  {240402} (\bibinfo {year} {2021})}\BibitemShut {NoStop}%
\bibitem [{\citenamefont {{Yu}}\ \emph {et~al.}(2022)\citenamefont {{Yu}},
  \citenamefont {{Yu}}, \citenamefont {{Zhang}}, \citenamefont {{Zhang}},
  \citenamefont {{Ouyang}}, \citenamefont {{Liu}}, \citenamefont {{Deng}},\
  and\ \citenamefont {{Duan}}}]{2022npjQI...8..116Y}%
  \BibitemOpen
  \bibfield  {author} {\bibinfo {author} {\bibfnamefont {Yefei}\ \bibnamefont
  {{Yu}}}, \bibinfo {author} {\bibfnamefont {Li-Wei}\ \bibnamefont {{Yu}}},
  \bibinfo {author} {\bibfnamefont {Wengang}\ \bibnamefont {{Zhang}}}, \bibinfo
  {author} {\bibfnamefont {Huili}\ \bibnamefont {{Zhang}}}, \bibinfo {author}
  {\bibfnamefont {Xiaolong}\ \bibnamefont {{Ouyang}}}, \bibinfo {author}
  {\bibfnamefont {Yanqing}\ \bibnamefont {{Liu}}}, \bibinfo {author}
  {\bibfnamefont {Dong-Ling}\ \bibnamefont {{Deng}}}, \ and\ \bibinfo {author}
  {\bibfnamefont {L.~M.}\ \bibnamefont {{Duan}}},\ }\bibfield  {title}
  {\enquote {\bibinfo {title} {{Experimental unsupervised learning of
  non-Hermitian knotted phases with solid-state spins}},}\ }\href {\doibase
  10.1038/s41534-022-00629-w} {\bibfield  {journal} {\bibinfo  {journal} {npj
  Quantum Information}\ }\textbf {\bibinfo {volume} {8}},\ \bibinfo {eid} {116}
  (\bibinfo {year} {2022})},\ \Eprint {http://arxiv.org/abs/2112.13785}
  {arXiv:2112.13785 [quant-ph]} \BibitemShut {NoStop}%
\bibitem [{\citenamefont {Che}\ \emph {et~al.}(2020)\citenamefont {Che},
  \citenamefont {Gneiting}, \citenamefont {Liu},\ and\ \citenamefont
  {Nori}}]{cheTopologicalQuantumPhase2020}%
  \BibitemOpen
  \bibfield  {author} {\bibinfo {author} {\bibfnamefont {Yanming}\ \bibnamefont
  {Che}}, \bibinfo {author} {\bibfnamefont {Clemens}\ \bibnamefont {Gneiting}},
  \bibinfo {author} {\bibfnamefont {Tao}\ \bibnamefont {Liu}}, \ and\ \bibinfo
  {author} {\bibfnamefont {Franco}\ \bibnamefont {Nori}},\ }\bibfield  {title}
  {\enquote {\bibinfo {title} {Topological quantum phase transitions retrieved
  through unsupervised machine learning},}\ }\href {\doibase
  10.1103/PhysRevB.102.134213} {\bibfield  {journal} {\bibinfo  {journal}
  {Phys. Rev. B}\ }\textbf {\bibinfo {volume} {102}},\ \bibinfo {pages}
  {134213} (\bibinfo {year} {2020})}\BibitemShut {NoStop}%
\bibitem [{\citenamefont {Kuo}\ and\ \citenamefont
  {Dehghani}(2021)}]{kuoUnsupervisedLearningSymmetry2021}%
  \BibitemOpen
  \bibfield  {author} {\bibinfo {author} {\bibfnamefont {En-Jui}\ \bibnamefont
  {Kuo}}\ and\ \bibinfo {author} {\bibfnamefont {Hossein}\ \bibnamefont
  {Dehghani}},\ }\bibfield  {title} {\enquote {\bibinfo {title} {Unsupervised
  {{Learning}} of {{Symmetry Protected Topological Phase Transitions}}},}\
  }\href {http://arxiv.org/abs/2111.08747} {\bibfield  {journal} {\bibinfo
  {journal} {arXiv:2111.08747 [cond-mat, physics:quant-ph]}\ } (\bibinfo {year}
  {2021})},\ \Eprint {http://arxiv.org/abs/2111.08747} {arXiv:2111.08747
  [cond-mat, physics:quant-ph]} \BibitemShut {NoStop}%
\bibitem [{\citenamefont {Lustig}\ \emph {et~al.}(2020)\citenamefont {Lustig},
  \citenamefont {Yair}, \citenamefont {Talmon},\ and\ \citenamefont
  {Segev}}]{lustigIdentifyingTopologicalPhase2020}%
  \BibitemOpen
  \bibfield  {author} {\bibinfo {author} {\bibfnamefont {Eran}\ \bibnamefont
  {Lustig}}, \bibinfo {author} {\bibfnamefont {Or}~\bibnamefont {Yair}},
  \bibinfo {author} {\bibfnamefont {Ronen}\ \bibnamefont {Talmon}}, \ and\
  \bibinfo {author} {\bibfnamefont {Mordechai}\ \bibnamefont {Segev}},\
  }\bibfield  {title} {\enquote {\bibinfo {title} {Identifying {{Topological
  Phase Transitions}} in {{Experiments Using Manifold Learning}}},}\ }\href
  {\doibase 10.1103/PhysRevLett.125.127401} {\bibfield  {journal} {\bibinfo
  {journal} {Phys. Rev. Lett.}\ }\textbf {\bibinfo {volume} {125}},\ \bibinfo
  {pages} {127401} (\bibinfo {year} {2020})}\BibitemShut {NoStop}%
\bibitem [{\citenamefont {Lidiak}\ and\ \citenamefont
  {Gong}(2020)}]{lidiakUnsupervisedMachineLearning2020}%
  \BibitemOpen
  \bibfield  {author} {\bibinfo {author} {\bibfnamefont {Alexander}\
  \bibnamefont {Lidiak}}\ and\ \bibinfo {author} {\bibfnamefont {Zhexuan}\
  \bibnamefont {Gong}},\ }\bibfield  {title} {\enquote {\bibinfo {title}
  {Unsupervised {{Machine Learning}} of {{Quantum Phase Transitions Using
  Diffusion Maps}}},}\ }\href {\doibase 10.1103/PhysRevLett.125.225701}
  {\bibfield  {journal} {\bibinfo  {journal} {Phys. Rev. Lett.}\ }\textbf
  {\bibinfo {volume} {125}},\ \bibinfo {pages} {225701} (\bibinfo {year}
  {2020})}\BibitemShut {NoStop}%
\bibitem [{\citenamefont {Gyawali}\ \emph {et~al.}(2022)\citenamefont
  {Gyawali}, \citenamefont {Ahmed}, \citenamefont {Aspling}, \citenamefont
  {{Ellert-Beck}},\ and\ \citenamefont
  {Lawler}}]{gyawaliRevealingMicrocanonicalPhase2022}%
  \BibitemOpen
  \bibfield  {author} {\bibinfo {author} {\bibfnamefont {Gaurav}\ \bibnamefont
  {Gyawali}}, \bibinfo {author} {\bibfnamefont {Mabrur}\ \bibnamefont {Ahmed}},
  \bibinfo {author} {\bibfnamefont {Eric}\ \bibnamefont {Aspling}}, \bibinfo
  {author} {\bibfnamefont {Luke}\ \bibnamefont {{Ellert-Beck}}}, \ and\
  \bibinfo {author} {\bibfnamefont {Michael~J.}\ \bibnamefont {Lawler}},\
  }\href {http://arxiv.org/abs/2211.01259} {\enquote {\bibinfo {title}
  {Revealing microcanonical phase diagrams of strongly correlated systems via
  time-averaged classical shadows},}\ } (\bibinfo {year} {2022}),\ \Eprint
  {http://arxiv.org/abs/2211.01259} {arXiv:2211.01259 [cond-mat,
  physics:quant-ph]} \BibitemShut {NoStop}%
\bibitem [{\citenamefont {Sornsaeng}\ \emph {et~al.}(2021)\citenamefont
  {Sornsaeng}, \citenamefont {Dangniam}, \citenamefont {Palittapongarnpim},\
  and\ \citenamefont {Chotibut}}]{sornsaengQuantumDiffusionMap2021}%
  \BibitemOpen
  \bibfield  {author} {\bibinfo {author} {\bibfnamefont {Apimuk}\ \bibnamefont
  {Sornsaeng}}, \bibinfo {author} {\bibfnamefont {Ninnat}\ \bibnamefont
  {Dangniam}}, \bibinfo {author} {\bibfnamefont {Pantita}\ \bibnamefont
  {Palittapongarnpim}}, \ and\ \bibinfo {author} {\bibfnamefont {Thiparat}\
  \bibnamefont {Chotibut}},\ }\bibfield  {title} {\enquote {\bibinfo {title}
  {Quantum diffusion map for nonlinear dimensionality reduction},}\ }\href
  {\doibase 10.1103/PhysRevA.104.052410} {\bibfield  {journal} {\bibinfo
  {journal} {Phys. Rev. A}\ }\textbf {\bibinfo {volume} {104}},\ \bibinfo
  {pages} {052410} (\bibinfo {year} {2021})}\BibitemShut {NoStop}%
\bibitem [{\citenamefont {Carleo}\ and\ \citenamefont
  {Troyer}(2017)}]{carleoSolvingQuantumManybody2017}%
  \BibitemOpen
  \bibfield  {author} {\bibinfo {author} {\bibfnamefont {Giuseppe}\
  \bibnamefont {Carleo}}\ and\ \bibinfo {author} {\bibfnamefont {Matthias}\
  \bibnamefont {Troyer}},\ }\bibfield  {title} {\enquote {\bibinfo {title}
  {Solving the quantum many-body problem with artificial neural networks},}\
  }\href {\doibase 10.1126/science.aag2302} {\bibfield  {journal} {\bibinfo
  {journal} {Science}\ }\textbf {\bibinfo {volume} {355}},\ \bibinfo {pages}
  {602--606} (\bibinfo {year} {2017})}\BibitemShut {NoStop}%
\bibitem [{\citenamefont {Gao}\ and\ \citenamefont
  {Duan}(2017)}]{gaoEfficientRepresentationQuantum2017}%
  \BibitemOpen
  \bibfield  {author} {\bibinfo {author} {\bibfnamefont {Xun}\ \bibnamefont
  {Gao}}\ and\ \bibinfo {author} {\bibfnamefont {Lu-Ming}\ \bibnamefont
  {Duan}},\ }\bibfield  {title} {\enquote {\bibinfo {title} {Efficient
  representation of quantum many-body states with deep neural networks},}\
  }\href {\doibase 10.1038/s41467-017-00705-2} {\bibfield  {journal} {\bibinfo
  {journal} {Nature Communications}\ }\textbf {\bibinfo {volume} {8}},\
  \bibinfo {pages} {662} (\bibinfo {year} {2017})}\BibitemShut {NoStop}%
\bibitem [{\citenamefont {Carleo}\ \emph {et~al.}(2018)\citenamefont {Carleo},
  \citenamefont {Nomura},\ and\ \citenamefont
  {Imada}}]{carleoConstructingExactRepresentations2018}%
  \BibitemOpen
  \bibfield  {author} {\bibinfo {author} {\bibfnamefont {Giuseppe}\
  \bibnamefont {Carleo}}, \bibinfo {author} {\bibfnamefont {Yusuke}\
  \bibnamefont {Nomura}}, \ and\ \bibinfo {author} {\bibfnamefont {Masatoshi}\
  \bibnamefont {Imada}},\ }\bibfield  {title} {\enquote {\bibinfo {title}
  {Constructing exact representations of quantum many-body systems with deep
  neural networks},}\ }\href {\doibase 10.1038/s41467-018-07520-3} {\bibfield
  {journal} {\bibinfo  {journal} {Nat Commun}\ }\textbf {\bibinfo {volume}
  {9}},\ \bibinfo {pages} {5322} (\bibinfo {year} {2018})}\BibitemShut
  {NoStop}%
\bibitem [{\citenamefont {Lu}\ \emph {et~al.}(2019)\citenamefont {Lu},
  \citenamefont {Gao},\ and\ \citenamefont
  {Duan}}]{luEfficientRepresentationTopologically2019a}%
  \BibitemOpen
  \bibfield  {author} {\bibinfo {author} {\bibfnamefont {Sirui}\ \bibnamefont
  {Lu}}, \bibinfo {author} {\bibfnamefont {Xun}\ \bibnamefont {Gao}}, \ and\
  \bibinfo {author} {\bibfnamefont {L.-M.}\ \bibnamefont {Duan}},\ }\bibfield
  {title} {\enquote {\bibinfo {title} {Efficient representation of
  topologically ordered states with restricted {{Boltzmann}} machines},}\
  }\href {\doibase 10.1103/PhysRevB.99.155136} {\bibfield  {journal} {\bibinfo
  {journal} {Phys. Rev. B}\ }\textbf {\bibinfo {volume} {99}},\ \bibinfo
  {pages} {155136} (\bibinfo {year} {2019})}\BibitemShut {NoStop}%
\bibitem [{\citenamefont {Sharir}\ \emph {et~al.}(2021)\citenamefont {Sharir},
  \citenamefont {Shashua},\ and\ \citenamefont
  {Carleo}}]{sharirNeuralTensorContractions2021}%
  \BibitemOpen
  \bibfield  {author} {\bibinfo {author} {\bibfnamefont {Or}~\bibnamefont
  {Sharir}}, \bibinfo {author} {\bibfnamefont {Amnon}\ \bibnamefont {Shashua}},
  \ and\ \bibinfo {author} {\bibfnamefont {Giuseppe}\ \bibnamefont {Carleo}},\
  }\bibfield  {title} {\enquote {\bibinfo {title} {Neural tensor contractions
  and the expressive power of deep neural quantum states},}\ }\href {\doibase
  10.48550/arXiv.2103.10293} {\  (\bibinfo {year} {2021}),\
  10.48550/arXiv.2103.10293}\BibitemShut {NoStop}%
\bibitem [{\citenamefont {Chen}\ \emph {et~al.}(2018)\citenamefont {Chen},
  \citenamefont {Cheng}, \citenamefont {Xie}, \citenamefont {Wang},\ and\
  \citenamefont {Xiang}}]{chenEquivalenceRestrictedBoltzmann2018}%
  \BibitemOpen
  \bibfield  {author} {\bibinfo {author} {\bibfnamefont {Jing}\ \bibnamefont
  {Chen}}, \bibinfo {author} {\bibfnamefont {Song}\ \bibnamefont {Cheng}},
  \bibinfo {author} {\bibfnamefont {Haidong}\ \bibnamefont {Xie}}, \bibinfo
  {author} {\bibfnamefont {Lei}\ \bibnamefont {Wang}}, \ and\ \bibinfo {author}
  {\bibfnamefont {Tao}\ \bibnamefont {Xiang}},\ }\bibfield  {title} {\enquote
  {\bibinfo {title} {Equivalence of restricted {{Boltzmann}} machines and
  tensor network states},}\ }\href {\doibase 10.1103/PhysRevB.97.085104}
  {\bibfield  {journal} {\bibinfo  {journal} {Phys. Rev. B}\ }\textbf {\bibinfo
  {volume} {97}},\ \bibinfo {pages} {085104} (\bibinfo {year}
  {2018})}\BibitemShut {NoStop}%
\bibitem [{\citenamefont
  {Nomura}(2022)}]{nomuraInvestigatingNetworkParameters2022}%
  \BibitemOpen
  \bibfield  {author} {\bibinfo {author} {\bibfnamefont {Yusuke}\ \bibnamefont
  {Nomura}},\ }\bibfield  {title} {\enquote {\bibinfo {title} {Investigating
  {{Network Parameters}} in {{Neural-Network Quantum States}}},}\ }\href
  {http://arxiv.org/abs/2202.01704} {\  (\bibinfo {year} {2022})},\ \Eprint
  {http://arxiv.org/abs/2202.01704} {arXiv:2202.01704 [cond-mat,
  physics:physics, physics:quant-ph]} \BibitemShut {NoStop}%
\bibitem [{\citenamefont {Deng}\ \emph
  {et~al.}(2017{\natexlab{a}})\citenamefont {Deng}, \citenamefont {Li},\ and\
  \citenamefont {Sarma}}]{dengQuantumEntanglementNeural2017}%
  \BibitemOpen
  \bibfield  {author} {\bibinfo {author} {\bibfnamefont {Dong-Ling}\
  \bibnamefont {Deng}}, \bibinfo {author} {\bibfnamefont {Xiaopeng}\
  \bibnamefont {Li}}, \ and\ \bibinfo {author} {\bibfnamefont {S~Das}\
  \bibnamefont {Sarma}},\ }\bibfield  {title} {\enquote {\bibinfo {title}
  {Quantum {{Entanglement}} in {{Neural Network States}}},}\ }\href@noop {} {\
  ,\ \bibinfo {pages} {17} (\bibinfo {year} {2017}{\natexlab{a}})}\BibitemShut
  {NoStop}%
\bibitem [{\citenamefont {Jia}\ \emph {et~al.}(2020)\citenamefont {Jia},
  \citenamefont {Wei}, \citenamefont {Wu}, \citenamefont {Guo},\ and\
  \citenamefont {Guo}}]{jiaEntanglementAreaLaw2020}%
  \BibitemOpen
  \bibfield  {author} {\bibinfo {author} {\bibfnamefont {Zhih-Ahn}\
  \bibnamefont {Jia}}, \bibinfo {author} {\bibfnamefont {Lu}~\bibnamefont
  {Wei}}, \bibinfo {author} {\bibfnamefont {Yu-Chun}\ \bibnamefont {Wu}},
  \bibinfo {author} {\bibfnamefont {Guang-Can}\ \bibnamefont {Guo}}, \ and\
  \bibinfo {author} {\bibfnamefont {Guo-Ping}\ \bibnamefont {Guo}},\ }\bibfield
   {title} {\enquote {\bibinfo {title} {Entanglement area law for shallow and
  deep quantum neural network states},}\ }\href {\doibase
  10.1088/1367-2630/ab8262} {\bibfield  {journal} {\bibinfo  {journal} {New J.
  Phys.}\ }\textbf {\bibinfo {volume} {22}},\ \bibinfo {pages} {053022}
  (\bibinfo {year} {2020})}\BibitemShut {NoStop}%
\bibitem [{\citenamefont {Cheng}\ \emph {et~al.}(2017)\citenamefont {Cheng},
  \citenamefont {Chen},\ and\ \citenamefont
  {Wang}}]{chengInformationPerspectiveProbabilistic2017}%
  \BibitemOpen
  \bibfield  {author} {\bibinfo {author} {\bibfnamefont {Song}\ \bibnamefont
  {Cheng}}, \bibinfo {author} {\bibfnamefont {Jing}\ \bibnamefont {Chen}}, \
  and\ \bibinfo {author} {\bibfnamefont {Lei}\ \bibnamefont {Wang}},\ }\href
  {\doibase 10.3390/e20080583} {\emph {\bibinfo {title} {Information
  {{Perspective}} to {{Probabilistic Modeling}}: {{Boltzmann Machines}} versus
  {{Born Machines}}}}},\ \bibinfo {type} {Tech. Rep.}\ (\bibinfo {year}
  {2017})\ \Eprint {http://arxiv.org/abs/1712.04144} {arXiv:1712.04144
  [cond-mat, physics:physics, physics:quant-ph, stat]} \BibitemShut {NoStop}%
\bibitem [{\citenamefont {Torlai}\ \emph {et~al.}(2018)\citenamefont {Torlai},
  \citenamefont {Mazzola}, \citenamefont {Carrasquilla}, \citenamefont
  {Troyer}, \citenamefont {Melko},\ and\ \citenamefont {Carleo}}]{QSTNatPhys}%
  \BibitemOpen
  \bibfield  {author} {\bibinfo {author} {\bibfnamefont {Giacomo}\ \bibnamefont
  {Torlai}}, \bibinfo {author} {\bibfnamefont {Guglielmo}\ \bibnamefont
  {Mazzola}}, \bibinfo {author} {\bibfnamefont {Juan}\ \bibnamefont
  {Carrasquilla}}, \bibinfo {author} {\bibfnamefont {Matthias}\ \bibnamefont
  {Troyer}}, \bibinfo {author} {\bibfnamefont {Roger}\ \bibnamefont {Melko}}, \
  and\ \bibinfo {author} {\bibfnamefont {Giuseppe}\ \bibnamefont {Carleo}},\
  }\bibfield  {title} {\enquote {\bibinfo {title} {Neural-network quantum state
  tomography},}\ }\href {\doibase 10.1038/s41567-018-0048-5} {\bibfield
  {journal} {\bibinfo  {journal} {Nature Physics}\ }\textbf {\bibinfo {volume}
  {14}},\ \bibinfo {pages} {447--450} (\bibinfo {year} {2018})}\BibitemShut
  {NoStop}%
\bibitem [{\citenamefont
  {Kitaev}(2003)}]{kitaevFaulttolerantQuantumComputation2003}%
  \BibitemOpen
  \bibfield  {author} {\bibinfo {author} {\bibfnamefont {A.~Yu}\ \bibnamefont
  {Kitaev}},\ }\bibfield  {title} {\enquote {\bibinfo {title} {Fault-tolerant
  quantum computation by anyons},}\ }\href {\doibase
  10.1016/S0003-4916(02)00018-0} {\bibfield  {journal} {\bibinfo  {journal}
  {Annals of Physics}\ }\textbf {\bibinfo {volume} {303}},\ \bibinfo {pages}
  {2--30} (\bibinfo {year} {2003})},\ \Eprint
  {http://arxiv.org/abs/quant-ph/9707021} {arXiv:quant-ph/9707021} \BibitemShut
  {NoStop}%
\bibitem [{\citenamefont {Deng}\ \emph
  {et~al.}(2017{\natexlab{b}})\citenamefont {Deng}, \citenamefont {Li},\ and\
  \citenamefont {Das~Sarma}}]{dengMachineLearningTopological2017}%
  \BibitemOpen
  \bibfield  {author} {\bibinfo {author} {\bibfnamefont {Dong-Ling}\
  \bibnamefont {Deng}}, \bibinfo {author} {\bibfnamefont {Xiaopeng}\
  \bibnamefont {Li}}, \ and\ \bibinfo {author} {\bibfnamefont {S.}~\bibnamefont
  {Das~Sarma}},\ }\bibfield  {title} {\enquote {\bibinfo {title} {Machine
  learning topological states},}\ }\href {\doibase 10.1103/PhysRevB.96.195145}
  {\bibfield  {journal} {\bibinfo  {journal} {Phys. Rev. B}\ }\textbf {\bibinfo
  {volume} {96}},\ \bibinfo {pages} {195145} (\bibinfo {year}
  {2017}{\natexlab{b}})}\BibitemShut {NoStop}%
\bibitem [{\citenamefont {Valenti}\ \emph {et~al.}(2021)\citenamefont
  {Valenti}, \citenamefont {Greplova}, \citenamefont {Lindner},\ and\
  \citenamefont {Huber}}]{valentiCorrelationEnhancedNeuralNetworks2021}%
  \BibitemOpen
  \bibfield  {author} {\bibinfo {author} {\bibfnamefont {Agnes}\ \bibnamefont
  {Valenti}}, \bibinfo {author} {\bibfnamefont {Eliska}\ \bibnamefont
  {Greplova}}, \bibinfo {author} {\bibfnamefont {Netanel~H.}\ \bibnamefont
  {Lindner}}, \ and\ \bibinfo {author} {\bibfnamefont {Sebastian~D.}\
  \bibnamefont {Huber}},\ }\bibfield  {title} {\enquote {\bibinfo {title}
  {Correlation-{{Enhanced Neural Networks}} as {{Interpretable Variational
  Quantum States}}},}\ }\href {http://arxiv.org/abs/2103.05017} {\  (\bibinfo
  {year} {2021})},\ \Eprint {http://arxiv.org/abs/2103.05017} {arXiv:2103.05017
  [cond-mat, physics:quant-ph]} \BibitemShut {NoStop}%
\bibitem [{\citenamefont {{Cian}}\ \emph {et~al.}(2022)\citenamefont {{Cian}},
  \citenamefont {{Hafezi}},\ and\ \citenamefont
  {{Barkeshli}}}]{2022arXiv220914302C}%
  \BibitemOpen
  \bibfield  {author} {\bibinfo {author} {\bibfnamefont {Ze-Pei}\ \bibnamefont
  {{Cian}}}, \bibinfo {author} {\bibfnamefont {Mohammad}\ \bibnamefont
  {{Hafezi}}}, \ and\ \bibinfo {author} {\bibfnamefont {Maissam}\ \bibnamefont
  {{Barkeshli}}},\ }\bibfield  {title} {\enquote {\bibinfo {title} {{Extracting
  Wilson loop operators and fractional statistics from a single bulk ground
  state}},}\ }\href@noop {} {\bibfield  {journal} {\bibinfo  {journal} {arXiv
  e-prints}\ } (\bibinfo {year} {2022})},\ \Eprint
  {http://arxiv.org/abs/2209.14302} {arXiv:2209.14302 [cond-mat.str-el]}
  \BibitemShut {NoStop}%
\bibitem [{\citenamefont {Hastings}(2007)}]{hastingsAreaLawOnedimensional2007}%
  \BibitemOpen
  \bibfield  {author} {\bibinfo {author} {\bibfnamefont {M.~B.}\ \bibnamefont
  {Hastings}},\ }\bibfield  {title} {\enquote {\bibinfo {title} {An area law
  for one-dimensional quantum systems},}\ }\href {\doibase
  10.1088/1742-5468/2007/08/P08024} {\bibfield  {journal} {\bibinfo  {journal}
  {J. Stat. Mech.}\ }\textbf {\bibinfo {volume} {2007}},\ \bibinfo {pages}
  {P08024--P08024} (\bibinfo {year} {2007})}\BibitemShut {NoStop}%
\bibitem [{\citenamefont {Verstraete}\ \emph {et~al.}(2006)\citenamefont
  {Verstraete}, \citenamefont {Wolf}, \citenamefont {{Perez-Garcia}},\ and\
  \citenamefont {Cirac}}]{verstraeteCriticalityAreaLaw2006}%
  \BibitemOpen
  \bibfield  {author} {\bibinfo {author} {\bibfnamefont {F.}~\bibnamefont
  {Verstraete}}, \bibinfo {author} {\bibfnamefont {M.~M.}\ \bibnamefont
  {Wolf}}, \bibinfo {author} {\bibfnamefont {D.}~\bibnamefont
  {{Perez-Garcia}}}, \ and\ \bibinfo {author} {\bibfnamefont {J.~I.}\
  \bibnamefont {Cirac}},\ }\bibfield  {title} {\enquote {\bibinfo {title}
  {Criticality, the {{Area Law}}, and the {{Computational Power}} of
  {{Projected Entangled Pair States}}},}\ }\href {\doibase
  10.1103/PhysRevLett.96.220601} {\bibfield  {journal} {\bibinfo  {journal}
  {Phys. Rev. Lett.}\ }\textbf {\bibinfo {volume} {96}},\ \bibinfo {pages}
  {220601} (\bibinfo {year} {2006})}\BibitemShut {NoStop}%
\bibitem [{\citenamefont {Wolf}\ \emph {et~al.}(2008)\citenamefont {Wolf},
  \citenamefont {Verstraete}, \citenamefont {Hastings},\ and\ \citenamefont
  {Cirac}}]{wolfAreaLawsQuantum2008}%
  \BibitemOpen
  \bibfield  {author} {\bibinfo {author} {\bibfnamefont {Michael~M.}\
  \bibnamefont {Wolf}}, \bibinfo {author} {\bibfnamefont {Frank}\ \bibnamefont
  {Verstraete}}, \bibinfo {author} {\bibfnamefont {Matthew~B.}\ \bibnamefont
  {Hastings}}, \ and\ \bibinfo {author} {\bibfnamefont {J.~Ignacio}\
  \bibnamefont {Cirac}},\ }\bibfield  {title} {\enquote {\bibinfo {title} {Area
  {{Laws}} in {{Quantum Systems}}: {{Mutual Information}} and
  {{Correlations}}},}\ }\href {\doibase 10.1103/PhysRevLett.100.070502}
  {\bibfield  {journal} {\bibinfo  {journal} {Physical Review Letters}\
  }\textbf {\bibinfo {volume} {100}} (\bibinfo {year} {2008}),\
  10.1103/PhysRevLett.100.070502}\BibitemShut {NoStop}%
\bibitem [{\citenamefont {Eisert}\ \emph {et~al.}(2010)\citenamefont {Eisert},
  \citenamefont {Cramer},\ and\ \citenamefont
  {Plenio}}]{eisertColloquiumAreaLaws2010}%
  \BibitemOpen
  \bibfield  {author} {\bibinfo {author} {\bibfnamefont {J.}~\bibnamefont
  {Eisert}}, \bibinfo {author} {\bibfnamefont {M.}~\bibnamefont {Cramer}}, \
  and\ \bibinfo {author} {\bibfnamefont {M.~B.}\ \bibnamefont {Plenio}},\
  }\bibfield  {title} {\enquote {\bibinfo {title} {{\emph{Colloquium}} :
  {{Area}} laws for the entanglement entropy},}\ }\href {\doibase
  10.1103/RevModPhys.82.277} {\bibfield  {journal} {\bibinfo  {journal} {Rev.
  Mod. Phys.}\ }\textbf {\bibinfo {volume} {82}},\ \bibinfo {pages} {277--306}
  (\bibinfo {year} {2010})}\BibitemShut {NoStop}%
\bibitem [{\citenamefont {Chen}\ \emph {et~al.}(2002)\citenamefont {Chen},
  \citenamefont {Fu}, \citenamefont {Ungar},\ and\ \citenamefont
  {Zhao}}]{chenAlternativeFidelityMeasure2002}%
  \BibitemOpen
  \bibfield  {author} {\bibinfo {author} {\bibfnamefont {Jing-Ling}\
  \bibnamefont {Chen}}, \bibinfo {author} {\bibfnamefont {Libin}\ \bibnamefont
  {Fu}}, \bibinfo {author} {\bibfnamefont {Abraham~A.}\ \bibnamefont {Ungar}},
  \ and\ \bibinfo {author} {\bibfnamefont {Xian-Geng}\ \bibnamefont {Zhao}},\
  }\bibfield  {title} {\enquote {\bibinfo {title} {Alternative fidelity measure
  between two states of an {{N}} -state quantum system},}\ }\href {\doibase
  10.1103/PhysRevA.65.054304} {\bibfield  {journal} {\bibinfo  {journal} {Phys.
  Rev. A}\ }\textbf {\bibinfo {volume} {65}},\ \bibinfo {pages} {054304}
  (\bibinfo {year} {2002})}\BibitemShut {NoStop}%
\bibitem [{\citenamefont {Mendon{\c c}a}\ \emph {et~al.}(2008)\citenamefont
  {Mendon{\c c}a}, \citenamefont {Napolitano}, \citenamefont {Marchiolli},
  \citenamefont {Foster},\ and\ \citenamefont
  {Liang}}]{mendoncaAlternativeFidelityMeasure2008}%
  \BibitemOpen
  \bibfield  {author} {\bibinfo {author} {\bibfnamefont {Paulo E. M.~F.}\
  \bibnamefont {Mendon{\c c}a}}, \bibinfo {author} {\bibfnamefont {Reginaldo
  d.~J.}\ \bibnamefont {Napolitano}}, \bibinfo {author} {\bibfnamefont
  {Marcelo~A.}\ \bibnamefont {Marchiolli}}, \bibinfo {author} {\bibfnamefont
  {Christopher~J.}\ \bibnamefont {Foster}}, \ and\ \bibinfo {author}
  {\bibfnamefont {Yeong-Cherng}\ \bibnamefont {Liang}},\ }\bibfield  {title}
  {\enquote {\bibinfo {title} {Alternative fidelity measure between quantum
  states},}\ }\href {\doibase 10.1103/PhysRevA.78.052330} {\bibfield  {journal}
  {\bibinfo  {journal} {Phys. Rev. A}\ }\textbf {\bibinfo {volume} {78}},\
  \bibinfo {pages} {052330} (\bibinfo {year} {2008})}\BibitemShut {NoStop}%
\bibitem [{\citenamefont {Pucha{\l}a}\ and\ \citenamefont
  {Miszczak}(2009)}]{puchalaBoundTraceDistance2009}%
  \BibitemOpen
  \bibfield  {author} {\bibinfo {author} {\bibfnamefont {Zbigniew}\
  \bibnamefont {Pucha{\l}a}}\ and\ \bibinfo {author} {\bibfnamefont
  {Jaros{\l}aw~Adam}\ \bibnamefont {Miszczak}},\ }\bibfield  {title} {\enquote
  {\bibinfo {title} {Bound on trace distance based on superfidelity},}\ }\href
  {\doibase 10.1103/PhysRevA.79.024302} {\bibfield  {journal} {\bibinfo
  {journal} {Phys. Rev. A}\ }\textbf {\bibinfo {volume} {79}},\ \bibinfo
  {pages} {024302} (\bibinfo {year} {2009})}\BibitemShut {NoStop}%
\bibitem [{\citenamefont {Miszczak}\ \emph {et~al.}(2008)\citenamefont
  {Miszczak}, \citenamefont {Pucha{\l}a}, \citenamefont {Horodecki},
  \citenamefont {Uhlmann},\ and\ \citenamefont
  {Zyczkowski}}]{miszczakSubSuperFidelity2008}%
  \BibitemOpen
  \bibfield  {author} {\bibinfo {author} {\bibfnamefont {J.~A.}\ \bibnamefont
  {Miszczak}}, \bibinfo {author} {\bibfnamefont {Z.}~\bibnamefont
  {Pucha{\l}a}}, \bibinfo {author} {\bibfnamefont {P.}~\bibnamefont
  {Horodecki}}, \bibinfo {author} {\bibfnamefont {A.}~\bibnamefont {Uhlmann}},
  \ and\ \bibinfo {author} {\bibfnamefont {K.}~\bibnamefont {Zyczkowski}},\
  }\href@noop {} {\enquote {\bibinfo {title} {Sub- and super-fidelity as bounds
  for quantum fidelity},}\ } (\bibinfo {year} {2008}),\ \Eprint
  {http://arxiv.org/abs/0805.2037} {arXiv:0805.2037 [quant-ph]} \BibitemShut
  {NoStop}%
\bibitem [{\citenamefont {Tupitsyn}\ \emph {et~al.}(2010)\citenamefont
  {Tupitsyn}, \citenamefont {Kitaev}, \citenamefont {Prokof'ev},\ and\
  \citenamefont {Stamp}}]{tupitsynTopologicalMulticriticalPoint2010}%
  \BibitemOpen
  \bibfield  {author} {\bibinfo {author} {\bibfnamefont {I.~S.}\ \bibnamefont
  {Tupitsyn}}, \bibinfo {author} {\bibfnamefont {A.}~\bibnamefont {Kitaev}},
  \bibinfo {author} {\bibfnamefont {N.~V.}\ \bibnamefont {Prokof'ev}}, \ and\
  \bibinfo {author} {\bibfnamefont {P.~C.~E.}\ \bibnamefont {Stamp}},\
  }\bibfield  {title} {\enquote {\bibinfo {title} {Topological multicritical
  point in the phase diagram of the toric code model and three-dimensional
  lattice gauge {{Higgs}} model},}\ }\href {\doibase
  10.1103/PhysRevB.82.085114} {\bibfield  {journal} {\bibinfo  {journal} {Phys.
  Rev. B}\ }\textbf {\bibinfo {volume} {82}},\ \bibinfo {pages} {085114}
  (\bibinfo {year} {2010})}\BibitemShut {NoStop}%
\bibitem [{\citenamefont {Trebst}\ \emph {et~al.}(2007)\citenamefont {Trebst},
  \citenamefont {Werner}, \citenamefont {Troyer}, \citenamefont {Shtengel},\
  and\ \citenamefont {Nayak}}]{trebstBreakdownTopologicalPhase2007}%
  \BibitemOpen
  \bibfield  {author} {\bibinfo {author} {\bibfnamefont {Simon}\ \bibnamefont
  {Trebst}}, \bibinfo {author} {\bibfnamefont {Philipp}\ \bibnamefont
  {Werner}}, \bibinfo {author} {\bibfnamefont {Matthias}\ \bibnamefont
  {Troyer}}, \bibinfo {author} {\bibfnamefont {Kirill}\ \bibnamefont
  {Shtengel}}, \ and\ \bibinfo {author} {\bibfnamefont {Chetan}\ \bibnamefont
  {Nayak}},\ }\bibfield  {title} {\enquote {\bibinfo {title} {Breakdown of a
  {{Topological Phase}}: {{Quantum Phase Transition}} in a {{Loop Gas Model}}
  with {{Tension}}},}\ }\href {\doibase 10.1103/PhysRevLett.98.070602}
  {\bibfield  {journal} {\bibinfo  {journal} {Phys. Rev. Lett.}\ }\textbf
  {\bibinfo {volume} {98}},\ \bibinfo {pages} {070602} (\bibinfo {year}
  {2007})}\BibitemShut {NoStop}%
\bibitem [{\citenamefont {Wu}\ \emph {et~al.}(2012)\citenamefont {Wu},
  \citenamefont {Deng},\ and\ \citenamefont
  {Prokof'ev}}]{wuPhaseDiagramToric2012}%
  \BibitemOpen
  \bibfield  {author} {\bibinfo {author} {\bibfnamefont {Fengcheng}\
  \bibnamefont {Wu}}, \bibinfo {author} {\bibfnamefont {Youjin}\ \bibnamefont
  {Deng}}, \ and\ \bibinfo {author} {\bibfnamefont {Nikolay}\ \bibnamefont
  {Prokof'ev}},\ }\bibfield  {title} {\enquote {\bibinfo {title} {Phase diagram
  of the toric code model in a parallel magnetic field},}\ }\href {\doibase
  10.1103/PhysRevB.85.195104} {\bibfield  {journal} {\bibinfo  {journal} {Phys.
  Rev. B}\ }\textbf {\bibinfo {volume} {85}},\ \bibinfo {pages} {195104}
  (\bibinfo {year} {2012})}\BibitemShut {NoStop}%
\bibitem [{\citenamefont {{Schuler}}\ \emph {et~al.}(2016)\citenamefont
  {{Schuler}}, \citenamefont {{Whitsitt}}, \citenamefont {{Henry}},
  \citenamefont {{Sachdev}},\ and\ \citenamefont
  {{L{\"a}uchli}}}]{Lauchlitoriccode}%
  \BibitemOpen
  \bibfield  {author} {\bibinfo {author} {\bibfnamefont {Michael}\ \bibnamefont
  {{Schuler}}}, \bibinfo {author} {\bibfnamefont {Seth}\ \bibnamefont
  {{Whitsitt}}}, \bibinfo {author} {\bibfnamefont {Louis-Paul}\ \bibnamefont
  {{Henry}}}, \bibinfo {author} {\bibfnamefont {Subir}\ \bibnamefont
  {{Sachdev}}}, \ and\ \bibinfo {author} {\bibfnamefont {Andreas~M.}\
  \bibnamefont {{L{\"a}uchli}}},\ }\bibfield  {title} {\enquote {\bibinfo
  {title} {{Universal Signatures of Quantum Critical Points from Finite-Size
  Torus Spectra: A Window into the Operator Content of Higher-Dimensional
  Conformal Field Theories}},}\ }\href {\doibase
  10.1103/PhysRevLett.117.210401} {\bibfield  {journal} {\bibinfo  {journal}
  {\prl}\ }\textbf {\bibinfo {volume} {117}},\ \bibinfo {eid} {210401}
  (\bibinfo {year} {2016})},\ \Eprint {http://arxiv.org/abs/1603.03042}
  {arXiv:1603.03042 [cond-mat.str-el]} \BibitemShut {NoStop}%
\bibitem [{\citenamefont {Bradbury}\ \emph {et~al.}(2018)\citenamefont
  {Bradbury}, \citenamefont {Frostig}, \citenamefont {Hawkins}, \citenamefont
  {Johnson}, \citenamefont {Leary}, \citenamefont {Maclaurin}, \citenamefont
  {Necula}, \citenamefont {Paszke}, \citenamefont {Vander{P}las}, \citenamefont
  {Wanderman-{M}ilne},\ and\ \citenamefont {Zhang}}]{jax2018github}%
  \BibitemOpen
  \bibfield  {author} {\bibinfo {author} {\bibfnamefont {James}\ \bibnamefont
  {Bradbury}}, \bibinfo {author} {\bibfnamefont {Roy}\ \bibnamefont {Frostig}},
  \bibinfo {author} {\bibfnamefont {Peter}\ \bibnamefont {Hawkins}}, \bibinfo
  {author} {\bibfnamefont {Matthew~James}\ \bibnamefont {Johnson}}, \bibinfo
  {author} {\bibfnamefont {Chris}\ \bibnamefont {Leary}}, \bibinfo {author}
  {\bibfnamefont {Dougal}\ \bibnamefont {Maclaurin}}, \bibinfo {author}
  {\bibfnamefont {George}\ \bibnamefont {Necula}}, \bibinfo {author}
  {\bibfnamefont {Adam}\ \bibnamefont {Paszke}}, \bibinfo {author}
  {\bibfnamefont {Jake}\ \bibnamefont {Vander{P}las}}, \bibinfo {author}
  {\bibfnamefont {Skye}\ \bibnamefont {Wanderman-{M}ilne}}, \ and\ \bibinfo
  {author} {\bibfnamefont {Qiao}\ \bibnamefont {Zhang}},\ }\href
  {http://github.com/google/jax} {\enquote {\bibinfo {title} {{JAX}: composable
  transformations of {P}ython+{N}um{P}y programs},}\ } (\bibinfo {year}
  {2018})\BibitemShut {NoStop}%
\bibitem [{\citenamefont {Metropolis}\ \emph {et~al.}(1953)\citenamefont
  {Metropolis}, \citenamefont {Rosenbluth}, \citenamefont {Rosenbluth},
  \citenamefont {Teller},\ and\ \citenamefont
  {Teller}}]{metropolisEquationStateCalculations1953}%
  \BibitemOpen
  \bibfield  {author} {\bibinfo {author} {\bibfnamefont {Nicholas}\
  \bibnamefont {Metropolis}}, \bibinfo {author} {\bibfnamefont {Arianna~W.}\
  \bibnamefont {Rosenbluth}}, \bibinfo {author} {\bibfnamefont {Marshall~N.}\
  \bibnamefont {Rosenbluth}}, \bibinfo {author} {\bibfnamefont {Augusta~H.}\
  \bibnamefont {Teller}}, \ and\ \bibinfo {author} {\bibfnamefont {Edward}\
  \bibnamefont {Teller}},\ }\bibfield  {title} {\enquote {\bibinfo {title}
  {Equation of {{State Calculations}} by {{Fast Computing Machines}}},}\ }\href
  {\doibase 10.1063/1.1699114} {\bibfield  {journal} {\bibinfo  {journal} {J.
  Chem. Phys.}\ }\textbf {\bibinfo {volume} {21}},\ \bibinfo {pages}
  {1087--1092} (\bibinfo {year} {1953})}\BibitemShut {NoStop}%
\bibitem [{\citenamefont {Kingma}\ and\ \citenamefont
  {Ba}(2014)}]{kingma2014adam}%
  \BibitemOpen
  \bibfield  {author} {\bibinfo {author} {\bibfnamefont {Diederik~P}\
  \bibnamefont {Kingma}}\ and\ \bibinfo {author} {\bibfnamefont {Jimmy}\
  \bibnamefont {Ba}},\ }\bibfield  {title} {\enquote {\bibinfo {title} {Adam: A
  method for stochastic optimization},}\ }\href
  {https://arxiv.org/abs/1412.6980} {\bibfield  {journal} {\bibinfo  {journal}
  {arXiv preprint arXiv:1412.6980}\ } (\bibinfo {year} {2014})}\BibitemShut
  {NoStop}%
\end{thebibliography}%
\end{document}